\documentclass[letterpaper,11pt]{article}
\usepackage[utf8]{inputenc}

\pdfoutput=1
\usepackage{jheppub}
\usepackage{graphicx,verbatim}
\usepackage{blindtext}
\usepackage[T1]{fontenc}
\usepackage{epsf}
\usepackage{mathtools}
\usepackage{lmodern}
\usepackage{dsfont}
\usepackage[mathscr]{euscript}

\usepackage{amsmath}
\usepackage{amsfonts}
\usepackage{amssymb}
\usepackage{mathrsfs}
\usepackage{slashed}
\usepackage{xcolor}
\usepackage{caption}
\usepackage{subcaption}
\definecolor{green}{RGB}{0,255,0}

\usepackage{ytableau,tikz,varwidth}
\usetikzlibrary{calc}
\usepackage{tikz-cd}
\usepackage{caption}
\usetikzlibrary{shapes.geometric}
\usetikzlibrary{decorations.markings}
\tikzset{->-/.style={decoration={
  markings,
  mark=at position #1 with {\arrow{>}}},postaction={decorate}}}

\def\qe{\mathfrak{q}}
\def\kq{\mathfrak{q}}

\def\fR{\mathfrak{R}}

\def\fgl{\mathfrak{gl}}

\def\fsl{\mathfrak{sl}}

\def\ii{\mathrm{i}}
\def\ri{\mathrm{i}}

\def\by{\mathbf{y}}
\def\bx{\mathbf{x}}
\def\ba{\mathbf{a}}

\def\bt{\mathbf{t}}
\def\bu{\mathbf{u}}

\def\bL{\mathbf{L}}

\def\BC{\mathbb{C}}
\def\BR{\mathbb{R}}
\def\BE{\mathbb{E}}
\def\BZ{\mathbb{Z}}

\def\BP{\mathbb{P}}

\def\CalR{\mathcal{R}}
\def\CalZ{\mathcal{Z}}
\def\CalH{\mathcal{H}}
\def\CalW{\mathcal{W}}

\def\CalS{\mathcal{S}}
\def\CalC{\mathcal{C}}

\def\qe{\mathfrak{q}}

\def\CalK{\mathcal{K}}

\def\ve{{\varepsilon}}

\def\EQ{\EuScript{Q}}

\def\EN{\EuScript{N}}
\def\EY{\EuScript{Y}}

\def\EB{\EuScript{B}}

\def\EM{\EuScript{M}}

\def\EC{\EuScript{C}}

\def\ED{\EuScript{D}}

 \def\p{\partial}
 
 \def\a{\alpha}
 \def\b{\beta}
 \def\g{\gamma}
 \def\d{\delta}

 \def\th{\theta}
 
 \def\k{\kappa}
 \def\l{\lambda}
 \def\m{\mu}

 \def\r{\rho}

 \def\s{\sigma}
 \def\t{\tau}
 \def\th{\theta}
 \def\z{\zeta }
 
 \def\G{\Gamma}
 \def\D{\Delta}

 \def\S{\Sigma}
 \def\L{\Lambda}
 \def\O{\Omega}
 \def\o{\omega }

 \def\bl{\boldsymbol{\lambda}}
 \def\bL{\boldsymbol{\Lambda}}

 \makeatletter
\newsavebox{\@brx}
\newcommand{\llangle}[1][]{\savebox{\@brx}{\(\m@th{#1\langle}\)}%
  \mathopen{\copy\@brx\kern-0.5\wd\@brx\usebox{\@brx}}}
\newcommand{\rrangle}[1][]{\savebox{\@brx}{\(\m@th{#1\rangle}\)}%
  \mathclose{\copy\@brx\kern-0.5\wd\@brx\usebox{\@brx}}}
\makeatother

\def\beq{\begin{equation}}
\def\eeq{\end{equation}}

\title{{Bispectral duality and separation of variables\\from surface defect transition}}
\author[a]{Saebyeok Jeong}
\author[b]{and Norton Lee}

\affiliation[a]{Department of Theoretical Physics, CERN, \\ 1211 Geneva 23, Switzerland}
\affiliation[b]{Center for Geometry and Physics, Institute for Basic Science (IBS),\\Pohang 37673, Republic of Korea}

\emailAdd{saebyeok.jeong@cern.ch}
\emailAdd{nortonxs@gmail.com}

\preprint{CERN-TH-2024-024, CGP24003}

\vspace{2cm}
\abstract{We study two types of surface observables $-$ the $\mathbf{Q}$-observables and the $\mathbf{H}$-observables $-$ of the 4d $\EN=2$ $A_1$-quiver $U(N)$ gauge theory obtained by coupling a 2d $\EN=(2,2)$ gauged linear sigma model. We demonstrate that the transition between the two surface defects manifests as a Fourier transformation between the surface observables. Utilizing the results from our previous works, which establish that the $\mathbf{Q}$-observables and the $\mathbf{H}$-observables give rise, respectively, to the $Q$-operators on the evaluation module over the Yangian $Y(\mathfrak{gl}(2))$ and the Hecke operators on the twisted $\widehat{\mathfrak{sl}}(N)$-coinvariants, we derive an exact duality between the spectral problems of the $\fgl(2)$ XXX spin chain with $N$ sites and the $\fsl(N)$ Gaudin model with 4 sites, both of which are defined on bi-infinite modules. Moreover, we present a dual description of the monodromy surface defect as coupling a 2d $\EN=(2,2)$ gauged linear sigma model. Employing this dual perspective, we demonstrate how the monodromy surface defect undergoes a transition to multiple $\mathbf{Q}$-observables or $\mathbf{H}$-observables, implemented through integral transformations between their surface observables. These transformations provide, respectively, $\hbar$-deformation and a higher-rank generalization of the KZ/BPZ correspondence. In the limit $\ve_2\to 0$, they give rise to the quantum separation of variables for the $\fgl(2)$ XXX spin chain and the $\fsl(N)$ Gaudin model, respectively.}

\begin{document}
\maketitle

\section{Introduction}
In this paper, we explain how transitions between surface defects in the 4d $\EN=2$ gauge theory reveal new features of its correspondence with a quantum integrable system. We consider the $\EN=2$ theory of class $\CalS$ \cite{gai1} of type $A_{N-1}$ associated with the cylinder $\BC^\times = \BP^1 \setminus \{0,\infty\}$, with $n$ marked points $S = \{p_1,p_2,\cdots, p_n \} \subset \BC^\times$. Here, we choose the marked point data such that the $\EN=2$ theory has a Lagrangian description as the linear $A_{n-1}$-quiver $SU(N)$ gauge theory in an appropriate S-duality frame. The associated integrable system is defined on the moduli space $\EM_H (SU(N), \BP^1; S)$ of Hitchin's equations on $\BP^1$ with (regular) ramification data at $S \cup \{0,\infty\}$ \cite{Gaiotto:2009hg}. We primarily restrict to the case $n=2$ (keeping $N$ arbitrary), but we set $n$ generic in the introduction to provide a comprehensive overview.

We study this $\EN=2$ gauge theory subject to the Donaldson-Witten twist \cite{cmp/1104161738} and further to the $\Omega_{\ve_1,\ve_2}$-deformation associated with the $U(1)^2$ isometry of spacetime $\BC^2$ \cite{Nekrasov:2002qd}. Compactifying along the torus fiber of $\BC^2 \simeq \S \tilde{\times} T^2$, we arrive at a two-dimensional $\EN=(2,2)$ sigma model of maps from the worldsheet $\S$ to the Hitchin moduli space target $\EM_H (SU(N),\BP^1;S)$ \cite{Bershadsky:1995qy,Harvey:1995tg}. In the cohomology of the preserved supercharge, the sigma model is either an $A$-model associated with a symplectic structure or a $B$-model associated with a complex structure, determined by the ratio $\k = -\frac{\ve_2}{\ve_1}$ of the $\O$-background parameters \cite{Nekrasov:2010ka,Jeong:2023qdr}.

In the context of the topological sigma model, the quantization of the moduli space $\EM_H (SU(N),\BP^1;S)$ is realized by open strings stretched between $A$-branes \cite{Gukov:2008ve}. The $(\EB_{cc},\EB_{cc})$-strings, where $\EB_{cc}$ is the space-filling canonical coisotropic brane \cite{Kapustin:2001ij}, form the twisted differential operators on the moduli space $\text{Bun}_{PGL(N)} (\BP^1;S)$ of parabolic $PGL(N)$-bundles. They provide a non-commutative deformation of the algebra of holomorphic functions on $\EM_H (SU(N),\BP^1;S)$. Then, any $A$-brane defines a twisted $\ED$-module on $\text{Bun}_{PGL(N)} (\BP^1;S)$, acted on by this non-commutative algebra through the action of joining the open strings.

In our previous work \cite{Jeong:2023qdr}, we studied how half-BPS surface defects in the $\EN=2$ gauge theory are associated with certain $A$-branes and functors acting on them. In particular, the regular monodromy surface defect descends to the brane of $\l$-connections, while the canonical surface defect, which we will refer to as the $\mathbf{H}$-observable in this work, leads to the Hecke operator. The \textit{magnetic} eigenbrane on which the Hecke operators act \textit{diagonally} was revealed to have its origin in the boundary condition at infinity.\footnote{It is called the magnetic eigenbrane since the Hecke operators are realized by the 't Hooft line defects in the GL-twisted $\EN=4$ theory \cite{Kapustin:2006pk}.} From the perspective of the geometric Langlands correspondence, these eigenbranes are associated with special $SL(N)$-local systems called opers \cite{BD1,BD2}, and we showed how these opers are parametrized by the Coulomb moduli of the $\EN=2$ theory \cite{Jeong:2018qpc,Jeong:2023qdr}. We showed that the vacuum expectation value of the regular monodromy surface defect in the limit $\ve_2 \to 0$ provides a distinguished basis for the sections of the twisted $\widehat{\fsl}(N)$-coinvariants on $\text{Bun}_{PGL(N)}(\BP^1;S)$, on which the Hecke operators act diagonally. We also proved that the same vev of the regular monodromy surface defect is a common eigenfunction of the quantum Hamiltonians of the associated $\fsl(N)$ Gaudin model. We suggested a string duality that relates our $\EN=2$ gauge theoretical formulation of the geometric Langlands correspondence (with ramifications) \cite{BD1,BD2,Frenkel:2005ef,Frenkel:2006nm} to the more conventional approach from the topologically twisted (GL-twisted) $\EN=4$ gauge theory \cite{Kapustin:2006pk,gukwit2,Gaiotto:2016hvd,Frenkel:2018dej}. See also \cite{Frenkel:2005pa,Teschner:2010je,Balasubramanian:2017gxc,Teschner:2017djr} for the vertex algebra approach to the geometric Langlands correspondence.

Meanwhile, the same Hitchin integrable system is approached from a dual perspective in the companion paper \cite{Jeong:2024hwf}. The moduli space $\EM_H (SU(N),\BP^1;S)$ is known to be isomorphic, as a hyper-K\"{a}hler space, to the moduli space of periodic $U(n)$-monopoles on $\BP^1 \times S^1$ with a framing at $\infty \in \BP^1$ and Dirac singularities at $D \times \{0\}$, where $D = \{m^+_0,m^+ _1,\cdots, m^+ _{N-1} \} \subset \BP^1 \setminus\{\infty\} = \BC$, related to each other by a Nahm transform \cite{Cherkis:2000cj,Cherkis:2000ft,Nekrasov:2012xe,Nikita-Pestun-Shatashvili,Elliott:2018yqm}. In one of the complex structures, it is viewed as the moduli space $\EM_{\text{mHiggs}} (GL(n),\BP^1;D)$ of \textit{multiplicative} $GL(n)$-Higgs bundles on $\BP^1$ with a framing at $\infty \in \BP^1$ and regular singularities at $D$ \cite{Cherkis:2000cj,charbonneau2011,Elliott:2018yqm}. The quantization of $\EM_{\text{mHiggs}} (GL(n),\BP^1;D)$ is known to lead to a representation of the Yangian of $\fgl(n)$, which we denote by $Y(\fgl(n))$ \cite{Nekrasov:2012xe,Nikita-Pestun-Shatashvili,Elliott:2018yqm}.

In the companion paper \cite{Jeong:2024hwf}, we found that the vacuum expectation value of the regular monodromy surface defect gives a distinguished basis for the evaluation module over the Yangian $Y(\fgl(n))$. We showed that the $\mathbf{Q}$-observable, defined as a surface defect in the $\EN=2$ gauge theory, acts on this Yangian module as the $Q$-operator. In particular, we derived the universal $\hbar$-oper difference equation (equivalently, Baxter's TQ equation) satisfied by the $\mathbf{Q}$-observables using the constraints of the surface defect correlation functions. The action of the $Q$-operators on the distinguished basis was shown to be diagonal due to the cluster decomposition of the surface defects. Combined with the universal $\hbar$-oper equation, it followed that the $Q$-eigenstates, constructed as the vevs of the regular monodromy surface defect, are also the common eigenstates of the quantum Hamiltonians of the $\fgl(n)$ XXX spin chain. This statement was regarded as the $\hbar$-deformation of the geometric Langlands correspondence (restricted to the locus of ($\hbar$-)opers) \cite{BD1,BD2}, called the \textit{$\hbar$-Langlands correspondence}.\footnote{The $\hbar$-Langlands correspondence (additive) is closely related to but different from the $q$-Langlands correspondence (multiplicative), since the latter is relevant to the moduli space of multiplicative Higgs bundles on $\BP^1$ with framing at $0,\infty \in \BP^1$. See \cite{frenkel1996,Aganagic:2017smx,Koroteev:2018jht, Frenkel:2020iqq, Haouzi:2023doo} for previous works on the $q$-Langlands correspondence.}

\subsection{$Q$-operator/Hecke operator transition and bispectral duality}
Let us revisit the isomorphism between the hyper-K\"{a}hler target spaces at the level of the topological sigma model. We note that any $A$-brane $\EB'$, which provides a twisted $\ED$-module on $\text{Bun}_{PGL(N)}(\BP^1;S)$, is expected to give a \textit{$\hbar$-difference module} on $\text{Bun}_{GL(n)}(\BP^1;D)$, since they are two dual interpretations of the same $(\EB_{cc},\EB')$-strings acted on by the $(\EB_{cc},\EB_{cc})$-strings, which quantize the algebra of holomorphic functions on the same moduli space.

Our findings in the companion paper \cite{Jeong:2024hwf} give an example of such a duality, realized at specific $A$-branes descending from the regular monodromy surface defect and the boundary condition at infinity of the $\EN=2$ gauge theory. In the context of the ordinary Langlands correspondence, the regular monodromy surface defect (at fixed monodromy parameters) gives rise to the $E$-twisted coinvariants of $\widehat{\fsl}(N)$-modules, where $E \in \text{Bun}_{PGL(N)}(\BP^1;S)$ is determined by the monodromy parameters \cite{Jeong:2023qdr}. In the context of the $\hbar$-Langlands correspondence, we explained in the companion paper \cite{Jeong:2024hwf} that it leads to elements in an evaluation module $\CalH$ over the Yangian $Y(\fgl(n))$. Moreover, for the magnetic eigenbrane $\textbf{F}_\ba$ descending from the boundary condition $\ba$ at infinity, the section of $(\EB_{cc},\textbf{F}_\ba)$-strings is obtained as the vacuum expectation value of the regular monodromy surface defect at the vacuum $\ba$ in the limit $\ve_2 \to 0$, with varying monodromy parameters. It is a section of the Hecke eigensheaf and a $Q$-eigenstate at the same time, corresponding to a $SL(N)$ oper and a $GL(n)$ $\hbar$-oper, respectively, determined by the same Coulomb moduli $\ba$.

Therefore, it is natural to investigate the relation between the $Q$-operator and the Hecke operator, since their eigen-object is realized by the same $A$-brane. This is the first goal of the present work. We describe a transition between the $Q$-operator and the Hecke operator using their realization as surface defects in the $\EN=2$ gauge theory, and study its implications for the associated integrable models.

Note that the $Q$-operator, which is the $\hbar$-analogue of the Hecke operator, is realized as a surface defect in the $\EN=2$ gauge theory created by coupling a two-dimensional $\EN=(2,2)$ gauged linear sigma model with the vev of the complex adjoint scalar turned on \cite{Jeong:2024hwf}. Meanwhile, recall that in our $\EN=2$ gauge theoretical formulation of the ordinary Langlands correspondence \cite{Jeong:2023qdr}, the Hecke operator follows from the $\mathbf{H}$-observable, which is defined by coupling the two-dimensional gauged linear sigma model with the same field contents but in the non-linear sigma model phase instead. Thus, we explain the relation between the $Q$-operator and the Hecke operator through a transition in the surface defect theory, where the two observables are distinguished by turning on either the vev of the complex adjoint scalar or the complexified FI parameter. This surface defect transition manifests as a Fourier transform between their observable expressions \cite{Jeong:2018qpc,Jeong:2023qdr}, which schematically reads
\begin{align}
    \mathbf{H}(y) = \oint_{\CalC} dy\, y^{-\frac{x}{\ve_1}} \mathbf{Q}(x), \qquad  y \in \BC^\times \setminus S.
\end{align}
This equation directly connects the $Q$-operator and the Hecke operator in their gauge theoretical realization. We emphasize that the contour $\CalC$ is explicitly given to pick up certain semi-infinite $\ve_1$-lattice of simple poles in the integrand. There are $N$ choices for the contour, corresponding to the $N$ choices of the vacuum at infinity of the two-dimensional non-linear sigma model, and also to the $N$ fundamental coweights of $PGL(N)$, each of which defines a Hecke operator. Additionally, the $\mathbf{Q}$-observable in the integrand is selected from $n$ inequivalent $\mathbf{Q}$-observables according to the domain of $y \in \BC^\times \setminus S$. In particular, the Fourier transform is not just a formal one but is evaluated as a concrete convergent series. 

The transformation explains the existence of the common eigen-objects in an explicit manner. Moreover, the (universal) $GL(n)$ $\hbar$-oper equation and the (universal) $SL(N)$ oper equation satisfied by them are shown to be exchanged with each other precisely under the transformation. In terms of the associated integrable models $-$ the $\fgl(n)$ XXX spin chain defined on $N$ bi-infinite modules and the $\fsl(N)$ Gaudin model defined on two (co-)Verma modules and $n$ bi-infinite modules $-$ the transition induces an exact duality between their spectral problems, giving a one-to-one matching of their quantum Hamiltonians as well as their quantum spectra in the $\EN=2$ gauge theoretical terms.

\subsection{Surface defect transition and separation of variables}
Next, we move on to another point of view on the moduli space $\EM_H (SU(N),\BP^1;S)$ of Hitchin's equations. When viewed as the moduli space of parabolic Higgs bundles $\EM_{\text{Higgs}} (PGL(N),\BP^1;S)$, it is birational to the affine deformation of the cotangent bundle over $\text{Bun}_{PGL(N)}(\BP^1;S)$. We observe that there is also another birational map \cite{Gorsky:1999rb}
\begin{align}
     \EM_{\text{Higgs}} (PGL(N),\BP^1;S) \longrightarrow (T^* \BC^\times  )^{[(n-1)(N-1)]},
\end{align}
which is holomorphic symplectic in open dense subsets. Here, $[(n-1)(N-1)]$ denotes the Hilbert scheme of points of length $(n-1)(N-1)$. Importantly, the map provides the classical separation of variables for the associated integrable model.

When we regard $(T^* \BC^\times)^{[(n-1)(N-1)]}$ as the target space of the topological sigma model compatible with the birational map ($\ve_2 = 0$ and $\k = 0$; the $A$-model in the symplectic structure $\o_K$ in the convention of \cite{Kapustin:2006pk}), the $A$-branes now give rise to $\ED$-modules on $\text{Sym}^{(n-1)(N-1)} \BC^\times$. In particular, the counterpart of the Hecke eigensheaf $\D_{\r}$ on $\text{Bun}_{PGL(N)} (\BP^1;S)$ associated with an oper $\r$ is identified as the $\ED$-module on $\text{Sym}^{(n-1)(N-1)} \BC^\times$ defined by the oper differential equation for $\r$ \cite{Frenkel95}. The sections of the two $\ED$-modules are related by a formal integral transform \cite{Frenkel95}. In other words, the section of the Hecke eigensheaf associated with the oper $\rho$ is expressed as an integral transform of the product of $(n-1)(N-1)$ independent oper solutions. Recalling that this Hecke eigensheaf is also the twisted $\ED$-module representing the spectral equations for the quantum Gaudin Hamiltonians, we observe that the integral transform precisely induces the quantum separation of variables for the Gaudin model.\footnote{The term \textit{quantum} sometimes indicates a $\ve_2 \neq 0$ deviation, but the \textit{quantum} separation of variables here is relevant to $\ve_1 = \hbar \neq 0$ and $\ve_2 = 0$, as opposed to the classical separation of variables at $\ve_1 = \ve_2 = 0$.}

In fact, there is a dual perspective here as well. The image of the birational map can simply be viewed as $(\BC \times \BC^\times)^{[(n-1)(N-1)]}$, since the cotangent bundle on $\BC^\times$ is trivial. By exchanging the base and the fiber, the $A$-branes should lead to $\hbar$-difference modules on $\text{Sym}^{(n-1)(N-1)} \BC$. By analogy with the dual case, we expect that the $A$-brane for the $Q$-eigenstate (magnetic eigenbrane) would be related to the $\hbar$-difference module on $\text{Sym}^{(n-1)(N-1)} \BC$ representing the $\hbar$-oper difference equation. The $Q$-eigenstate would be expressed as an integral transform of the product of $(n-1)(N-1)$ independent $\hbar$-oper solutions, i.e., $Q$-functions. Recalling that a $Q$-eigenstate is also a common eigenstate of the Yangian transfer matrices represented on the relevant evaluation module, such a transformation would lead to the quantum separation of variables for the associated integrable model in the bispectral dual perspective; namely, the XXX spin chain.

The second goal of the present work is to elucidate these separation of variable transformations as transitions of the surface defect in the $\EN=2$ gauge theory; specifically, from a regular monodromy surface defect to multiple $\mathbf{Q}$-observables or $\mathbf{H}$-observables. In fact, the $\EN=2$ gauge theoretical formulation enhances the previous analysis in two aspects. Firstly, we establish the surface defect transition at generic values of $\ve_2$ (namely, $\k \neq 0$), and the integral transformations between the correlation functions are valid even away from $\ve_2 = 0$. Secondly, the integral transformations are not just formal ones but are obtained as definite contour integrals with explicit contour prescriptions.

We begin by giving a dual description of the monodromy surface defect as coupling to a two-dimensional $\EN=(2,2)$ sigma model, by which the monodromy of the fields can be thought of as the remnant of integrating out two-dimensional degrees of freedom \cite{Gukov:2014gja}. The target space of this sigma model is the total space of a vector bundle over a flag variety, and the complexified K\"{a}hler parameters are identified with the monodromy parameters. The duality exhibits the surface defect observable as an equivariant integral over a hand-saw quiver variety \cite{FR2010,Nakajima:2011yq} modified by the effect of the coupling to the four-dimensional theory. We then derive a crossing formula relating the equivariant integrals performed in two different stability chambers of the modified hand-saw quiver variety. The observable expression in the other (negative) stability chamber is shown to admit two contour integral representations. One has the product of $\mathbf{Q}$-observables in the integrand with a contour of Barnes type. The other has the product of $\mathbf{H}$-observables in the integrand with a contour of Pochhammer type. Namely, upon taking the vacuum expectation values of the two relations, we arrive at
\begin{align}
        \Big\langle \Psi(\bu) \Big\rangle_\ba = \int_{\CalC} d \bx \, \CalK(\bu;\bx) \left\langle \prod_{i=1} ^{(n-1)(N-1)} \mathbf{Q}(x_i) \right\rangle_\ba = \int_{\text{PC}} d \by \, \EuScript{K}(\bu;\by) \left\langle \prod_{i=1} ^{(n-1)(N-1)} \mathbf{H}(y_i) \right\rangle_\ba.
\end{align}
We emphasize again that the integral transformations are derived at generic values of $\ve_1$ and $\ve_2$. The correlation functions of the surface observables in the integrands are not factorized in this general case. In the second integral transformation, the vacuum expectation value of the regular monodromy surface defect is identified with a twisted coinvariant of two (co-)Verma modules and $n$ bi-infinite $\widehat{\fsl}(N)$-modules \cite{Nikita:V,Nekrasov:2020qcq,Jeong:2021bbh}, while the correlation function of the canonical surface defects on the right-hand side is a $\CalW_N$-algebra conformal block with two generic vertex operators, $n$ semi-degenerate vertex operators, and $(n-1)(N-1)$ fully degenerate vertex operators at $(y_i)_{i=1}^{(n-1)(N-1)}$ \cite{Jeong:2017pai,Jeong:2018qpc,Jeong:2020uxz,Nikita:V}. In this sense, it provides higher-rank generalizations of the KZ/BPZ correspondence studied in \cite{Frenkel95,Stoyanovsky2000ARB,Ribault:2005wp}.\footnote{The first integral transformation is expected to be interpreted as the \textit{$\hbar$-KZ/$\hbar$-BPZ correspondence}, relating the $\hbar$-conformal blocks of the quantum vertex algebras for the Yangian double and the $\hbar$-deformed $\CalW$-algebra. See section \ref{sec:discussion} for a discussion. See \cite{Nieri:2019mdl} for a $\EN=2$ theory realization of the $\hbar$-deformed $\CalW$-algebra.} See \cite{Ribault:2008si,Creutzig:2015hla} for a vertex algebra approach to the higher-rank generalization of the KZ/BPZ correspondence, and \cite{Frenkel:2015rda} for a vertex algebra approach to the surface defect transition and the separation of variables.

In the limit $\ve_2 \to 0$, the correlation functions in the integrand factorize as the product of individual vacuum expectation values due to the cluster decomposition of the surface defects, since they are local on the topological $\BC_2$-plane. Recall that in the limit $\ve_2 \to 0$, the vacuum expectation value of the $\mathbf{Q}$-observable is a $GL(n)$ $\hbar$-oper solution \cite{Jeong:2024hwf}, while the vacuum expectation value of the $\mathbf{H}$-observable is an $SL(N)$ oper solution \cite{Jeong:2018qpc,Jeong:2023qdr}. Therefore, we precisely obtain the quantum separation of variables for the XXX spin chain and the Gaudin model, respectively, from the two integral transformations. Note that the procedure provides the separation of variables not just at a formal level but is equipped with definite contour prescriptions for the integral transformations.

\subsection{Twisted M-theory perspective}
The $\EN=2$ gauge theory with its half-BPS surface defect can be constructed in the IIB string theory for the gauge origami \cite{Nikita:I,Nikita:IV,Nikita:V}. While our primary methodology involves the use of supersymmetric localization in the IIB gauge origami setup, it is instructive to consider an alternative duality frame to enrich our understanding.

It was suggested in \cite{Jeong:2023qdr} that the IIB gauge origami is related to the twisted M-theory \cite{Costello:2016nkh} by a string duality, after unrefining the $\Omega$-background so that the number of independent parameters is reduced from three to two. We are led to the twisted M-theory defined on the 11-dimensional worldvolume composed of the topological part $\BC_{\ve_1}\times \BC_{\ve_2} \times \BC_{\ve_4} \times \BR$ and the holomorphic part $\BC \times \BC^\times$. The $\Omega$-background associated with the $U(1)^2$ isometry of $\BC_{\ve_1}\times \BC_{\ve_2} \times \BC_{\ve_4}$ is turned on, with two independent parameters ($\ve_1+\ve_2+\ve_4 = 0$) to keep the supersymmetry.\footnote{We used the convention that the unrefinement $\ve_3 = 0$ was made in the IIB gauge origami. See \cite{Jeong:2023qdr}.}

To begin with, we consider two stacks of M5-branes supported on $\BC_{\ve_1} \times \BC_{\ve_2}$ in the topological directions: a stack of $N$ M5-branes wrapping $\BC^\times$ and another stack of $n$ M5-branes wrapping $\BC$ in the holomorphic directions. This is precisely the M-theory uplift of the IIA brane setup where the four-dimensional $\EN=2$ linear $A_{n-1}$-quiver $SU(N)$ gauge theory is engineered on the worldvolume of $N$ D4-branes stretched between $n$ NS5-branes and infinity \cite{Witten1997}, augmented by the $\O$-background (the first two rows in Table \ref{table:twM}). Putting differently, we build the $\EN=2$ theory of class $\CalS$ from the six-dimensional $\EN=(0,2)$ theory of type $A_{N-1}$ on the worldvolume of the $N$ M5-branes, viewing the holomorphic cylinder $\BC^\times$ as the associated Riemann surface. The other $n$ M5-branes are transverse to $\BC^\times$, located at $S = \{p_1,p_2,\cdots, p_n\} \subset \BC^\times$. They provide codimension-two defects in the six-dimensional $\EN=(0,2)$ theory, thereby determining the contents of the $\EN=2$ gauge theory. In the associated Hitchin moduli space, these defects are responsible for the ramifications at the marked points $S$ \cite{gukwit2}.

\begin{table}[h!]
    \centering
    \begin{tabular}{ c||c|c|c|c|c|c|c|c|c|c||c } 
        \text{IIA/M branes} & 0 & 1 & 2 & 3 & 4 & 5 & 6 & 7 & 8 & 9 & 10 \\ \hline\hline
         D4/M5 & \rm{x} & \rm{x}& \rm{x}& \rm{x} & \rm{x} & & & & & & \rm{x} \\  NS5/M5 & \rm{x} & \rm{x} & \rm{x} & \rm{x} & & & & & \rm{x}  & \rm{x}   & \\ \hline KK5${}_l$/KK6${}_l$ & \rm{x} & \rm{x} & &  &\rm{x}& \rm{x}  & & & \rm{x} &  \rm{x} &\rm{x} \\ D2/M2 & \rm{x} & \rm{x} & & &  & \rm{x} && &&
    \end{tabular}
    \caption{Twisted M-theory with M-branes and $A_{l-1}$ singularity}
    \label{table:twM}
\end{table}

\begin{table}[h!]
    \centering
    \begin{tabular}{c|c|c|c|c|c}
        $\BC_{\ve_1}$ & $\BC_{\ve_2}$ & $\BC_{\ve_4}$ &  $\mathbb{R}$ & $\BC$ & $ \BC^\times$   \\ \hline
        $x^0,x^1$ & $x^2,x^3$ & $x^6, x^7$ & $x^5$  & $x^8,x^9 $ &  $x^4, x^{10}$ 
    \end{tabular}
    \caption{Spacetime of the twisted M-theory}
    \label{table:spacetime}
\end{table}

The moduli space of multiplicative Higgs bundles emerges in a T-dual IIB frame \cite{Cherkis:2000cj,Cherkis:2000ft}. We may compactify $x^3$ and T-dualize along this direction (at least in the absence of the $\O$-background). In turn, there are $N$ D3-branes suspended between $n$ NS5-branes and infinity. The D3-branes stretched between NS5-branes precisely give periodic monopoles in the effective theory on the transverse part $\BC \times S^1$ of the worldvolume of the NS5-branes. The $N$ D3-branes stretched between the NS5-brane and infinity give Dirac singularities at their positions $D= \{m_0 ^+ ,m_1 ^+,\cdots, m^+_{N-1} \} \subset \BC$.  This duality, which exchanges the underlying effective theory between the one on the $N$ M5-branes wrapping $\BC^\times$ and the other on the $n$ M5-branes wrapping $\BC$, conceptually explains the isomorphism between the moduli space of Higgs bundles on $\BC^\times $ and the moduli space of multiplicative Higgs bundles on $\BC$ with the rank of the bundle and the number of regular singularities swapped between $N$ and $n$.

Next, let us turn to the surface defects in the class $\CalS$ theory. There are two different origins of the surface defects in the twisted M-theory setup: the $A_{l-1}$-singularity and the M2-branes. In this work, we consider M2-branes wrapping $\BC_{\ve_1} \times \BR$, which are local on the holomorphic planes $\BC \times \BC^\times$ (the fourth row of Table \ref{table:twM}). They can end on either stack of the M5-branes, the one supported on $\BC^\times$ or the one supported on $\BC$. We refer to the surface defect constructed from the former (resp. the latter) as the $\mathbf{H}$-observable (resp. the $\mathbf{Q}$-observable). They respectively give rise to the Hecke operator and the $Q$-operator in our $\EN=2$ gauge theoretical implementation of the ($\hbar$-)Langlands correspondence, as we extensively study in \cite{Jeong:2023qdr,Jeong:2024hwf}. We provide more details of the $\mathbf{Q}$-observable in section \ref{sec:Qopssrf}.

The $A_{l-1}$-singularity is introduced in the twisted M-theory by replacing the topological part of the 11-dimensional worldvolume by $\BC_{\ve_1} \times (\BC_{\ve_2}\times \BC_{\ve_4})/\BZ_l \times \BR$ (the third row in Table \ref{table:twM}) \cite{Costello:2016nkh}. It amounts to add Kaluza-Klein 6-monopoles supported on the transverse worldvolume, intersecting the stack of $N$ M5-branes along $\BC_{\ve_1} \times \BC^\times$. Hence it leads to a codimension-two defect in the six-dimensional $\EN=(0,2)$ theory, which reduces to a monodromy surface defect lying on $\BC_{\ve_1}$ in the four-dimensional $\EN=2$ theory \cite{K-T,Nikita:IV}. We will provide more details of the implementation of the monodromy surface defect by the $\BZ_l$-orbifold in section \ref{sec:Qopssrf}. 

\begin{figure}[h!]
    \centering
       \begin{tikzpicture}[inner sep=0in,outer sep=0in]
            \draw (0,0) -- (0,2);  \draw (0,0) -- (2,0); \draw (0,0) -- (-1,-1); \draw (-1,-1) -- (2,-1);
            \draw (-1,-1) -- (-3,-2); \draw (-3,-2) -- (2,-2);
            \draw (-3,-2) -- (-6,-3); \draw (-6,-3) -- (2,-3); \draw (-6,-3) -- (2,-3); \draw (-6,-3) -- (-7.6,-3.4);

            \draw  (1,1)  node {{$N$}}; \draw  (1,-1.4)  node {{$\vdots$}}; \draw  (1,-2.5)  node {{$r_0$}}; \draw  (1,-0.5)  node {{$r_{l-2}$}};
       \end{tikzpicture}
    \caption{Resolution of $\BC_{\ve_1} \times (\BC_{\ve_2} \times \BC_{\ve_4})/\BZ_l$ and M5-branes wrapping $\BC^\times$.}
    \label{fig:ltb}
\end{figure}
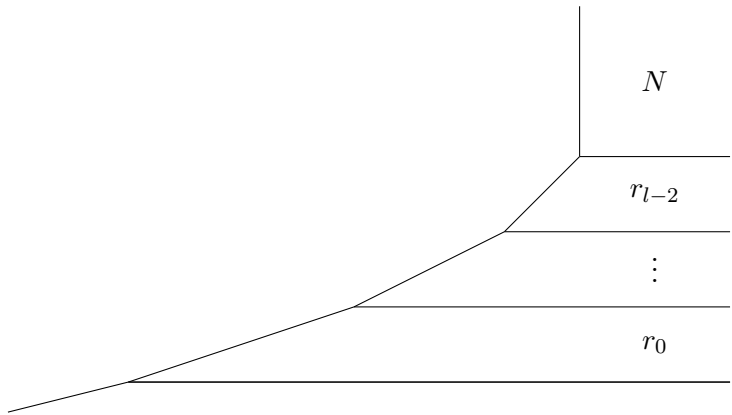

Here, we discuss dual description and transition of the monodromy surface defect that lead to the separation of variables. First, we make a continuous deformation of the metric to bring the $A_{l-1}$-singularity to the $l$-centered Taub-NUT space. The singularity is resolved by $l-1$ copies of $\BP^1$, and the M5-branes wrap these compact cycles according to the $\BZ_l$-charges that they were assigned before resolution. We assign $\BZ_l$-charge $0 $ for all the $n$ M5-branes wrapping $\BC$, while having $\BZ_l$-charges of the $N$ M5-branes wrapping $\BC^\times$ at our disposal. Let $r_\o$ denote the number of these M5-branes carrying $\BZ_l$-charge equal to or less than $\o$ for $\o=0,1,\cdots, l-1$ ($r_{l-1} = N$). Then, we may visualize the M5-brane configuration after the resolution by the toric diagram in Figure \ref{fig:ltb}.

Note that the M5-branes wrapping the compact cycles intersect the worldvolume of the four-dimensional $\EN=2$ gauge theory along $\BC_{\ve_1}$. Thus, it is expected that they create a surface defect by coupling an effective two-dimensional $\EN=(2,2)$ linear $A_{l-1}$-quiver gauge theory on their worldvolume, providing a dual description for the monodromy surface defect (see Figure \ref{fig:2dquiver}, for our main case of $n=2$). We will confirm this explicitly by matching their observable expressions in section \ref{sec:SoV}.

Now, let us restrict to the case of the regular monodromy surface defect where we set $l=N$ and $r_\o = \o+1$, $\o=0,1,\cdots, N-1$. In the IIA frame, we can initiate consecutive D2-brane creations by pulling out the D4-branes wrapping the compact cycles along the topological $\BR$-direction ($x^5$), starting from the bottom of the toric diagram. The D4-branes wrapping adjacent compact cycles can be joined across their junctions, so that they can be pulled out together while creating $n-1$ D2-brane ending on $n-1$ out of $n$ NS5-branes supported on the non-compact worldvolume $\BC_{\ve_1} \times \BC_{\ve_2} \times \BC$. The net result of the consecutive transitions is $(n-1)(N-1)$ D2-branes stretched between $n-1$ of the NS5-branes and distinct D4-branes, on top of the usual IIA brane setup for the underlying $\EN=2$ gauge theory. This is precisely the IIA brane engineering of the $\mathbf{Q}$-observable surface defect \cite{Jeong:2023qdr} (see section \ref{subsec:Qobser}). Thus, the brane transition exhibits the surface defect transition from the regular monodromy surface defect to $(n-1)(N-1)$ $\mathbf{Q}$-observables. Using the transition from the $\mathbf{Q}$-observable to the $\mathbf{H}$-observable (see section \ref{sec:Qopssrf}), we also achieve the surface defect transition from the regular monodromy surface defect to multiple $\mathbf{H}$-observables. We will establish these surface defect transitions as concrete integral transformations between their correlation functions in section \ref{sec:SoV}.

The M-theory on a toric Calabi-Yau threefold is dual to the $(p,q)$-web of fivebranes in IIB theory, where the $(p,q)$-web is given by the toric diagram \cite{Leung:1997tw}. Applying the duality to our setting, the two stacks of M5-branes become two stacks of D3-branes filling the face between the NS5-brane and the D5-brane(s), wrapping $\BC^\times$ and $\BC$ in the holomorphic direction respectively. In the presence of the $A_{l-1}$-singularity, the number of D5-branes increases from $1$ to $l$. The D3-branes end on these D5-branes according to the $\BZ_l$-charge that the M5-branes were assigned. After resolving the singularity, the $l$ D5-branes are separated as in the $(p,q)$-web of Figure \ref{fig:ltb}, each of which forms a junction with 
a $(1,\o)$-brane and a $(1,\o+1)$-brane, $\o=0,1,\cdots, l-1$. Meanwhile, the M2-branes wrapping $\BC_{\ve_1}$ become D1-branes attached to the D5-brane(s). In the effective GL-twisted $\EN=4$ theory on the worldvolume of the D3-branes wrapping $\BC^\times$, the NS5-brane gives a (deformed) Neumann boundary, while the D5-brane(s) gives a (deformed) Nahm pole boundary \cite{Gaiotto:2008sa}. The D1-branes attached to the D5-branes become the 't Hooft line defects superposed on the Nahm pole boundary. The transition from the $A_{N-1}$-singularity to multiple M2-branes, which leads to the separation of variables, would be translated to a transition of the Dirichlet boundary into multiple 't Hooft line defects attached to the regular Nahm pole boundary.

Another version of M-brane transition, leading to the surface defect transition and the separation of variables, was also presented in \cite{Frenkel:2015rda}. We anticipate that these two perspectives are connected through a string duality. In particular, the monodromy surface defect is realized by M5-branes there, and they are expected to arise from the M-theory uplift of another IIA theory related to the one that we described earlier by two T-dualities, mapping the KK5-monopoles to D4-branes. See also \cite{Prochazka:2018tlo} where the resolution yields gluing of corner vertex algebras \cite{Gaiotto:2017euk}.

\subsection{Outline}
 In section \ref{sec:Qopssrf}, we explain that the $Q$-operator and the Hecke operator are realized by the same two-dimensional $\EN=(2,2)$ gauged linear sigma model coupled to the four-dimensional $\EN=2$ theory, where the vev of the complex adjoint scalar is turned on in the former while the complexified FI parameter is turned on in the latter. We achieve the transition between the two as the Fourier transform between their surface defect observables. In section \ref{sec:hbarlang}, we recap how the $\mathbf{Q}$-observables and the $\mathbf{H}$-observables give rise to the $Q$-operators and the Hecke operators, respectively. Then, using the Fourier transformation between them, we derive the bispectral duality between the $\fgl(2)$ XXX spin chain and the $\fsl(N)$ Gaudin model. In section \ref{sec:SoV}, we elucidate the monodromy surface defect has a dual description as coupling a two-dimensional $\EN=(2,2)$ sigma model. We establish the crossing formula between the observable expressions in two different stability chambers of the modified hand-saw quiver variety. Using the expression in the negative chamber, we re-express the monodromy defect observable into two different contour integrals, exhibiting its transition into multiple $\mathbf{Q}$-observables or $\mathbf{H}$-observables. In particular, we obtain definite expressions for the integral kernels and the contour prescriptions. We show the integral transformation leads to the quantum separation of variables for the XXX spin chain and the Gaudin model in the limit $\ve_2\to 0$. We also confirm that, when applied to some simple cases, the separation of variable transformations we found reproduce the known prescription of Sklyanin. Finally, we conclude in section \ref{sec:discussion} with discussions. The appendices contain a brief review on the gauge origami constructions of the $\EN=2$ theory and its surface defects and some computational details.

\paragraph{Acknowledgement}
The authors thank Kevin Costello, Chris Elliott, Alba Grassi, Nathan Haouzi, Nafiz Ishtiaque, Shota Komatsu, Nikita Nekrasov, Jihwan Oh, Miroslav Rap\v{c}\'{a}k, and Yehao Zhou for discussions and collaboration on related subjects. SJ is grateful to Du Pei for helpful discussion and support during his visit to Center for Quantum Mathematics at University of Southern Denmark, where a part of the work was done. The work of SJ is supported by CERN and CKC fellowship. The work of NL is supported by IBS project IBS-R003-D1.

\section{$\mathbf{Q}$- and $\mathbf{H}$-observables as surface defects} \label{sec:Qopssrf}
The contents of this section is mainly a review of the $\mathbf{Q}$-observables, the $\mathbf{H}$-observables, and the monodromy surface defect in the $\EN=2$ supersymmetric $A_1$-quiver $U(N)$ gauge theory. Nevertheless, we reorganize the known facts from the perspective of the bispectral duality. In particular, we give a description of the first two surface defects as coupling the same two-dimensional $\EN=(2,2)$ gauged linear sigma model with either the vev of the complex adjoint scalar or the complexified FI parameter turned on. The variation of the respective parameters is visualized in the IIA brane engineering of the two surface defects. We also provide higgsing constructions of the two surface defects which is in accordance with the suggested IIA brane picture.

\subsection{$\mathbf{Q}$-observables} \label{subsec:Qobser}
In the IIB gauge origami setup, the $\mathbf{Q}$-observables are built by inserting an additional D3-brane that intersects the worldvolume of the original stack of D3-branes for the gauge theory along a surface. To avoid duplication with the companion paper \cite{Jeong:2024hwf}, we will present the gauge origami construction of the $\mathbf{Q}/\tilde{\mathbf{Q}}$-observables in the appendix \ref{subsec:qobsapp} and \ref{subsec:dualqobsapp}.

Here, we explain that they are these surface defects created by coupling a $\EN=(2,2)$ two-dimensional gauged linear sigma model whose complexified FI parameter is turned off and the vev of the complex adjoint scalar is turned on. The coupling is implemented by gauging a part of the flavor group by the bulk gauge field restricted to the surface. We will also give a higgsing construction of $\mathbf{Q}/\tilde{\mathbf{Q}}$-observables which is in accordance with the IIA brane picture.

\subsubsection{$\mathbf{Q}$-observables from coupling two-dimensional $\EN=(2,2)$ theory} \label{subsubsec:qobscouple}
Due to the supersymmetric localization, the path integral of the $\EN=2$ $A_1$-quiver $U(N)$ gauge theory localizes into a finite-dimensional equivariant integral over the moduli space of instantons. Further by the equivariant localization, any gauge theory observable produces an observable on the ensemble over the fixed points $\{\bl\} = \{(\l^{(0)},\cdots, \l^{(N-1)}) \}$ of the moduli space of instantons enumerated by $N$ partitions.

On this ensemble over the fixed points $\{\bl\}$, the $\mathbf{Q}$-observable is expressed as \eqref{eq:qobsappp},
\begin{align} \label{eq:Qobs}
    \EQ(x)[{\boldsymbol\lambda}]  =  \BE \left[ \frac{e^{x} (M^+ -S[\bl])^*}{P_1^*}  \right],
\end{align}
where $x \in \BC$ is a parameter of the $\mathbf{Q}$-observable, $M^+$ is a flavor bundle, $S$ is the universal sheaf of the instantons, and $\BE[\cdots]$ is the symbol that converts the equivariant Chern character into the product of equivariant weights. To avoid duplication, we refer to the companion paper \cite{Jeong:2024hwf} and appendix \ref{sec:ori} for the details of the notations and conventions. We sometimes discard the $\G$-functions involving the flavor bundle, writing
\begin{align}
    \mathbf{Q}(x) = \BE \left[ - \frac{e^x S[\bl] ^*}{P_1 ^*} \right].
\end{align}

Note that the $\mathbf{Q}$-observable \eqref{eq:Qobs} reduces to a product of $\G$-functions in the decoupling limit $\qe\to 0$, where the universal sheaf $S$ becomes a flavor bundle $ N$. In this limit, the $\mathbf{Q}$-observable \eqref{eq:Qobs} can be viewed as the 1-loop fluctuation of two sets of $N$ 2d $\EN=(2,2)$ chiral multiplets supported on the $\BC_1$-plane, with the twisted masses assigned as $m^+ _\a -x$ and $a_\a -x$, $\a=0,1,\cdots, N-1$. Namely, the $x$-parameter is introduced as an overall shift of the twisted masses. Equivalently, we may instead consider the two-dimensional $\EN=(2,2)$ supersymmetric gauged linear sigma model, whose field contents are
\begin{itemize}
    \item $U(1)$ gauge multiplet
    \item $N$ chiral multiplets of charge $+1$ (flavor group $M^+$)
    \item $N$ chiral multiplets of charge $-1$ (flavor group $N$)
\end{itemize}
with the complexified FI parameter turned off and the vev of the complex adjoint scalar in the twisted chiral multiplet turned on to be $x \in \BC$ \cite{Gaiotto:2013sma}. There is no non-perturbative sector of the field configuration since the complexified FI parameter is turned off, and the partition function is precisely given by the product of $\G$-functions from the 1-loop fluctuations of the chiral multiplets (see \cite{Hori:2013ika} for instance). 

When coupled to the four-dimensional theory, the flavor symmetry group $N$ is gauged by the four-dimensional gauge field so that the associated flavor bundle gets uplifted to the universal sheaf $S$ for the four-dimensional instantons, yielding the surface observable \eqref{eq:Qobs}. Therefore, the $\mathbf{Q}$-observable \eqref{eq:Qobs} can indeed be viewed as the surface defect engineered by coupling the two-dimensional $\EN=(2,2)$ sigma model supported on the $\BC_1$-plane, where the complexified FI parameter turned off and the vev of the complex adjoint scalar turned on.

\begin{figure}[h!]
    \centering
    \begin{subfigure}[b]{0.45\textwidth}
       \centering
       \begin{tikzpicture}[inner sep=0in,outer sep=0in]
            \node[inner sep=0pt] (brane1) at (0,0)  {\includegraphics[width=\textwidth]{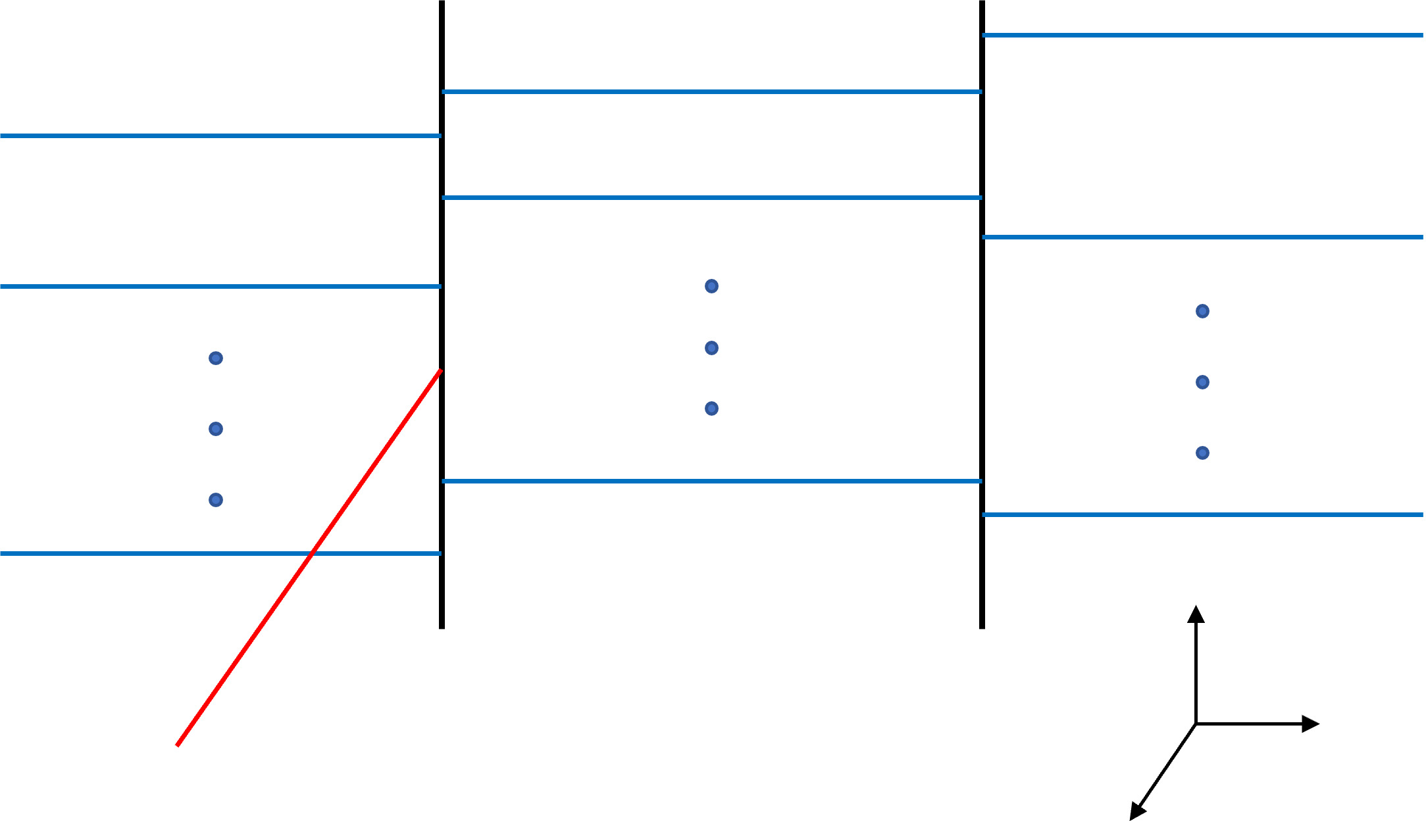}};
            \draw  (-1.3,2.4)  node {\footnotesize{NS5}};
            \draw  (3.2,2.2)  node {\footnotesize{D4}};
            \draw  (-2.85,-1.85)  node {\footnotesize{D2}};

            \draw  (2.7,-0.8)  node {\scriptsize{$x^{8,9}$}};
             \draw  (3.2,-1.45)  node {\scriptsize{$x^{4}$}};
             \draw  (1.85,-1.9)  node {\scriptsize{$x^{5}$}};
       \end{tikzpicture}
       \caption{}
       \end{subfigure}
       \quad\quad\quad
    \begin{subfigure}[b]{0.45\textwidth}
       \centering
       \begin{tikzpicture}[inner sep=0in,outer sep=0in]
            \node[inner sep=0pt] (brane2) at (0,0) {\includegraphics[width=\textwidth]{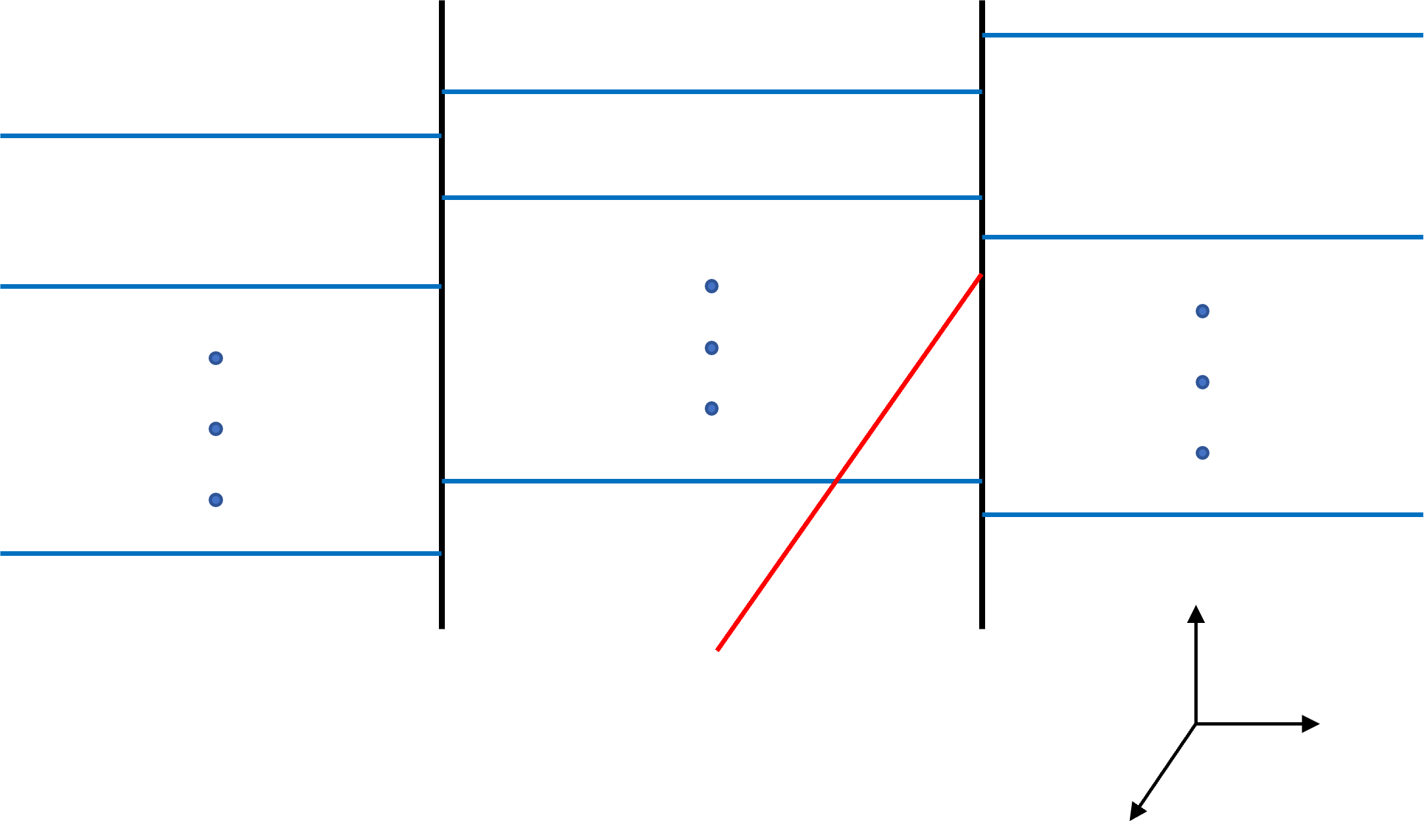}};
            \draw  (-1.3,2.4)  node {\footnotesize{NS5}};
            \draw  (3.2,2.2)  node {\footnotesize{D4}};
            \draw  (0,-1.4)  node {\footnotesize{D2}};

            \draw  (2.7,-0.8)  node {\scriptsize{$x^{8,9}$}};
             \draw  (3.2,-1.45)  node {\scriptsize{$x^{4}$}};
             \draw  (1.85,-1.9)  node {\scriptsize{$x^{5}$}};
       \end{tikzpicture}
       \caption{}
       \end{subfigure}
    \caption{IIA brane engineering of the $\mathbf{Q}$-observables, in our case $n=2$; (a) $\mathbf{Q}$-observable and (b) $\tilde{\mathbf{Q}}$-observable (see Table \ref{table:twM} and \ref{table:spacetime})}
    \label{fig:iiabrane}
\end{figure}
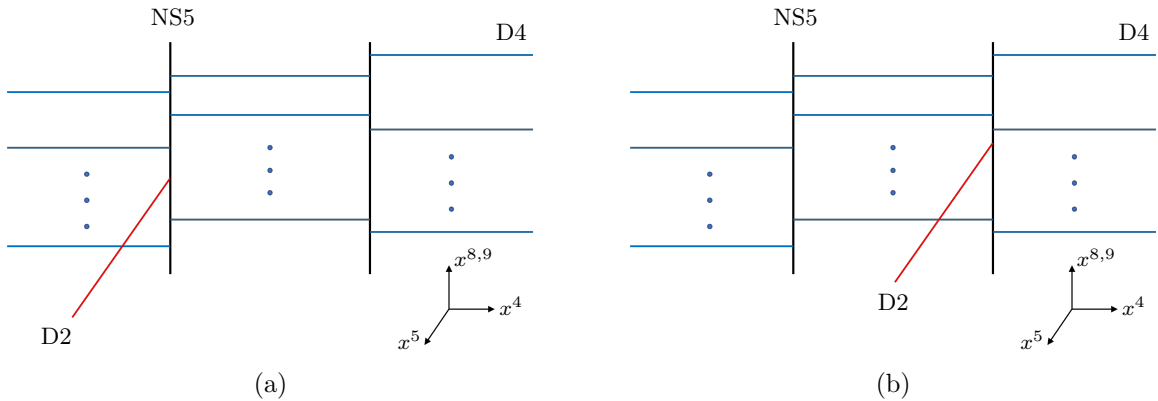

Recall that the IIB gauge origami setup (see appendix \ref{sec:ori}) is dualized to IIA theory, where the D3-brane supported on $\BC^2 _{13}$ becomes a D2-brane (see Figure \ref{fig:iiabrane}, and Table \ref{table:twM} and \ref{table:spacetime}). The D2-brane ends on one of the $n$ NS5-branes according to the $\BZ_{n+1}$-charge that the D3-brane carried. We will restrict our focus to cases where the D2-branes end on the NS5-branes from the positive $x^5$-direction, as the opposite direction can be treated symmetrically. In our work, we mainly consider the case $n=2$, so that there are $\mathbf{Q}$- and $\tilde{\mathbf{Q}}$-observables which are inequivalent to each other.\footnote{If we had $\BZ_{n+1}$-orbifold in the IIB gauge origami, the resulting $\EN=2$ gauge theory is the linear $A_{n-1}$-quiver gauge theory whose IIA brane realization is the $N$ D4-branes stretched between $n$ NS5-branes and infinity. Therefore, there are $n$ inequivalent $\mathbf{Q}$-observables in general case.} We choose our convention that the D3-brane carrying $\BZ_3$-charge $1$ is dualized to the D2-brane ending on the NS5-brane on the left, while the D3-brane carrying $\BZ_3$-charge $0$ is dualized to the D2-brane ending on the NS5-brane on the right. In the point of view of the low-energy effective theory, the D2-brane engineers a half-BPS surface defect by coupling its two-dimensional worldvolume theory to the $\EN=2$ theory on the worldvolume of the D4-branes. The effective field theory on worldvolume of the D2-brane can easily be read off from the IIA brane picture to be the already described two-dimensional $\EN=(2,2)$ gauged linear sigma model, where the adjoint scalar vev $x \in \BC$ is identified with the position of the D2-brane on the $(x^8, x^9)$-plane \cite{Gaiotto:2013sma}. We call these surface defects the $\mathbf{Q}$-observable and the $\tilde{\mathbf{Q}}$-observable, respectively. We recall that the number matches the order ($n=2$) of the universal $\hbar$-oper equation for $Y(\fgl(2))$, which is solved by two inequivalent $Q$-operators \cite{Jeong:2024hwf}.

When we take the limit $\ve_2 \to 0$ of turning off one of the $\O$-background parameters, the four-dimensional $\EN=2$ gauge theory is effectively described by a $\EN=(2,2)$ gauge theory on the 
topological plane $\BC_2$. The $\mathbf{Q}$-observable, being a surface defect supported on the $\BC_1$-plane, becomes a local observable in this effective two-dimensional theory. Thus, its vacuum expectation value gives
\begin{align} \label{eq:Qobsns}
    \lim_{\ve_2 \to 0} \Big\langle \mathbf{Q}(x) \Big\rangle_\ba = e^{\frac{\widetilde{\EuScript{W}}(\ba;\qe) }{\ve_2}} Q(\ba;x),
\end{align}
where we abuse the notation a bit and denote the normalized vacuum expectation value of the $\mathbf{Q}$-observable in the limit $\ve_2 \to 0$ by $Q(\ba;x)$. The crucial point here is that the effective twisted superpotential $\widetilde{\EuScript{W}}(\ba;\qe)$ is not affected by the insertion of the $\mathbf{Q}$-observable, since the $\mathbf{Q}$-observable is local in the effective two-dimensional theory.

The IIA brane picture (see Figure \ref{fig:iiabrane}) suggests that $n$ inequivalent $\mathbf{Q}$-observables exist as surface defects of $A_{n-1}$-quiver gauge theory. Therefore, in our main case of $n=2$, we expect to have another $\tilde{\mathbf{Q}}$-observable which is inequivalent to \eqref{eq:Qobs}. We present the gauge origami configuration for the $\mathbf{Q}$-observable in the appendix \ref{subsec:dualqobsapp}. The $\tilde{\mathbf{Q}}$-observable is given by
\begin{align} \label{eq:dualQ}
    \tilde{\mathbf{Q}} (x) := \qe^{\frac{x}{\ve_1}} \sum_{d=0} ^{\infty} \qe^d \prod_{i=1} ^d \frac{i \ve_1 + \ve_2}{i \ve_1}  \frac{Q (x) M(x  +d \ve_1)}{Q (x+ d  \ve_1) Q(x +\ve_{2} +(d+1) \ve_1)}.
 \end{align}

Since the $\tilde{\mathbf{Q}}$-observable is a surface defect of the four-dimensional $\EN=2$ theory supported on the $\BC_1$-plane, it becomes a local observable in the effective two-dimensional theory in the limit $\ve_2\to 0$ on the $\BC_2$-plane. Thus, the vacuum expectation value in this limit gives
\begin{align}
    \Big\langle \tilde{\mathbf{Q}}(x) \Big\rangle_\ba = e^{\frac{\widetilde{\EuScript{W}}(\ba;\qe) }{\ve_2}} \tilde{Q}(\ba;x).
\end{align}
where we abuse the notation a bit and denote the normalized vacuum expectation value of the $\tilde{\mathbf{Q}}$-observable in the limit $\ve_2 \to 0$ by $\tilde{Q}(\ba;x)$. More explicitly, from \eqref{eq:dualQ} we have
\begin{align} \label{eq:dualQns}
    \tilde{Q}(\ba;x) = \qe^{\frac{x}{\ve_1}} \sum_{d=0} ^\infty \qe^d \frac{Q(\ba;x) M(x+d\ve_1)}{Q(\ba;x+d\ve_1)Q(\ba;x+(d+1)\ve_1)}.
\end{align}
Just like the $\mathbf{Q}$-observable, the insertion of $\tilde{\mathbf{Q}}$-observable does not alter the effective twisted superpotential $\widetilde{\EuScript{W}}(\ba;\qe)$ in the limit $\ve_2 \to 0$, since the $\tilde{\mathbf{Q}}$-observable is local in the effective two-dimensional theory.

\subsubsection{$\mathbf{Q}$-observables from higgsing} \label{subsubsec:qhiggs}
We have constructed the $\mathbf{Q}/\tilde{\mathbf{Q}}$-observables as surface defects in the four-dimensional $\EN=2$ gauge theory created by coupling a two-dimensional $\EN=(2,2)$ theory. Here, we show that these surface defects can be engineered through partial Higgsing of a higher-rank gauge theory. We will first provide a qualitative description of this higgsing construction within the IIA brane setup, and then implement it at the level of exact partition functions. It should be emphasized that the IIA brane picture is ignorant of the $\Omega$-background and should be regarded only as a qualitative description once it is turned on.\\

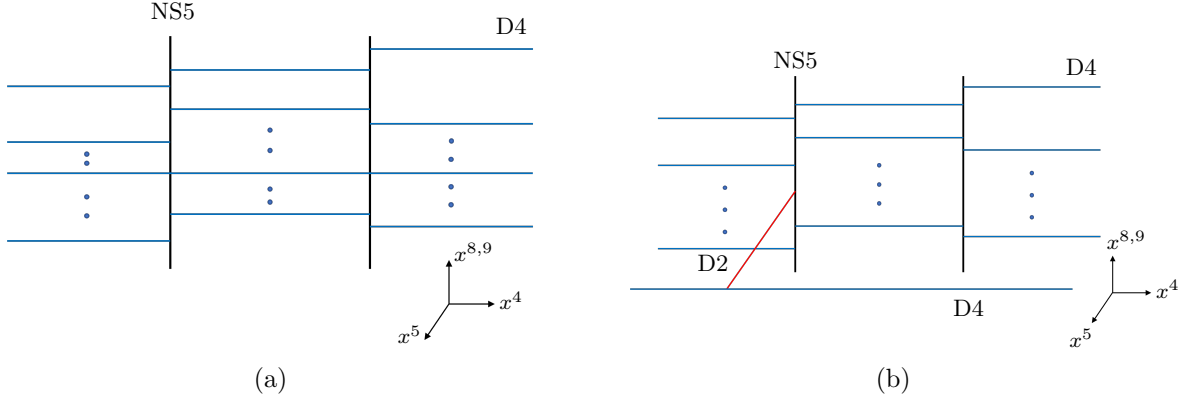
\begin{figure}[h!]
    \centering
    \begin{subfigure}[b]{0.45\textwidth}
       \centering
       \begin{tikzpicture}[inner sep=0in,outer sep=0in]
            \node[inner sep=0pt] (brane1) at (0,0)  {\includegraphics[width=\textwidth]{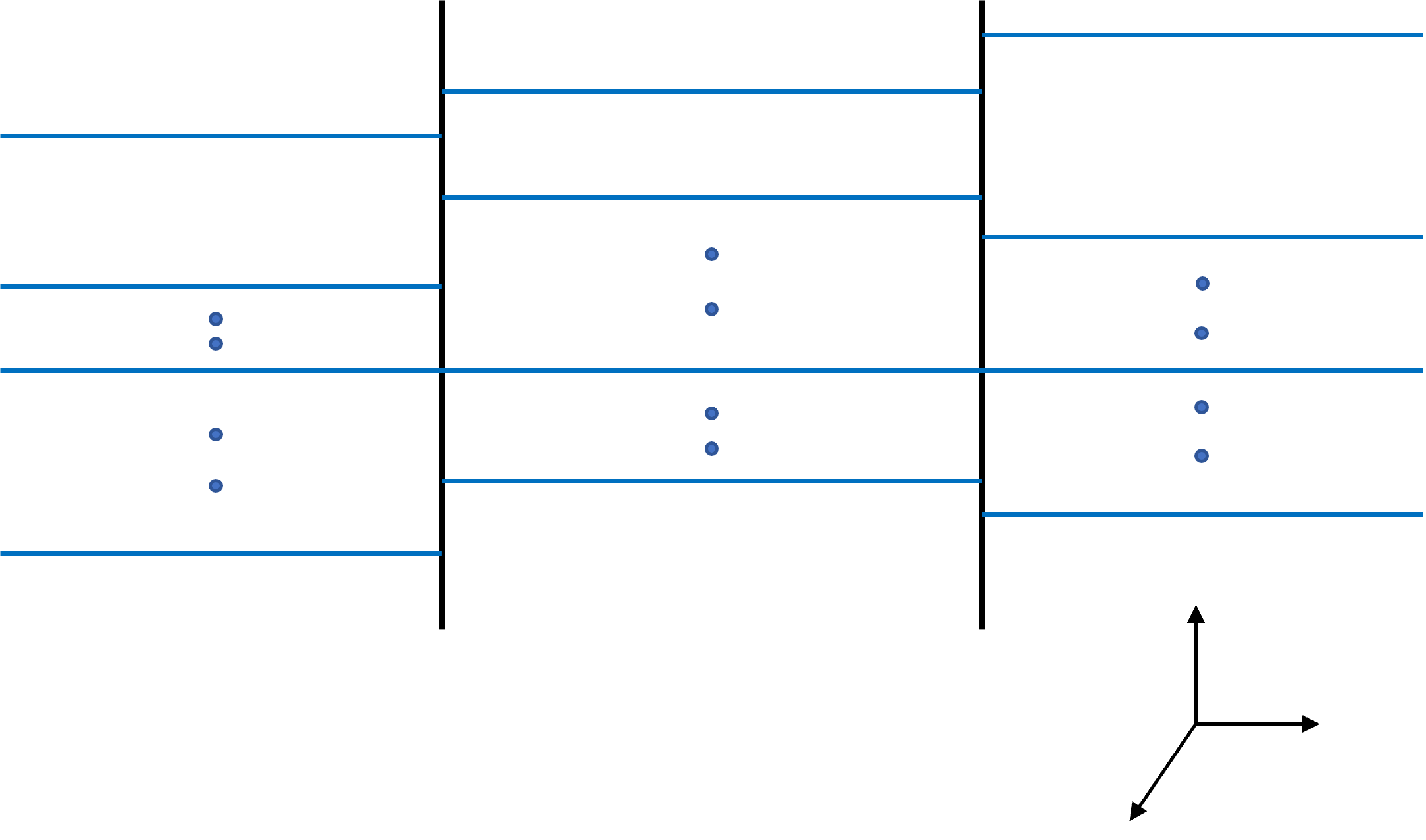}};
            \draw  (-1.3,2.4)  node {\footnotesize{NS5}};
            \draw  (3.2,2.2)  node {\footnotesize{D4}};

            \draw  (2.7,-0.8)  node {\scriptsize{$x^{8,9}$}};
             \draw  (3.2,-1.45)  node {\scriptsize{$x^{4}$}};
             \draw  (1.85,-1.9)  node {\scriptsize{$x^{5}$}};
       \end{tikzpicture}
       \caption{}
       \end{subfigure}
       \quad\quad\quad
    \begin{subfigure}[b]{0.45\textwidth}
       \centering
       \begin{tikzpicture}[inner sep=0in,outer sep=0in]
            \node[inner sep=0pt] (brane2) at (0,0) {\includegraphics[width=\textwidth]{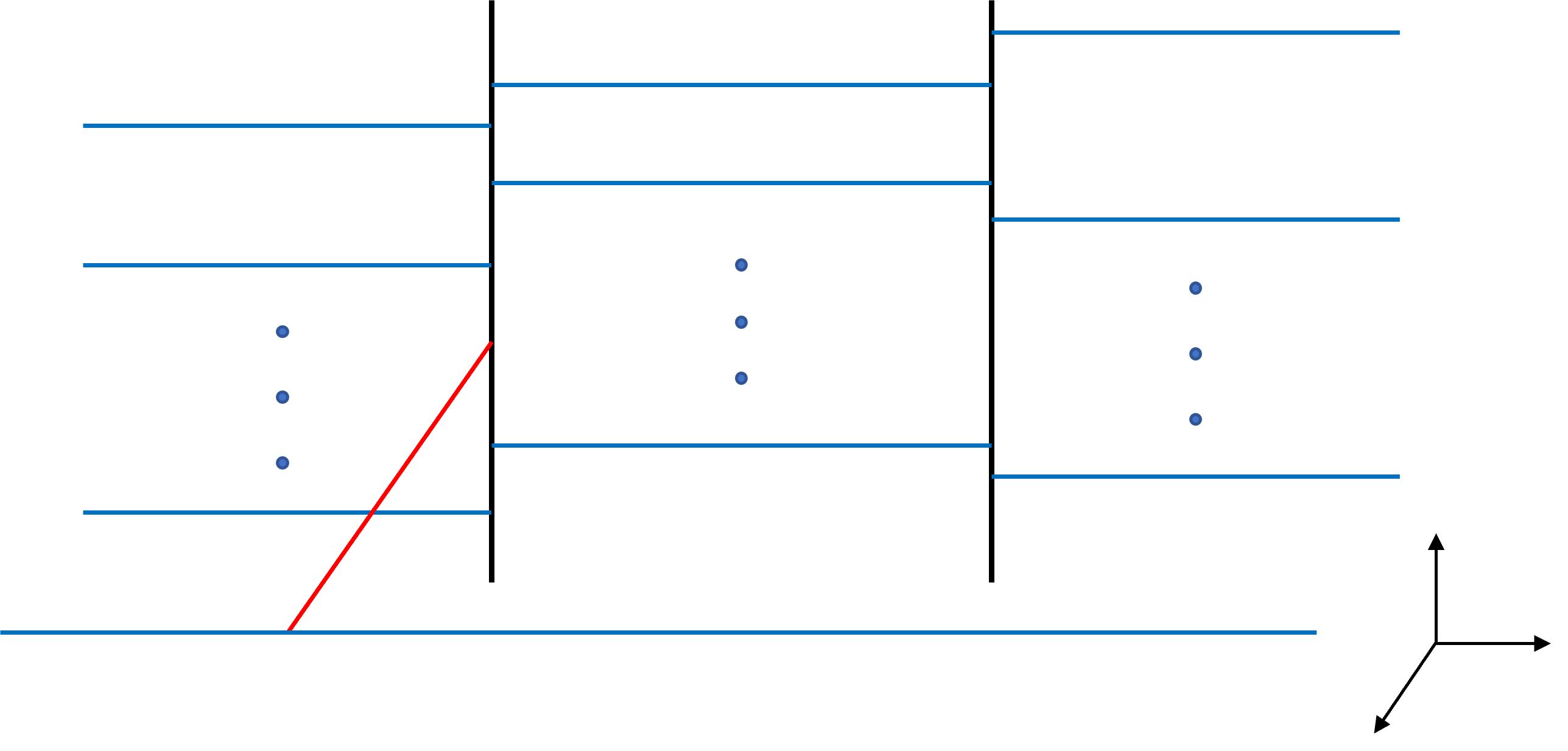}};
            \draw  (-1.3,1.9)  node {\footnotesize{NS5}};
            \draw  (2.5,1.8)  node {\footnotesize{D4}};
            \draw  (-2.4,-0.8)  node {\footnotesize{D2}};
            \draw  (1,-1.4)  node {\footnotesize{D4}};
            
            \draw  (3.1,-0.5)  node {\scriptsize{$x^{8,9}$}};
             \draw  (3.65,-1.15)  node {\scriptsize{$x^{4}$}};
             \draw  (2.5,-1.8)  node {\scriptsize{$x^{5}$}};
       \end{tikzpicture}
       \caption{}
       \end{subfigure}
    \caption{D2-brane creation and higgsing construction of $\mathbf{Q}$-observable. (a) One of $N+1$ D4-branes is aligned across the two NS5-branes, enabling the D4-brane to move along the $x^5$-direction. (b) A D2-brane ending on one of the NS5-branes is created as a result of the transition.}
    \label{fig:higgsQ}
\end{figure}

Let us consider having $N+1$ D4-branes stretched between two NS5-branes and infinity. The low energy effective theory is the $SU(N+1)$ gauge theory with $N+1$ fundamental hypermultiplets and $N+1$ anti-fundamental hypermultiplets (see Figure \ref{fig:higgsQ}). Now, consider aligning the position (on the $(x^8 ,x^9)$-plane) of one of the D4-branes across the two NS5-branes. This alignment enables the D4-brane to move along the $x^5$-direction (positive or negative), creating (possibly multiple) D2-branes, each of which is stretched between the D4-brane and one of the NS5-branes. Thus, after sending the D4-brane to the infinity in the $x^5$-direction, we are left with the original stack of $N$ D4-branes across two NS5-branes augmented by D2-branes ending on the NS5-branes. This is precisely the IIA brane engineering of the $\mathbf{Q}/\tilde{\mathbf{Q}}$-observables that we explained earlier in section \ref{subsubsec:qobscouple}. 

In terms of the $\EN=2$ gauge theory in the low-energy, the alignment corresponds to tune the masses of one of fundamental hypermultiplets and one of anti-fundamental hypermultiplets to zero. Then we can partially higgs the gauge group $SU(N+1)$ down to $SU(N)$ by giving vacuum expectation values for these hypermultiplets, proportional to the deviation of the D4-brane in the $x^5$-direction. The relevant $U(1)$ part of the gauge field is squeezed into a two-dimensional plane, giving a two dimensional $\EN=(2,2)$ gauged linear sigma model with the vev of the adjoint scalar coupled to the remnant four-dimensional $\EN=2$ gauge theory with the gauge group $SU(N)$. Note that the number of the created D2-branes ending on the respective NS5-branes, as well as their orientations, are unspecified within the brane picture; however, in the presence of the $\O$-background the implementation at the level of the partition function will reveal that the higgsing constraints single out those numbers and orientations, lifting all the other possible configurations. \\

Now, in the presence of the $\O$-background, let us confirm the higgsing construction by implementing it in exact computation of the partition function, both for the $\mathbf{Q}$-observable and the $\tilde{\mathbf{Q}}$-observable. Let us begin with the partition function of the four-dimensional $\EN=2$ gauge theory with the gauge group $U(N+1)$ and $N+1$ fundamental and $N+1$ anti-fundamental hypermultiplets. The equivariant Chern characters of the framing and flavor bundles are set to be
\begin{align} \label{eq:higgs1}
\begin{split}
    &N = \sum_{\a=0} ^{N-1} e^{a_\a} + e^{x+\ve_1+\ve_2} ,\quad M^+ = \sum_{\a=0} ^{N-1} e^{m^+ _\a } + e^{x+\ve_1+\ve_2}, \quad M^- = \sum_{\a=0} ^{N-1} e^{m^- _\a} + e^{x -\ve_2},
\end{split}
\end{align}
where we identified the $(N+1)$-th Coulomb moduli and the masses up to $\ve_{1,2}$-deviation. Note that the constraint $a_{N} = m_{N} ^+ = m^- _N + \ve_1 + 2\ve_2$ indeed makes one of the fundamentals and one of the antifundamentals massless up to an $\ve_{1,2}$-deviation. It turns out that this constraint engineers a single D2-brane supported on the $\BC_1$-plane ending on the left NS5-brane; namely, a single $\mathbf{Q}$-observable.\footnote{More precisely, the number of the created D2-branes is determined by the integer coefficients of the $\O$-background parameters in the constraint. The orientation is determined by whether it is $\ve_1$ or $\ve_2$ that is used to deviate from the exact massless condition.}

The partition function of the $U(N+1)$ gauge theory under the higgsing constraint gives
\begin{align}
\begin{split}
    &\CalZ_{U(N+1)} (\mathbf{a},\mathbf{m}^\pm ,a_N  -\ve_1 -\ve_2= m^+ _N -\ve_1-\ve_2 = m^- _N -\ve_1 = x  ; \qe ) 
 \\ & = \sum_{\bl} \qe^{\vert \bl \vert} \BE \left[ \frac{-(S+e^{a_N})(S^* + e^{-a_N}) + (M^+ + e^{m^+ _N}) (S^* +e^{-a_N} ) +q_{12} ^{-1} (S+ e^{a_N})((M^-)^* + e^{-m^- _N} ) }{P_{12} ^*} \right] \\
 &=  \sum_{\bl} \qe^{\vert \bl \vert} \BE \left[ \frac{-SS^*+M^+ S^* + q_{12} ^{-1} S (M^-)^*}{P_{12}^*} \right] \times \BE \left[ -\frac{Se^{-x}}{P_1} \right] \BE \left[ \frac{M^+ e^{-x}}{P_{12}} + \frac{e^x (M^-)^* + q_2  }{P_{12}^*} \right] \\
 & =\BE \left[ \frac{M^+ e^{-x}}{P_{12}} + \frac{e^x (M^-)^* + q_2}{P_{12}^*} \right] \times \Big\langle Q(x) \Big \rangle_\ba .
\end{split}
\end{align}
The additional 1-loop contribution in front is from the open strings stretching between the distinguished D4-brane and the rest of $N$ D4-branes extended to infinity. It decouples from the $\EN=2$ gauge theory and thus has to be discarded. Hence, we confirm that the partition function of the partially higgsed $U(N+1)$ gauge theory reproduces the vacuum expectation value of the $\mathbf{Q}$-observable defined in the $U(N)$ gauge theory, as expected.

Next, we construct the $\tilde{\mathbf{Q}}$-observable by imposing a different higgsing condition to the $U(N+1)$ gauge theory. The equivariant Chern character of the framing bundle is now set as
\begin{align} \label{eq:higgs2}
\begin{split}
    &N = \sum_{\a=0} ^{N-1} e^{a_\a} + e^{x+\ve_1} ,\quad M^+ = \sum_{\a=0} ^{N-1} e^{m^+ _\a} + e^{x+\ve_1+\ve_2}, \quad M^- = \sum_{\a=0} ^{N-1} e^{m^- _\a} + e^{x-\ve_2}.
\end{split}
\end{align}
The higgsing condition makes the relevant instanton configuration squeezed into the $\BC_1$-plane, constraining the corresponding universal sheaf as
\begin{align}
    K_N = e^{a_N} \sum_{i=1} ^d q_1 ^{i-1}, \qquad S_N = e^{a_N} (q_2 + P_2 q_1 ^d),
\end{align}
where we set $\vert K_N \vert = d$. The partition function of the $U(N+1)$ gauge theory subject to the higgsing constraint is computed as
\begin{align}
\begin{split}
        &\CalZ_{U(N+1)} (\mathbf{a},\mathbf{m}^\pm ,a_N -\ve_1= m^+ _N -\ve_1-\ve_2 = m^- _N +\ve_2 = x  ; \qe ) \\ 
    & = \sum_{\boldsymbol\lambda} \sum_{d=0}^\infty \kq^{|{\boldsymbol\lambda}|+d} \, \BE \left[ - \frac{(S+e^{a_{N}}(q_2+P_2q_1^d))(S+e^{a_{N}}(q_2+P_2q_1^d))^*}{P_{12}^*} \right. \\
    & \qquad \qquad \quad \left. + \frac{(M^++e^{m_{N}^+})(S+e^{a_{N}}(q_2+P_2q_1^d))^* + q_{12} ^{-1} (S+e^{a_{N}}(q_2+P_2q_1^d)) (M^- + e^{m_{N}^- })^* }{P_{12}^*} \right] \\
    & = \sum_{\boldsymbol\lambda} \kq^{|{\boldsymbol\lambda}|} \left[ \frac{-SS^* + M^+ S^* + q_{12} ^{-1} S (M^-)^*}{P_{12}^*} \right] \\ 
    &\quad \times \sum_{d=0}^\infty \kq^d \prod_{i=1}^d \frac{i\ve_1+\ve_2}{i\ve_1} \BE\left[ \frac{e^{x} q_1^{d+1} q_2 S^* }{P_{1}^*} + \frac{e^{-x} S (1-q_1^{-d}) }{P_{1}} \right] \BE \left[ -\frac{e^{-x} q_1 ^{-d} M^+  }{P_1} - \frac{e^{x} q_1^{d} (M^-)^* }{P_1^*} \right] \\
    & \qquad  \times \BE \left[ \frac{M^+e^{-x}}{P_{12}} + \frac{e^x (M^-)^* + q_2}{P_{12}^*} \right] \\
    & = \BE \left[ \frac{M^+e^{-x}}{P_{12}} + \frac{e^x (M^-)^* + q_2}{P_{12}^*} \right] \times \Big\langle \tilde{Q}(x) \Big\rangle_\ba.
\end{split}    
\end{align}
By the same reason as above, the extra 1-loop part in front has to be discarded. Thus, the partition function of the partially higgsed $U(N+1)$ gauge theory gives the vacuum expectation value of the $\tilde{\mathbf{Q}}$-observable, as suggested.

\paragraph{Remark} We stress that we have chosen the higgsing constraints at which the positive vevs of hypermultiplets can be turned on (namely, in Figure \ref{fig:higgsQ}, deviation of the D4-brane to the positive $x^5$-direction). Alternatively, we could have imposed different Higgsing constraints, resulting in the D4-brane deviating in the negative $x^5$-direction. Indeed, consider imposing
\begin{subequations}
    \begin{align}
         &a_N = x + \ve_1, \ m_N^+ = x - \ve_2, \ m_N^- = x + \ve_1 +\ve_2, \\
         & a_N = m_N^- = x+\ve_1+\ve_2 , \ m_N^+ = x-\ve_2.
    \end{align}
\end{subequations}
Under the constraint, the partition function of the higher-rank gauge theory produces a single $\tilde{\mathbf{Q}}$-observable and a single $\mathbf{Q}$-observable, respectively. The former should be thought of as a D2-brane ending on the left NS5-brane from the negative $x^5$-direction, while the latter should be thought of as a D2-brane ending on the right NS5-brane from the negative $x^5$-direction. The confirmation of these higgsing constructions at the level of partition functions are straightforward as the cases we discussed in detail.

\subsubsection{$\mathbf{Q}$-observables across stability chambers}
Even though the $\mathbf{Q}$-observable and the $\tilde{\mathbf{Q}}$-observable have similar origin in the IIA brane picture, their observable expressions look fairly distinct. We illuminate the two expressions are in fact connected to each other across different stability chambers of the four-dimensional instanton moduli space.

In the other (negative) stability chamber of the instanton moduli space, the fixed points enumerated by partitions $\bl$ carry different equivariant weights under the symmetry group action,
\begin{align}
    K = \sum_{\a} \sum_{(i,j) \in \l^{(\a)}} e^{a_\a - i\ve_1 -j\ve_2}.
\end{align}

Let us proceed with the partial higgsing \eqref{eq:higgs2} of the $U(N+1)$ gauge theory that led to the $\tilde{\mathbf{Q}}$-observable in the original chamber. The higgsing condition now enforces the relevant instanton configuration to be trivial, reducing the associated universal sheaf to framing bundle of the resultant two-dimensional theory. The partition function of the partially higgsed $U(N+1)$ gauge theory is computed as
\begin{align}
\begin{split}
        &\CalZ_{U(N+1)} (\mathbf{a},\mathbf{m}^\pm ,a_N -\ve_1= m^+ _N -\ve_1-\ve_2 = m^- _N +\ve_2 = x  ; \qe )^{\text{<0}} \\ 
    & = \sum_{\boldsymbol\lambda} \kq^{|{\boldsymbol\lambda}|} \CalZ_{U(N)}[{\boldsymbol\lambda}]|_{<0} \BE \left[ \frac{e^{x}q_{12} S^*}{P_{1}^*} \right] \BE \left[ \frac{M^+e^{-x}q_2}{P_{12}} + \frac{e^xq_2^*(M^-)^*}{P_{12}^*} \right] \BE \left[ \frac{q_2}{P_{12}^*} \right] \\
    & =  \BE \left[ \frac{M^+e^{-x}q_2}{P_{12}} + \frac{e^xq_2^*(M^-)^*}{P_{12}^*} \right]\BE \left[ \frac{q_2}{P_{12}^*} \right]  \times \Big \langle Q(x+\ve_1 + \ve_2) \Big\rangle^{<0} _\ba,
\end{split}
\end{align}
where the notation $<0$ indicates the other stability chamber. We discard the extra 1-loop contribution in front that decouples from the $\EN=2$ gauge theory. Thus, we confirm that the $\tilde{\mathbf{Q}}$-observable in the original chamber passes to the $\mathbf{Q}$-observable in the other chamber.

Similarly, we can start from the partial higgsing \eqref{eq:higgs1} which gave the $\mathbf{Q}$-observable in the original chamber. The instanton configuration for the $(N+1)$-th $U(1)$ gauge field is no longer trivial, but is squeezed into the two-dimensional plane $\BC_1$. The associated universal sheaf is constrained as
\begin{align}
    K_{N} = \sum_{\ri=1}^k e^{a_{N}-\ri\ve_1-\ve_2} = e^x \frac{1-q_1^{-k}}{1-q_1^{-1}}, \qquad S_{N} = e^{a_{N}} (q_2^{-1} -P_2^*q_1^{-k}).
\end{align}
As a result, the partition function of the partially higgsed $U(N+1)$ gauge theory is computed as
\begin{align}
\begin{split}
    &\CalZ_{U(N+1)} (\mathbf{a},\mathbf{m}^\pm ,a_N  -\ve_1 -\ve_2= m^+ _N -\ve_1-\ve_2 = m^- _N -\ve_1 = x  ; \qe ) ^{<0} \\
        & = \sum_{\boldsymbol\lambda}\kq^{|\boldsymbol\lambda|} \CalZ_{U(N)}[\boldsymbol\lambda]\vert_{<0} \sum_{k=0}^\infty \kq^k \prod_{j=1}^k \frac{Q(x+\ve_1+\ve_2)M^-(x-k\ve_1)^{-1}M^+(x-k\ve_1)^{-1}}{Q(x-k\ve_1)Q(x+\ve_1+\ve_2-k\ve_1)} \prod_{j=1}^k \frac{j\ve_1+\ve_2}{j\ve_1} \\
    & \qquad \times \BE \left[ \frac{M^+e^{-a_{N+1}}q_2}{P_{12}^*} + \frac{e^{a_{N+1}}q_2^*(M^-)^*}{P_{12}^*} \right] \BE \left[ \frac{q_2}{P_{12}^*} \right] \\
    & =  \BE \left[ \frac{M^+e^{-x}q_2}{P_{12}} + \frac{e^{x}q_2^*(M^-)^*}{P_{12}^*} \right] \BE \left[ \frac{q_2}{P_{12}^*} \right] \times \kq^{\frac{x}{\ve_1}} \Big\langle \tilde{Q}(x+\ve_1+\ve_2) \Big\rangle^{<0} _\ba , 
\end{split}
\end{align}
where the extra 1-loop part in front has to be discarded by the same reason above. Thus, we also verify that the $\mathbf{Q}$-observable in the original chamber translates into the $\tilde{\mathbf{Q}}$-observable in the other chamber.

\subsection{$\mathbf{H}$-observables}
We have seen that the $\mathbf{Q}/\tilde{\mathbf{Q}}$-observables are the surface defects created by coupling a two-dimensional $\EN=(2,2)$ theory with the vev of the complex adjoint scalar. Here, we consider the surface defect constructed by coupling the same gauged linear sigma model with the complexified FI parameter $\log y = r-i \theta$ turned on. It is called the canonical surface defect in \cite{Gaiotto:2011tf,Jeong:2018qpc,Jeong:2023qdr}, which we use interchangeably with the $\mathbf{H}$-observable in this work.

The real FI parameter $r$ can be either positive or negative (namely, $\vert y \vert >1 $ or $\vert y \vert<1$). In either case, the gauged linear sigma model flows to the non-linear sigma model with the target being the total space of
\begin{align} \label{eq:target}
    \mathcal{O}(-1) \otimes \BC^N \longrightarrow \BP^{N-1},
\end{align}
where the complexified FI parameter becomes the complexified K\"{a}hler parameter. It has $N$ discrete vacua. When coupled to the four-dimensional $\EN=2$ theory, the $U(N)$ flavor symmetry of rotating the fiber gets gauged by the bulk gauge field restricted to the surface. 

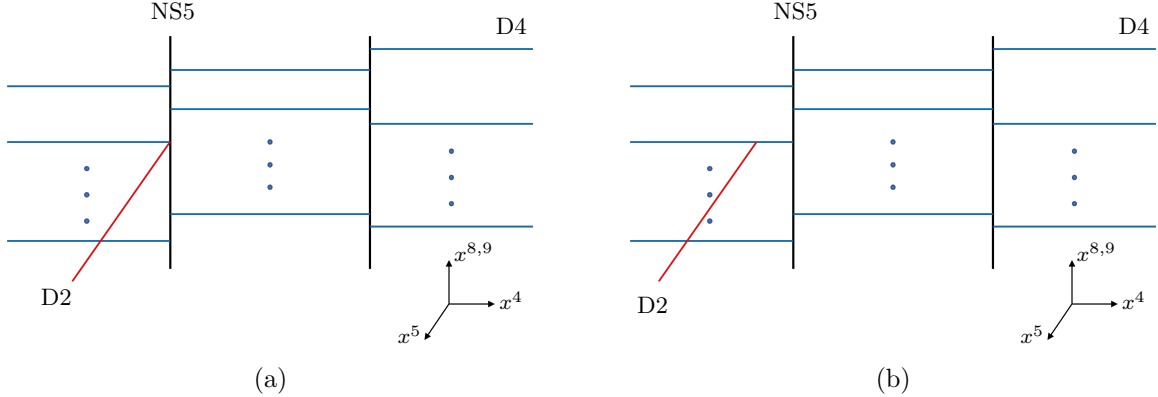
\begin{figure}[h!]
    \centering
    \begin{subfigure}[b]{0.45\textwidth}
       \centering
       \begin{tikzpicture}[inner sep=0in,outer sep=0in]
            \node[inner sep=0pt] (brane1) at (0,0)  {\includegraphics[width=\textwidth]{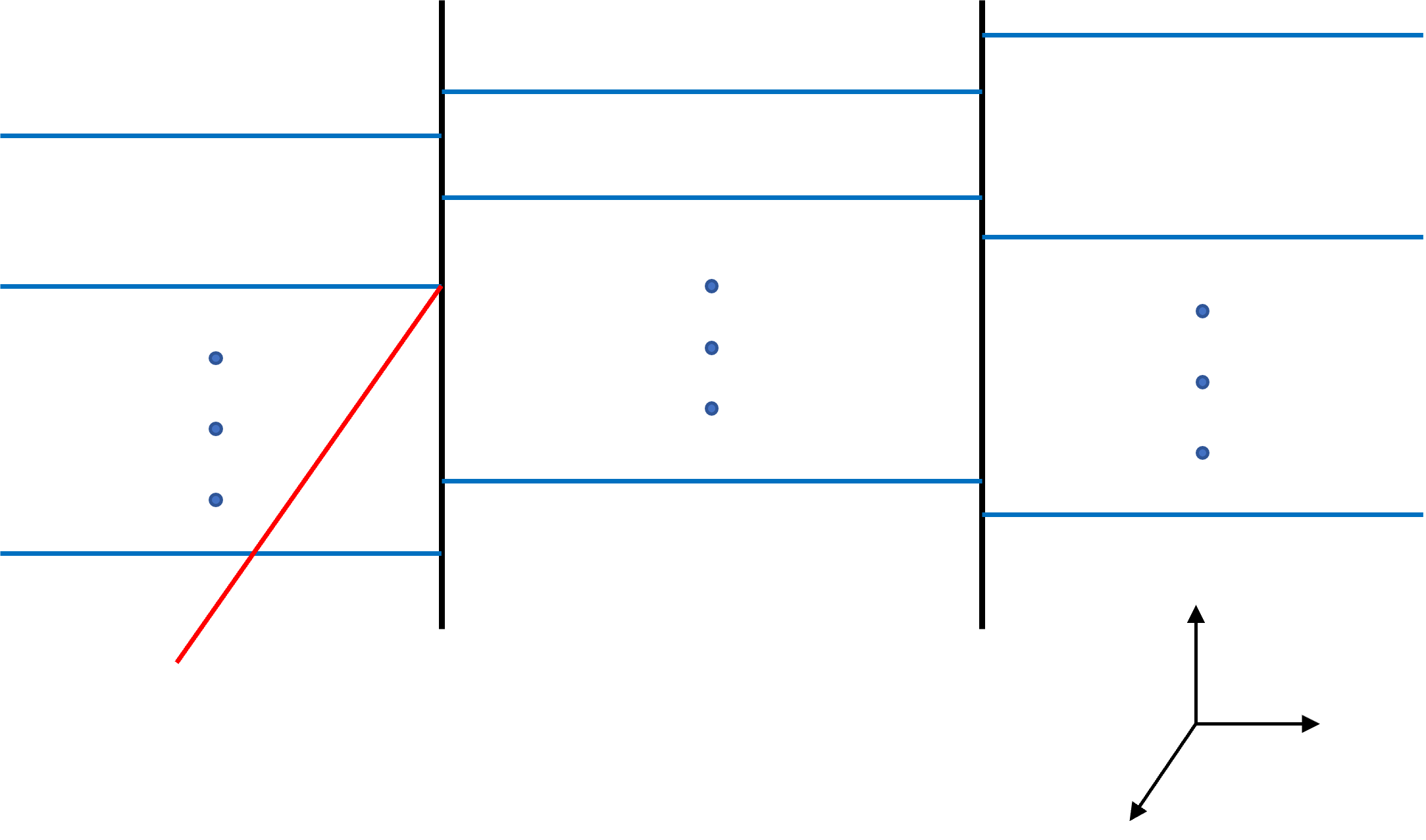}};
            \draw  (-1.3,2.4)  node {\footnotesize{NS5}};
            \draw  (3.2,2.2)  node {\footnotesize{D4}};
            \draw  (-2.85,-1.4)  node {\footnotesize{D2}};

            \draw  (2.7,-0.8)  node {\scriptsize{$x^{8,9}$}};
             \draw  (3.2,-1.45)  node {\scriptsize{$x^{4}$}};
             \draw  (1.85,-1.9)  node {\scriptsize{$x^{5}$}};
       \end{tikzpicture}
       \caption{}
       \end{subfigure}
       \quad\quad\quad
    \begin{subfigure}[b]{0.45\textwidth}
       \centering
       \begin{tikzpicture}[inner sep=0in,outer sep=0in]
            \node[inner sep=0pt] (brane2) at (0,0) {\includegraphics[width=\textwidth]{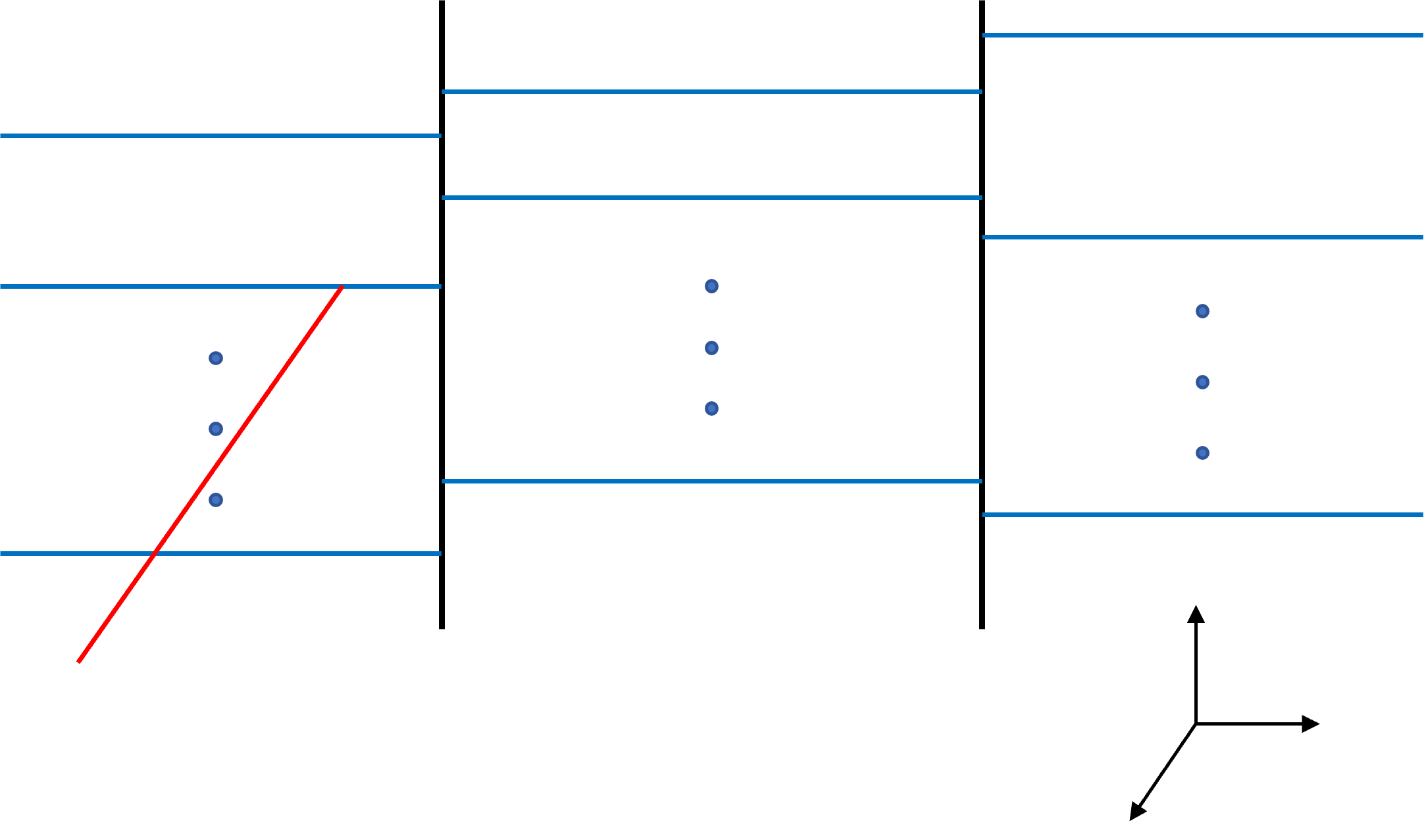}};
            \draw  (-1.3,2.4)  node {\footnotesize{NS5}};
            \draw  (3.2,2.2)  node {\footnotesize{D4}};
            \draw  (-3.2,-1.5)  node {\footnotesize{D2}};

            \draw  (2.7,-0.8)  node {\scriptsize{$x^{8,9}$}};
             \draw  (3.2,-1.45)  node {\scriptsize{$x^{4}$}};
             \draw  (1.85,-1.9)  node {\scriptsize{$x^{5}$}};
       \end{tikzpicture}
       \caption{}
       \end{subfigure}
    \caption{Transition between $\mathbf{Q}$-observable and $\mathbf{H}$-observable in IIA brane picture. (a) D2-brane meets one of the D4-branes on the left (resp. right) by tuning vev of the complex adjoint scalar (b) D2-brane moves toward left (resp. right) along the D4-brane, turning on positive (resp. negative) real FI parameter.}
    \label{fig:transition1}
\end{figure}

The transition can be intuitively understood in the IIA brane picture, at least in the absence of the $\O$-background (see Figure \ref{fig:transition1}). The D2-brane ending on one of the NS5-branes can be brought to the position (on the $(x^8, x^9)$-plane) of one of the $2N$ D4-branes ending on the same NS5-brane. There are $N$ D4-branes on the left and $N$ D4-branes on the right. Depending on which D4-brane it meets, the D2-brane can be moved toward left or right, deviating the real FI parameter from the locus $y=1$. In either case, there are $N$ choices of D4-brane that the D2-brane can end on, which correspond to the choice of vacuum at infinity among $N$ discrete vacua of the non-linear sigma model. Away from the locus $y=1$, the two non-linear sigma models are smoothly connected to each other, undergoing a flop transition of the target space across the zero K\"{a}hler parameter.

Implementing the $\O$-background, the $\mathbf{H}$-observable gives rise to an observable defined on the ensemble of fixed points of the instanton moduli space enumerated by the partitions $\{\bl\}$. The transition between the $\mathbf{Q}$- and $\mathbf{H}$-observables manifests at the level of their observable expressions, in the contour integral presentation of the latter \cite{Jeong:2018qpc} which generalizes the \textit{Coulomb branch presentation} \cite{Hori:2013ika} of the vortex partition function of two-dimensional $\EN=(2,2)$ gauged linear sigma model. When the real FI parameter is positive ($\vert y \vert >1$), the contour integral presentation is given by
\begin{align} \label{eq:transition}
    \mathbf{H} _{\text{left}} ^{(\a)} (y)[\bl] =\oint_{\CalC_\a} dx \, y^{-\frac{x}{\ve_1}}  \EQ(x)[\bl]
, \qquad  \begin{split} &\a=0,1,\cdots, N-1, \\ & \vert y\vert>1 \end{split}
\end{align}
where the integrand is the 1-loop fluctuation of the chiral multiplets at the vev of the complex adjoint scalar $x \in \BC$, which is essentially the $\mathbf{Q}$-observable (with the $\G$-functions from the 1-loop part of the $\mathbf{Q}$-observable). The integral kernel $y^{-\frac{x}{\ve_1}} = e^{- \frac{x\log y}{\ve_1}}$ is the classical contribution, with the complexified FI parameter $\log y$. There are $N$ semi-infinite $\ve_1$-lattices of simple poles due to the $\G$-functions, and the contour $\CalC_\a$ is chosen to enclose only the poles of single lattice, $x = m^+ _\a + k \ve_1$, $k \geq 0$. The resulting expressions are series in $y^{-1}$, converging in the domain $0<\vert \qe \vert <1  < \vert y \vert$ of $y \in \BP^1 \setminus \{0,\qe,1,\infty\}$. The $N$ choices for the contour exactly correspond to the choice of a vacuum at infinity among the $N$ discrete vacua of the non-linear sigma model. 

In the limit $\ve_2 \to 0$, the $\mathbf{H}$-observable gives a local observable in the effective two-dimensional theory on the $\BC_2$-plane. Accordingly, the vacuum expectation value in the limit becomes
\begin{align} \label{eq:hvac}
    \Big\langle \mathbf{H}^{(\a)} (y) \Big\rangle_\ba = e^{\frac{\widetilde{\EuScript{W}}(\ba;\qe) }{\ve_2}} \chi_\a (\ba;y), \qquad \a=0,1,\cdots, N-1,
\end{align}
where we denoted the normalized vacuum expectation value of the $\mathbf{H}$-observable in the limit $\ve_2 \to 0$ by $\chi_\a (\ba;y)$. Just as the $\mathbf{H}$-observables themselves, the vev $\chi_\a(\ba;y)$, which is given as a series which is a convergent in a specific domain in $\BP^1 \setminus \{0,\qe,1,\infty\}$, analytically continues to the other domains of $\BP^1 \setminus \{0,\qe,1,\infty\}$ \cite{Jeong:2018qpc}.

\subsubsection{$\mathbf{H}$-observables across domains of convergence}
Due to the coupling to the four-dimensional theory, the parameter space ($y$-space) for the $\mathbf{H}$-observable extends to the Riemann surface associated to the class $\CalS$ theory \cite{Gaiotto:2009fs}. In the present case of the $A_1$-quiver theory, it was explicitly shown in \cite{Jeong:2018qpc} that the vacuum expectation value of the $\mathbf{H}$-observables \eqref{eq:transition} analytically continues across different domains of convergence in the $y$-space $\BP^1 \setminus\{0,\qe,1,\infty\}$. We will not repeat the derivation of the analytic continuation here, but only gather some necessary facts to present the complete map of transitions between the $\mathbf{Q}$-observables and the $\mathbf{H}$-observables.

As noted earlier, the non-linear sigma model experiences a flop transition of the target across (but away from) $y=1$. At the level of the $\mathbf{H}$-observable, the flop transition manifests as the analytic continuation from one domain (say, $0<\vert \qe \vert<1 <\vert y \vert$) to the other ($0< \vert \qe \vert < \vert y \vert <1$). Once analytically continued, each of $N$ $\mathbf{H}$-observables \eqref{eq:transition} can be re-expressed as a linear combination of 
\begin{align} \label{eq:hmiddle1}
    \mathbf{H} _{\text{middle}} ^{(\a)} (y) = \sum_{x \in L_\a} y^{-\frac{x}{\ve_1}} \EQ (x), \qquad \begin{split}
        &\a=0,1,\cdots, N-1, \\ & 0 < \vert \qe \vert< \vert y \vert <1
    \end{split}
\end{align}
where $L_\a = a_\a + \ve_1 \BZ$ is the $\ve_1$-lattice centered at the Coulomb parameter $a_\a$. Due to the zeros of the inverse $\G$-functions in $\EQ(x)$, the non-zero contribution to the summation truncates from above and $\frac{x}{\ve_1} $ grows only to the negative infinity. It follows that the series converges in the domain $0 < \vert \qe \vert <\vert y \vert < 1$ \cite{Jeong:2018qpc}. See \cite{Jeong:2018qpc} for the exact connection formulas across the two domains.

It is apparent from the IIA brane picture (Figure \ref{fig:transition1}) that a D2-brane ending on a D4-brane in the middle can originate from a D2-brane ending on either the left NS5-brane or the right NS5-brane. In the gauge theory perspective, this implies the $\mathbf{H}$-observables in the domain $0<\vert \qe \vert < \vert y 
 \vert < 1$ can also be obtained by a transition from the $\tilde{\mathbf{Q}}$-observable. In the field theory terms, it suggests that there is an alternative weakly coupled description of the 2d/4d system where the 2d flavor group is identified with 4d flavor group in a different way (relevant to the masses $(m^-_\a)_{\a=0} ^{N-1}$ instead of $(m^+_\a)_{\a=0}^{N-1}$) and the complexified K\"{a}hler parameter is given by $\frac{\qe}{y}$ instead of $y$. The corresponding $\mathbf{H}$-observables are obtained as
\begin{align} \label{eq:hmiddle2}
    \tilde{\mathbf{H}} _{\text{middle}}^{(\a)} (y) = \oint_{\tilde{\CalC}_\a} dx \, y^{-\frac{x}{\ve_1}} \tilde{\EQ}(x) ,\qquad \begin{split}
        &\a=0,1,\cdots, N-1, \\ & 0 < \vert \qe \vert< \vert y \vert <1
    \end{split},
\end{align}
where
\begin{align}
    \tilde{\EQ}(x) = \frac{\tilde{\mathbf{Q}}(x)}{M^+(x)}= \qe^{\frac{x}{\ve_1}} \sum_{d=0} ^\infty \qe^d \left( \prod_{j= 1}^d \frac{j\ve_1 + \ve_2}{j \ve_1} \frac{P^+ (x+j\ve_1)}{\EY(x+j\ve_1)} \right)  \frac{M^- (x+d\ve_1)}{Q(x+ \ve_2 + (d+1)\ve_1)},
\end{align}
and the contour $\tilde{\CalC}_\a$ encloses the simple poles at $x = a_\a + (l(\l^{(\a)}) +l -d-1)\ve_1 -\ve_2$, $l\geq 0$. The vevs of the two $\mathbf{H}$-observables, \eqref{eq:hmiddle1} and \eqref{eq:hmiddle2}, in the same domain $0 <\vert \qe \vert < \vert y \vert <1$ are related to each other by a $\ve_2$-shift of a Coulomb parameter \cite{Jeong:2018qpc}.

Finally, the $\mathbf{H}$-observables \eqref{eq:hmiddle2} can be analytically continued to the last domain $0< \vert y \vert < \vert \qe \vert <1$ to be linear combinations of
\begin{align}
    \tilde{\mathbf{H}}_{\text{right}}^{(\a)} (y) = \sum_{x \in \tilde{L}_\a} y^{-\frac{x}{\ve_1}} \tilde{\EQ}(x),\qquad \begin{split}
        &\a=0,1,\cdots, N-1, \\
        &0< \vert y \vert < \vert \qe \vert <1
    \end{split},
\end{align}
where $\tilde{L}_\a = m^- _\a + \ve_1 \BZ$ is the $\ve_1$-lattice centered at the mass parameter $m^- _\a$. Due to the inverse $\G$-functions in $\tilde{\EQ} (x)$, the non-zero contribution to the summation truncates from above, guaranteeing the convergence of the series in the domain $0< \vert y \vert < \vert \qe \vert <1$. See \cite{Jeong:2018qpc} for the exact connection formulas across the two domains.

\subsubsection{$\mathbf{H}$-observables from higgsing}
The $\mathbf{H}$-observable built by coupling the non-linear sigma model with the target \eqref{eq:target} can also be realized by a partial higgsing \cite{Hanany:2004ea,Dorey:2011pa,Nikita:IV,Jeong:2018qpc}. In this construction, $N$ hypermultiplets are brought to be nearly massless, admitting a non-zero vacuum expectation value. After the vev is turned on, the BPS field configuration is squeezed into the $\BC_1$-plane, forming a vortex string coupled to the 4d theory. See \cite{Jeong:2018qpc, Nikita:IV}, for instance, for the implementation of the partial higgsing construction of the $\mathbf{H}$-observable at the level of the partition function.

\begin{figure}[h!]
    \centering
    \begin{subfigure}[b]{0.45\textwidth}
       \centering
       \begin{tikzpicture}[inner sep=0in,outer sep=0in]
            \node[inner sep=0pt] (brane1) at (0,0)  {\includegraphics[width=\textwidth]{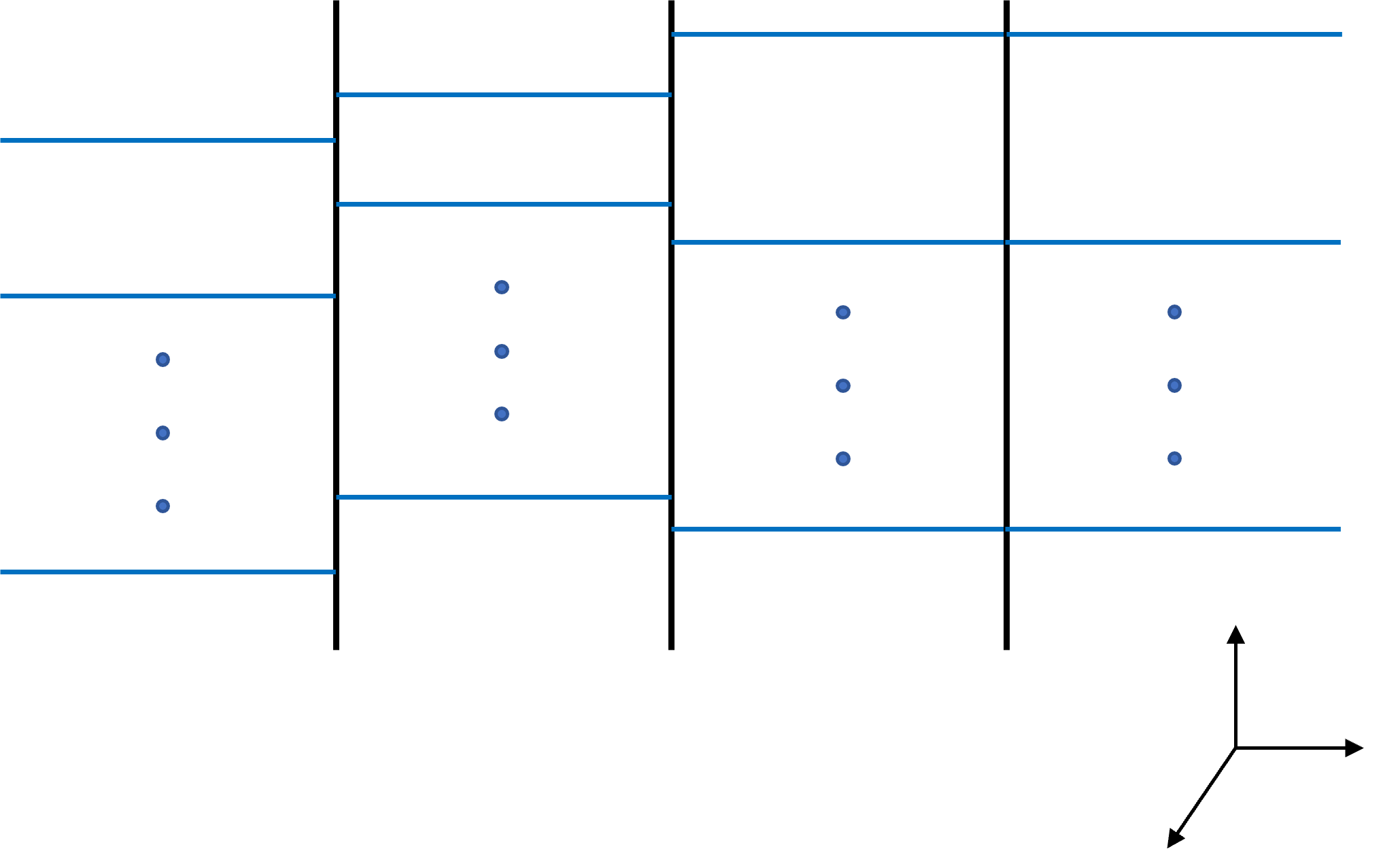}};
            \draw  (-1.8,2.4)  node {\footnotesize{NS5}};
            \draw  (3,2.25)  node {\footnotesize{D4}};

            \draw  (2.7,-0.8)  node {\scriptsize{$x^{8,9}$}};
             \draw  (3.4,-1.35)  node {\scriptsize{$x^{4}$}};
             \draw  (2.2,-2)  node {\scriptsize{$x^{5}$}};
       \end{tikzpicture}
       \caption{}
       \end{subfigure}
       \quad\quad\quad
    \begin{subfigure}[b]{0.45\textwidth}
       \centering
       \begin{tikzpicture}[inner sep=0in,outer sep=0in]
            \node[inner sep=0pt] (brane2) at (0,0) {\includegraphics[width=\textwidth]{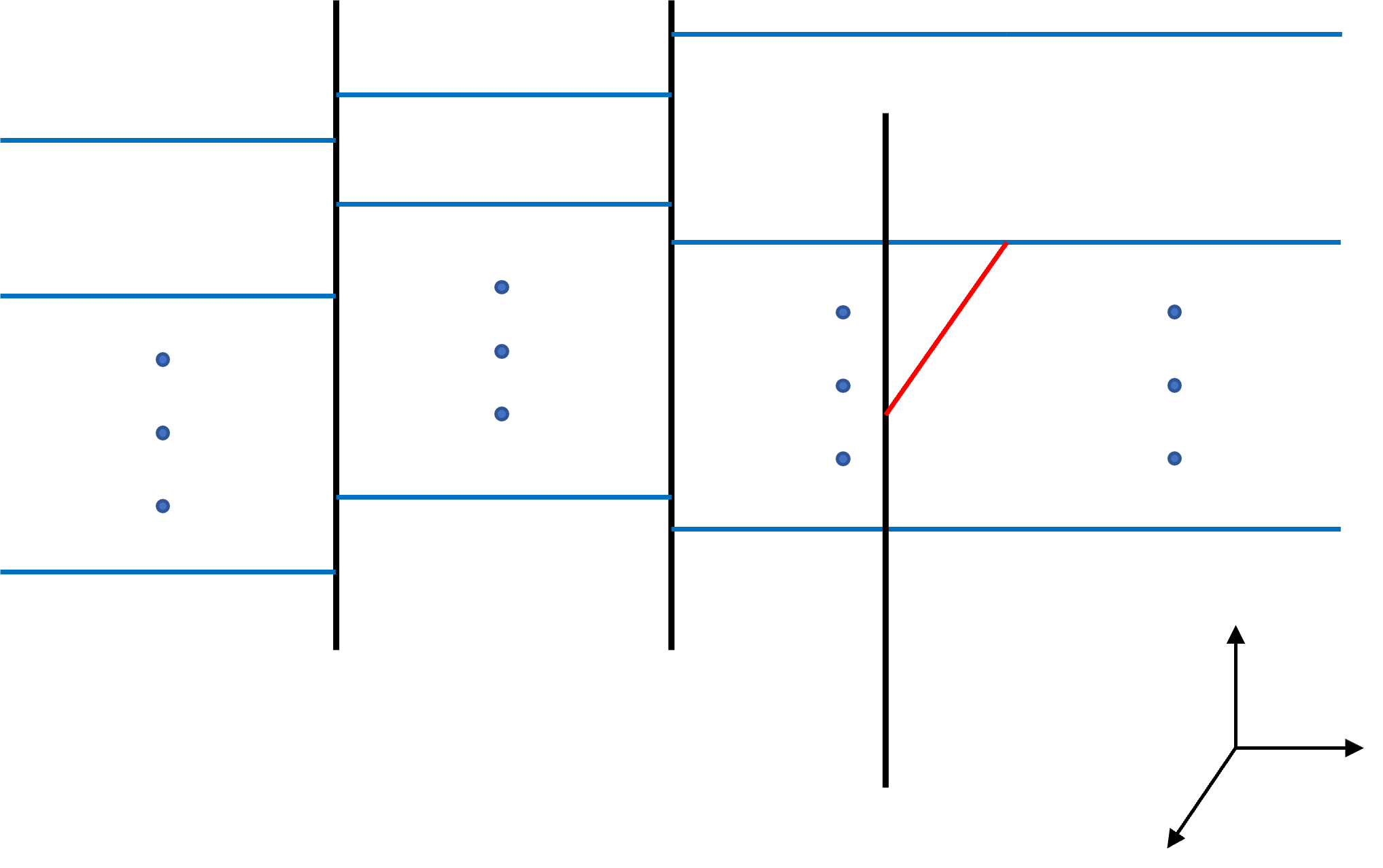}};
            \draw  (-1.8,2.4)  node {\footnotesize{NS5}};
            \draw  (3,2.25)  node {\footnotesize{D4}};
            \draw  (1.5,0.35)  node {\footnotesize{D2}};
            \draw  (0.6,-1.5)  node {\footnotesize{NS5}};
            
            \draw  (2.7,-0.8)  node {\scriptsize{$x^{8,9}$}};
             \draw  (3.4,-1.35)  node {\scriptsize{$x^{4}$}};
             \draw  (2.2,-2)  node {\scriptsize{$x^{5}$}};
       \end{tikzpicture}
       \caption{}
       \end{subfigure}
    \caption{D2-brane creation and higgsing construction of the $\mathbf{H}$-observable. (a) All $N$ D4-branes are aligned across one of the NS5-branes, enabling the NS5-brane to move along the $x^5$-direction. (b) A D2-brane ending on one of the D4-branes is created as a result of the higgsing.}
    \label{fig:higgscan}
\end{figure}
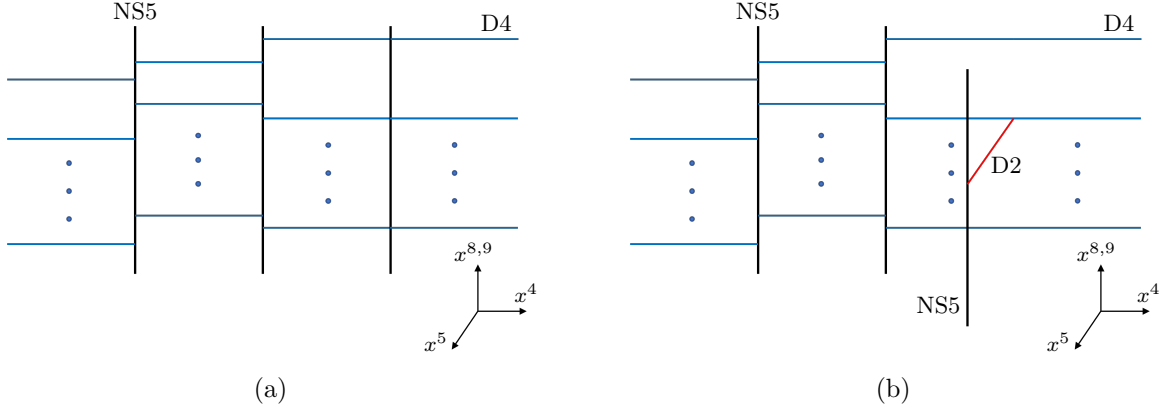

The higgsing construction of the $\mathbf{H}$-observables can easily visualized in the IIA brane setup (see Figure \ref{fig:higgscan}) \cite{Hanany:2004ea,Dorey:2011pa}. We begin with a higher-rank quiver gauge theory with one more gauge node, which is realized by the IIA branes with an additional NS5-brane. A D2-brane creation can be initiated by aligning all the D4-branes across one of the NS5-branes, by pulling out the relevant NS5-brane along the $x^5$-direction. The worldvolume theory on the D2-brane is precisely the non-linear sigma model with the target \eqref{eq:target}, which couples to the remaining 4d theory by gauging its flavor symmetry.

We refer to \cite{Hanany:2004ea,Dorey:2011pa,Nikita:IV, Jeong:2018qpc} more details on the higgsing construction of the $\mathbf{H}$-observable surface defect. Here, we only remind that the higgsing construction of the $\mathbf{H}$-observables exactly parallels the higgsing construction of the $\mathbf{Q}$-observables in section \ref{subsubsec:qhiggs}. Here, \textit{all} the D4-branes are aligned across \textit{one} of the NS5-brane, enabling the NS5-brane to move along the $x^5$-direction. A D2-brane ending on one of the D4-branes is created, engineering a $\mathbf{H}$-observable in the remnant $\EN=2$ gauge theory. In higgsing construction of the $\mathbf{Q}$-observables, instead, only \textit{one} of the D4-branes is aligned across \textit{all} the NS5-branes, enabling the D4-brane to move along the $x^5$-direction. The newly created D2-brane ends on one of the NS5-branes, giving a $\mathbf{Q}$-observable.

\subsection{Monodromy surface defect} \label{subsec:monodef}
The monodromy surface defect in the four-dimensional $\EN=2$ gauge theory is engineered by assigning singular behavior of the gauge field along a surface \cite{gukwit}. The singular behavior of the gauge field can be modelled by placing the $\EN=2$ gauge theory on an orbifold \cite{K-T}, $\BC_1 \times (\BC_2 /\BZ_l)$. As we review in appendix \ref{subsec:orimono}, the construction can be embedded into the IIB origami setup by replacing a part of the worldvolume to the orbifold.

\subsubsection{Construction by orbifold}
 To model the singularity of the gauge field, we consider the $\EN=2$ gauge theory supported on the orbifold $\BC_1 \times (\BC_2 /\BZ_l)$, where the $\BZ_l$-action is given by \cite{K-T},
\begin{align}
    (z_1, z_2) \mapsto (z_1 , \zeta z_2 ),\qquad \z = \exp \frac{2\pi i}{l}.
\end{align}
The map $z_2 \mapsto z_2 ^l$ sends the $\EN=2$ gauge theory supported on the $\BZ_l$-orbifold to the one supported on the ordinary $\BC_1 \times \BC_2$, compensating the presence of the orbifold by singularity of fields along the locus $\{z_2 = 0\}$. Replacing the support of the worldvolume of the $\EN=2$ gauge theory to the $\BZ_l$-orbifold engineers a monodromy surface defect in this sense.

The monodromy surface defects are classified by the Levi subgroups of the gauge group and the flavor group it preserves \cite{gukwit}, which we encode in \textit{coloring functions}. In our case of the $A_1$-quiver $U(N)$ gauge theory, the coloring functions $(c,c_f,c_{af})$ assign $\BZ_l$-charges to the framing and the flavor bundles by
\begin{align} \label{eq:colormono}
\begin{split}
    &c: N \longrightarrow \{0,1,\cdots, l-1\} \\
    &c_f : M^+ \longrightarrow \{0,1,\cdots, l-1\} \\
    &c_{af} : M^- \longrightarrow \{0,1,\cdots, l-1\}.
\end{split}
\end{align}
The preserved Levi subgroups are determined to be $S\left(\bigtimes_{\o=0} ^{l-1} U(\vert c^{-1}(\o) \vert)\right)$, etc. 

In the special case where $l =N$ and the coloring functions $(c,c_f,c_{af})$ are one-to-one functions, we may set $c(\a) = c_f (\a) = c_{af} (\a) = \a$) without loss of generality. We call the corresponding monodromy surface defect to be regular. In the present work, we will mainly restrict to the regular monodromy surface defect, even though we will give a dual description of monodromy surface defect defined in the most general case.

\subsubsection{Monodromy surface observable}
To avoid duplication with the companion paper \cite{Jeong:2024hwf}, we do not present the derivation of the observable expression for the monodromy surface defect here. See \cite{Jeong:2023qdr,Jeong:2024hwf} and appendix \ref{subsec:orimono} for the detail of the definitions and the conventions. The observable expression of the regular monodromy surface defect labelled by the coloring function $\mathbf{c}$ is given by (cf. \eqref{eq:monoapp})
\begin{align} \label{eq:monoobser}
\begin{split}
    &\Psi _{\mathbf{c}} (\mathbf{u}) [\bl] =\sum_{\hat{\bl} \in \r^{-1} (\bl)} \prod_{\o=0} ^{l-2} \qe_\o ^{k_\o - k_{l-1}}  \BE\left[ \frac{S \S_0 ^*}{P_1 ^*} + \sum_{\o=0} ^{l-2} \frac{- \S_\o (\S_\o - \S_{\o+1})^* - M^+ _\o \S_\o ^* + q_1 ^{-1} (M^- _{\o}) ^* \S_\o}{P_1 ^*} \right].
\end{split}
\end{align}
There are $l-1$ defect parameters $(\qe_\o)_{\o=0} ^{l-2}$ encoding the singularity of the gauge field and the magnetic flux along the support $\BC_1$. 

We will re-parameterize these defect parameters using $\mathbf{u} = (u_\o)_{\o=0} ^{l-1}$,
\begin{align}
    \qe_\o = \frac{u_{\o+1}}{u_\o},\qquad \o=0,1,\cdots, l-2,
\end{align}
with a redundancy of overall scaling. Also, we will incorporate the classical contribution to the observable \eqref{eq:monoobser}, given by
\begin{align} \label{eq:classcal}
    \Psi_{\mathbf{c}} (\bu) ^{\text{classical}} = \prod_{\o=0} ^{l-1} u_\o ^{ \sum_{\a \in c^{-1} (\o)} \frac{m^+ _\a - a_\a }{\ve_1} }.
\end{align}

\subsubsection{Space of regular monodromy surface defects}
For our discussion on the ($\hbar$-)Langlands correspondence and the associated bispectral duality, we will restrict to the case of the regular monodromy surface defect. 

It is immediate that the vacuum expectation value of the monodromy surface defect in the space of degree-zero Laurent polynomials in the monodromy defect parameters $\mathbf{u}= (u_\o)_{\o=0} ^{N-1}$; namely,
\begin{align} \label{eq:psiele}
   \Big\langle \Psi(\mathbf{u}) \Big\rangle_\ba \in \CalH ^{\BC^\times},\qquad \CalH = \bigotimes_{\o=0}^{N-1} \CalH_\o ,
\end{align}
where $\CalH_\o = u_\o ^{ \frac{m^+ _\o - a_\o }{\ve_1} } \BC(u_\o )$ is the space of Laurent polynomials in $u_\o$. Here, the degree-zero condition $\CalH^{\BC^\times}$ 
really means being degree-zero except the factor in front given by \eqref{eq:classcal}; namely, $\CalH^{\BC^\times} =\prod_{\o=0} ^{N-1} u_\o ^{  \frac{m^+ _\o - a_\o }{\ve_1} } \BC(u_0,u_1,\cdots, u_{l-1})^{\BC^\times}$. In the companion paper \cite{Jeong:2024hwf}, we viewed $\CalH^{\BC^\times}$ as the degree-zero subspace of the evaluation module over the Yangian $Y(\fgl(2))$, defined upon $N$ bi-infinite $\fgl(2)$-modules.\footnote{These modules are said to be of bi-infinite type, since they are not of highest-weight (nor lowest-weight) so that the weights grow toward both infinities. See \cite{Jeong:2023qdr} for precise definitions.} We refer to the companion paper \cite{Jeong:2024hwf} for the precise definition of the $\fgl(2)$-module structure on $\CalH_\o$, and the mapping between the module parameters and the $\EN=2$ gauge theory parameters. In particular, the associated $\fgl(2)$ XXX spin chain was formulated on this bi-infinite Yangian module.

Meanwhile, let us briefly recap from \cite{Nekrasov:2021tik, Jeong:2021bbh, Jeong:2023qdr} that the same vev of the regular monodromy surface defect is valued in the space of $E$-twisted coinvariants of four $\widehat{\fsl}(N)$-modules, where $E \in \text{Bun}_{PGL(N)} (\BP^1;S)$. Here, the four $\widehat{\fsl}(N)$-modules are induced from two (co-)Verma $\fsl(N)$-modules ($V_{\boldsymbol\z}$ and $\tilde{V}_{\tilde{\boldsymbol\z}}$) and two bi-infinite $\fsl(N)$-modules ($H_{\boldsymbol\t - \boldsymbol\z,\s}$ and $H_{\boldsymbol\t - \tilde{\boldsymbol\z},\tilde{\s}}$). The space of twisted $\widehat{\fsl}(N)$-coinvariants is isomorphic to the space of $\fsl(N)$-invariants \cite{Feigin:1994in}, so that
\begin{align} \label{eq:slinv}
     \Big\langle \Psi(\mathbf{u}) \Big\rangle_\ba \in M ^{\fsl(N)},\qquad M = V_{\boldsymbol{\z}} \otimes H_{\boldsymbol{\t} - \boldsymbol{\z},\s} \otimes H_{\boldsymbol{\t} - \tilde{\boldsymbol{\z}},\tilde{\s}} \otimes \tilde{V}_{\tilde{\boldsymbol\z}},
\end{align}
where the monodromy parameters $\mathbf{u}$ give holomorphic coordinates on $\text{Bun}_{PGL(N)} (\BP^1;S)$. We refer to our previous work \cite{Jeong:2023qdr} for the precise definitions of the (co-)Verma $\fsl(N)$-modules ($V_{\boldsymbol\z}$ and $\tilde{V}_{\tilde{\boldsymbol\z}}$) and the bi-infinite $\fsl(N)$-modules ($H_{\boldsymbol\t - \boldsymbol\z,\s}$ and $H_{\boldsymbol\t - \tilde{\boldsymbol\z},\tilde{\s}}$), and the mapping between the module parameters and the $\EN=2$ gauge theory parameters. In particular, the associated $\fsl(N)$ Gaudin model was defined on these (bi-)infinite $\fsl(N)$-modules.

At this point, let us split the Coulomb moduli $(a_\a)_{\a=0} ^{N-1}$ into two parts: the reference values $\frac{a_\a}{\ve_1}$ mod $\BZ$ and the $\ve_1$-integral shifts. We regard the reference values $\frac{a_\a}{\ve_1}$ mod $\BZ$ to be fixed, so that the vacuum expectation value \eqref{eq:psiele} and \eqref{eq:slinv} always gives an element of the same spaces $\CalH^{\BC^\times}$ and $M^{\fsl(N)}$. Namely,
\begin{align}
     \Big\langle \Psi(\mathbf{u}) \Big\rangle_{\ba + \ve_1 \mathbf{n} } \in \CalH ^{\BC^\times},\qquad  \Big\langle \Psi(\mathbf{u}) \Big\rangle_{\ba + \ve_1 \mathbf{n} } \in M ^{\fsl(N)}.
\end{align}
There are $\BZ^{N-1}$ amount of remaining degrees of freedom, $\mathbf{n}$, for the $\ve_1$-shifts of $N$ Coulomb moduli up to the vanishing trace constraint $\sum_{\a=0} ^{N-1} a_\a =0$. The vacuum expectation value of the monodromy surface defect provides a basis of the spaces $\CalH ^{\BC^\times}$ and $M^{\fsl(N)}$ enumerated by the $\ve_1$-integral shifts in the Coulomb moduli $\ba$.

In the limit $\ve_2 \to 0$ where the four-dimensional theory is described by the effective two-dimensional theory on the $\BC_2$-plane, the monodromy surface defect becomes a local defect since it is supported on the $\BC_1$-plane. Its vacuum expectation value becomes
\begin{align} \label{eq:monodefns}
    \lim_{\ve_2 \to 0} \Big\langle \Psi (\mathbf{u}) \Big\rangle_\ba = e^{\frac{\widetilde{\EuScript{W}}(\ba;\qe)}{\ve_2}} \psi (\ba;\mathbf{u};\qe),
\end{align}
where we denote the normalized vacuum expectation value of the monodromy surface defect in the limit $\ve_2 \to 0$ as $\psi (\ba;\mathbf{u};\qe)$. Note the effective twisted superpotential $\widetilde{\EuScript{W}}(\ba;\qe)$ does not depend on the defect parameters $\mathbf{u}$, and the normalized vacuum expectation value $\psi (\ba;\mathbf{u};\qe)$, with $\ve_1$-integral shifts in the Coulomb moduli $\ba$, still provides a distinguished basis of the spaces $\CalH^{\BC^\times}$ and $M^{\fsl(N)}$.

\section{Bispectral duality from $\mathbf{Q}/\mathbf{H}$ surface defect transition} \label{sec:hbarlang}
In this section, we explain that the bispectral duality between the associated XXX spin chain and the Gaudin model follows from the transition between the $\mathbf{Q}$-observables and the $\mathbf{H}$-observables. First, we recap how $\mathbf{Q}$- and $\mathbf{H}$-observables give rise to the $Q$-operators and the Hecke operators. The transition between the two observables immediately yields the Fourier transform between the universal $\hbar$-oper equation and the universal oper equation satisfied by the respective operators. The one-to-one mapping between the spectral equations of the $\fgl(2)$ XXX spin chain with $N$ sites and the $\fsl(N)$ Gaudin model with $4$ sites is obtained as a direct consequence.

\subsection{Surface defects as operators}
We consider the configuration where the $\mathbf{Q}$-observable or the $\mathbf{H}$-observable is inserted on the same $\BC_1$-plane on top of the regular monodromy surface defect. In the gauge origami setup, such a configuration descends from adding a stack of $N$ D3-branes to the stack of D3-branes engineering the $\EN=2$ gauge theory with the monodromy surface defect, each of which carries a $\BZ_N$-charge $\o=0,1,\cdots, N-1$ (see \cite{Jeong:2023qdr, Jeong:2024hwf}).

\subsubsection{$\EN=2$ gauge theory formulation of ($\hbar$-)Langlands correspondence}
Let us first briefly recap the role of such parallel surface defect configurations in the $\EN=2$ gauge theoretical account for the ordinary \cite{Jeong:2023qdr} and the $\hbar$-deformed Langlands correspondence \cite{Jeong:2024hwf}, at our main example of the $A_1$-quiver $U(N)$ theory.

We studied the geometric Langlands correspondence on the Riemann sphere $\EC = \BP^1$ with ramifications at four marked points $S = \{0,\qe,1,\infty\} \subset \BP^1$.\footnote{The marked points at $0$ and $\infty$ were taken to be \textit{maximal} while the ones at $\qe$ and $1$ were \textit{minimal}. For their precise definitions, see \cite{Jeong:2023qdr}.} On the automorphic side, we had coinvariants of four $\widehat{\fsl}(N)$-modules twisted by parabolic $PGL(N)$-bundles $E \in \text{Bun}_{PGL(N)} (\BP^1;S)$, which form sections of a twisted $\ED$-module on $\text{Bun}_{PGL(N)} (\BP^1;S)$ by the Ward identities. The relevant $\widehat{\fsl}(N)$-modules were the ones induced from $\fsl(N)$-modules by evaluation homomorphism, two of which are (co-)Verma modules (attached to $0$ and $\infty$) and the other two were of bi-infinite type (attached to $\qe$ and $1$). We constructed a distinguished basis $\psi(\ba;\mathbf{u})$, enumerated by the Coulomb moduli $\ba$, of these sections as the normalized vacuum expectation value of the regular monodromy surface defect in the limit $\ve_2 \to 0$. In particular, the monodromy defect parameters $\mathbf{u}$ were identified with holomorphic coordinates on an open patch in $\text{Bun}_{PGL(N)} (\BP^1;S)$.

On the Galois side, we considered the oper submanifold in the moduli space of parabolic $SL(N)$ local systems on $\BP^1$ with four regular singularities at $S=\{0,\qe,1,\infty\}$, $\text{Op}_{SL(N)} (\BP^1;S) \subset \EM_{\text{loc}} (SL(N),\BP^1;S)$. We realized an oper $\r_\ba \in \text{Op}_{SL(N)} (\BP^1;S)$, enumerated by the Coulomb moduli $\ba$, as a $N$-th order differential equation of the quantized chiral ring for the $\EN=2$ gauge theory coupled to the surface defect theory for the $\mathbf{H}$-observable. In particular, $N$ independent solutions to the oper differential equation were obtained as the normalized vacuum expectation values $\chi_\a (\ba;y)$ of the $\mathbf{H}$-observable, where $y \in \BP^1 \setminus \{0,\qe,1,\infty\}$ and $\a=0,1,\cdots, N-1$ labels the choice of the vacuum at infinity.

To establish the geometric Langlands correspondence in the $\EN=2$ theory context, we showed that the action of inserting a $\mathbf{H}$-observable on top of the regular monodromy surface defect defines a Hecke operator on the twisted $\ED$-module. The action was proven to be diagonal on the distinguished basis element $\psi(\ba;\mathbf{u})$ due to cluster decomposition of the two surface defects in the limit $\ve_2 \to 0$. Thus, the normalized vacuum expectation value $\psi(\ba;\mathbf{u})$ of the regular monodromy defect was shown to be a section of the Hecke eigensheaf corresponding to the oper $\r_\ba$, with the corresponding \textit{eigenvalue} being precisely the oper solution $\chi_\a (\ba;y)$, the normalized vacuum expectation value of the $\mathbf{H}$-observable. 

The $\EN=2$ gauge theoretical account of the $\hbar$-Langlands correspondence that we have established in the companion paper \cite{Jeong:2024hwf} completely parallels the prior study \cite{Jeong:2023qdr}. In particular, we showed that the action of inserting a $\mathbf{Q}$-observable on top of the same regular monodromy surface defect defines a $Q$-operator on the module over the Yangian. The action is shown to be diagonal on the distinguished basis element $\psi(\ba;\mathbf{u})$ by the cluster decomposition of the two surface defects in the limit $\ve_2 \to 0$. Accordingly, the normalized vev $\psi(\ba;\mathbf{u})$ of the regular monodromy surface defect is shown to be a $Q$-eigenstate corresponding to a $\hbar$-oper, with the eigenvalue being the $Q$-function, i.e., the normalized vev of the $\mathbf{Q}$-observable.

\subsubsection{$\mathbf{Q}$-observables as $Q$-operators} \label{subsubsec:paralleldefs}
In the companion paper \cite{Jeong:2024hwf}, we viewed the $\mathbf{Q}$-observables as $Q$-operators acting on the space $\CalH^{\BC^\times}$ as follows. Even with the additional insertion of the $\mathbf{Q}$-observable, the correlation function of the $\mathbf{Q}$-observable and the monodromy surface defect is valued in the space of degree-zero Laurent polynomials in the monodromy defect parameters $\mathbf{u}=(u_\o)_{\o=0} ^{N-1}$,
\begin{align} \label{eq:correl}
    \mathbf{Q}(x)\cdot \Big\langle \Psi(\bu) \Big\rangle_\ba := \Big \langle \mathbf{Q}(x) \Psi(\mathbf{u}) \Big\rangle_\ba  \in \CalH ^{\BC^\times} = \left( \bigotimes_{\o=0}^{N-1} \CalH_\o \right)^{\BC^\times}, \quad \text{also for } \tilde{\mathbf{Q}}(x),
\end{align}
just as the vacuum expectation value of the regular monodromy surface defect itself. In other words, we can view the $\mathbf{Q}$-observable also as an operator
\begin{align} \label{eq:Qoper}
\mathbf{Q}(x), \tilde{\mathbf{Q}}(x) \in \text{End}\left(\CalH ^{\BC^\times} \right),
\end{align}
whose action on the basis element $\langle \Psi(\mathbf{u}) \rangle_\ba $ is given by the above correlation function. Under this context, we refer to $\mathbf{Q}(x)$ and $\tilde{\mathbf{Q}}(x)$ as the \textit{$Q$-operators}. Since the contribution at each partition $\bl$ is modified by the insertion of the $\mathbf{Q}$-observable, the action of $Q$-operator is highly non-trivial.

One of the main results of the companion paper \cite{Jeong:2024hwf} is that the so-defined $Q$-operators satisfy the universal $\hbar$-oper difference equation in the limit $\ve_2 \to 0$,
\begin{align}\label{eq:univhop}
\begin{split}
   & 0 = \left[1- \hat{t}(x) e^{-\ve_1 \p_x} + \qe P(x) e^{-2 \ve_1 \p_x} \right]\mathbf{Q}(x) \\
   &0 = \left[1- \hat{t}(x) e^{-\ve_1 \p_x} + \qe P(x) e^{-2 \ve_1 \p_x} \right]\tilde{\mathbf{Q}}(x)
\end{split},
\end{align}
where $\hat{t}(x) \in \text{End}(\CalH^{\BC^\times})$ is the transfer matrix of the Yangian $Y(\fgl(2))$, i.e., the trace of the monodromy matrix whose coefficients generate the Bethe subalgebra of $Y(\fgl(2))$, represented on $\CalH^{\BC^\times}$. The universal $\hbar$-oper equation is nothing but the Baxter's TQ equation for the $Q$-operators and the transfer matrix for the $\fgl(2)$ XXX spin chain. This explains our terminology of referring to $\mathbf{Q}$-observables as the $Q$-operators.

\subsubsection{$\mathbf{H}$-observables as Hecke operators}
Meanwhile, in our previous work \cite{Jeong:2023qdr}, we defined the action of the $\mathbf{H}$-observables on the twisted coinvariants $\left\langle \Psi(\mathbf{u}) \right\rangle_\ba \in M^{\fsl(N)}$ by its insertion into the correlation function in the limit $\ve_2 \to 0$. Namely,
\begin{align} \label{eq:heigen}
    \mathbf{H}^{(\a)} (y)\cdot \Big\langle \Psi(\mathbf{u}) \Big\rangle_\ba =  \Big\langle 
\mathbf{H}^{(\a)} (y) \Psi(\mathbf{u}) \Big\rangle_\ba \in M^{\fsl(N)},\quad \a=0,1,\cdots, N-1.
\end{align}
More precisely, the action of the Hecke modifications on the twisted $\widehat{\fsl}(N)$-coinvariants is realized by the insertion of the $\mathbf{H}$-observable and a generating function of additional defect $0$-observables, with the extra fugacities. The extra fugacities precisely parametrize the space $\BP^{N-1}$ of Hecke modifications. Then, the Hecke operator was defined by a contour integral on this $\BP^{N-1}$ in the limit $\ve_2 \to 0$, which picks up the insertion of the $\mathbf{H}$-observable without any further insertion of the defect $0$-observable.

Using the constraints of the correlation function of the parallel surface defects, it was shown that the Hecke operator satisfies the universal oper equation
\begin{align} \label{eq:univop}
    0 = \left[\p_y ^N + \hat{\mathrm{t}}_2 (y) \p_y ^{N-2} + \cdots + \hat{\mathrm{t}}_{N-1} (y) \p_y + \hat{\mathrm{t}}_N (y) \right] \hat{\mathbf{H}}^{(\a)} (y),\quad \a=0,1,\cdots, N-1,
\end{align}
where we incorporated a classical part and decoupled an abelian $U(1)$ factor,
\begin{align} \label{eq:redef}
    \hat{\mathbf{H}} (y) = y^{\frac{\bar{m}^-}{\ve_1}+ \frac{N-3}{2}} (y-\qe)^{- \frac{\bar{m}^- -\bar{a} }{\ve_1}} (y-1) ^{1+ \frac{\bar{m}^+ -\bar{a}}{\ve_1}} \mathbf{H}(y),
\end{align}
so that the universal oper has vanishing coefficient of $\p_y ^{N-1}$. The universal oper equation is indeed the characteristic feature of the Hecke operator \cite{Etingof:2021eeu,Etingof:2021eub,Etingof:2023drx}. Here, the operator-valued meromorphic functions $\hat{\mathrm{t}}_k (y)$ have poles of order $\text{ord}_p (\hat{\mathrm{t}}_k (y)) = -k$ at the marked points $S = \{0,\qe,1,\infty\}$. Their Laurent coefficients are precisely the Gaudin Hamiltonians (valued in $\text{End}(M^{\fsl(N)})$), given by twisted differential operators on $\text{Bun}_{PGL(N)} (\BP^1;S)$. See our previous work \cite{Jeong:2023qdr} for more detail on the $\EN=2$ gauge theoretical construction of the universal oper and its application to the geometric Langlands correspondence.

\subsection{Eigenvalue properties and bispectral duality}
In our previous work \cite{Jeong:2023qdr} and the companion paper \cite{Jeong:2024hwf}, it was shown that the vevs of the regular monodromy surface defects in the limit $\ve_2 \to 0$, on which the Hecke operators and $Q$-operators act diagonally, give common eigenfunctions of the quantum Hamiltonians of the Gaudin model and the XXX spin chain, respectively. Here, we show that the two spectral problems are equivalent precisely through the transformation between the $\mathbf{Q}$-observables and the $\mathbf{H}$-observables.

\subsubsection{Eigenvalue properties for $Q$-operators and Hecke operators}
In \cite{Jeong:2023qdr}, we showed that insertion of a $\mathbf{H}$-observable in the limit $\ve_2 \to 0$ gives the action of a Hecke operator on the twisted covinvariants of $\widehat{\fsl}(N)$-modules. On the distinguished basis $\psi(\ba;\mathbf{u})$ constructed as the normalized vacuum expectation value of the regular monodromy surface defect, the action of Hecke operator was shown to be diagonal, 
\begin{align}
    \mathbf{H} ^{(\a)} (y) \cdot \psi(\ba;\mathbf{u}) = \lim_{\ve_2 \to 0}\Big\langle \mathbf{H} ^{(\a)} (y) \Psi(\mathbf{u}) \Big\rangle_\ba = \chi_\a (\ba;y) \psi(\ba;\mathbf{u}), \quad \a=0,1,\cdots, N-1,
\end{align}
due to the cluster decomposition of the $\mathbf{H}$-observable and the regular monodromy surface defect in the limit $\ve_2 \to 0$. Here, the eigenvalue $\chi_\a (\ba;y)$ is precisely the $\a$-th solution to the oper differential equation, constructed as the normalized vacuum expectation value \eqref{eq:hvac} of the $\mathbf{H}$-observable with the $\a$-th vacuum at infinity \cite{Jeong:2018qpc,Jeong:2023qdr}. 

In the point of view of the transition \eqref{eq:transition} between the $\mathbf{Q}$-observable and the $\mathbf{H}$-observable, this Hecke eigensheaf property is equivalent to the \textit{$Q$-eigenstate property} of the normalized vacuum expectation value $\psi(\ba;\bu)$ of the regular monodromy surface defect, now viewed as an element in the Yangian module $\psi(\ba;\mathbf{u}) \in \CalH^{\BC^\times}$. Namely, upon the action of the $Q$-operators in the limit $\ve_2 \to 0$, we have
\begin{align} \label{eq:qeigen}
    \mathbf{Q}(x) \cdot \psi(\ba;\mathbf{u}) = \lim_{\ve_2 \to 0}\Big \langle \mathbf{Q}(x) \Psi(\mathbf{u}) \Big\rangle_\ba = Q(\ba;x) \psi(\ba;\mathbf{u}),\quad \text{same for  }\, \tilde{\mathbf{Q}}(x),
\end{align}
again by the cluster decomposition. Thus, just as the Hecke operators, the $Q$-operators also act diagonally on the basis $\psi(\ba;\mathbf{u})$ of the space $\CalH ^{\BC^\times}$ enumerated by the $\ve_1$-integral shifts of the Coulomb moduli $\ba$, with the eigenvalue being the $Q$-functions $Q(\ba;x)$ (and $\tilde{Q}(\ba;x)$) solving the $\hbar$-oper difference equation \cite{Jeong:2024hwf}.

In this sense, the $Q$-operator and the Hecke operator are related to each other by the transition in the surface defect theory, turning on either the vev of the complex adjoint scalar or the complexified FI parameter. In particular, their eigenvalues $-$ $Q(\ba;x)$ (or $\tilde{Q}(\ba;x)$) and $\chi_\a (\ba;y)$ $-$ are related precisely by the Fourier transformation \eqref{eq:transition} and its analogues in other convergence domains of $\BP^1 \setminus \{0,\qe,1,\infty\}$. We emphasize again that the surface defect realization of the $Q$-operator and the Hecke operator in our $\EN=2$ theory setup gives a simple intuitive picture for these eigenvalue properties: the cluster decomposition of the correlation function of surface defects supported on the $\BC_1$-plane in the limit $\ve_2 \to 0$.

\subsubsection{Bispectral duality of XXX spin chain and Gaudin model} \label{subsubsec:bispectral}
Finally, let us investigate the implication of the transition between the $\mathbf{Q}$-observables and the $\mathbf{H}$-observables on the associated integrable models. Due to the eigenvalue property \eqref{eq:qeigen}, the universal $\hbar$-oper equation \eqref{eq:univhop} satisfied by the $\mathbf{Q}$-observables gives
\begin{align} \label{eq:univhoperre}
\begin{split}
       & 0 = \left[1 - \hat{t} (x) e^{-\ve_1 \p_x} +  \qe P(x) e^{-2\ve_1 \p_x} \right] Q(\ba;x+\ve_1) \psi(\ba;\mathbf{u};\qe), \\
                &0 = \left[1 - \hat{t} (x) e^{-\ve_1 \p_x} +  \qe P(x) e^{-2\ve_1 \p_x} \right] \tilde{Q}(\ba;x+\ve_1) \psi(\ba;\mathbf{u};\qe),
\end{split}
\end{align}
at each basis element $\psi(\ba;\mathbf{u};\qe)$. Recall that the coefficients of the degree $N$ polynomials $\hat{t} (x) \in \text{End}(\CalH^{\BC^\times})$ are quantum Hamiltonians of the $\fgl(2)$ XXX spin chain with $N$ sites. By the transition \eqref{eq:transition}, we can convert the $Q$-function $Q(\ba;x)$ into the oper solution $\chi_\a (\ba;y)$ by a Fourier transformation. Since the kernel of the Fourier transformation is $y^{-\frac{x}{\ve_1}}$, the shift operator $e^{-\ve_1 \p_x}$ gets compensated by a multiplication by $y$ while the multiplication by $x$ is converted into the differential operator $-\ve_1 y \p_y$. Note that 
 $y \p_y$ obviously commutes with itself so that polynomials in $y\p_y$ are well-defined. All in all, the above equation is brought into an operator-valued $N$-th order differential equation \cite{Jeong:2018qpc, Jeong:2021bbh, Jeong:2023qdr}
\begin{align}
    0 = \left[\p_y ^N + \hat{\mathrm{t}}_2 (y) \p_y ^{N-2} + \cdots + \hat{\mathrm{t}}_{N-1} (y) \p_y + \hat{\mathrm{t}}_N (y) \right] \hat{\chi}_\a (\ba;y) \psi(\ba;\mathbf{u};\qe), \quad \a=0,1,\cdots, N-1,
\end{align}
at each basis element $\psi(\ba;\mathbf{u};\qe)$. This is nothing but the universal oper equation \eqref{eq:univop} with the eigenvalue property \eqref{eq:heigen} applied. Here, the coefficient of the next-to-highest order differential $\p_y ^{N-1}$ is absent due to the redefinition \eqref{eq:redef} of the $\mathbf{H}$-observable.

By construction above, the coefficients of the Laurent series $\hat{\mathrm{t}}_k (y)$ are merely rearrangements of the coefficients of the transfer matrix $\hat{t}(x)$ in the universal $\hbar$-oper \eqref{eq:univhoperre}. Hence, we conclude that the quantum Hamiltonians of the XXX spin chain (defined on $N$ bi-infinite modules over $\fgl(2)$) and the Gaudin model (defined on two (co-)Verma modules and two bi-infinite modules over $\fsl(N)$) are exactly identical as differential operators in the monodromy defect parameters.\footnote{Recall that, in the companion paper \cite{Jeong:2024hwf}, we presented the quantum Hamiltonians of the XXX spin chain both as differential operators in the monodromy defect parameters $(u_\o)_{\o=0} ^{N-1} $ and as $\ve_1$-difference operators in the Fourier dual variables $(w_\o)_{\o=0} ^{N-1}$.}

We can proceed in the same way for the quantum spectra of the integrable models. Consider the $\hbar$-oper equation
\begin{align}
\begin{split}
    &0 = \left[ 1 - t(\ba;x) e^{-\ve_1 \p_x} + \qe P(x) e^{-2 \ve_1 \p_x} \right]Q(\ba;x+\ve_1), \\
    &0 = \left[ 1 - t(\ba;x) e^{-\ve_1 \p_x} + \qe P(x) e^{-2 \ve_1 \p_x} \right]\tilde{Q}(\ba;x+\ve_1),
\end{split}
\end{align}
where the coefficients of the degree $N$ polynomial $t(\ba;x)$ are now normalized vacuum expectation values of the local observables in the limit $\ve_2 \to 0$. Under the transition \eqref{eq:transition}, the equation becomes a scalar-valued $N$-th order differential equation,
\begin{align}
    0 = \left[\p_y ^N + \mathrm{t}_2 (\ba;y) \p_y ^{N-2} + \cdots + \mathrm{t}_{N-1} (\ba;y) \p_y + \mathrm{t}_N (\ba;y) \right] \hat{\chi}_\a (\ba;y), \quad \a=0,1,\cdots, N-1.
\end{align}
This is precisely the oper equation derived in \cite{Jeong:2018qpc, Jeong:2023qdr}. By construction, the Laurent coefficients of $\mathrm{t}_k (\ba;y)$, $k=2,3,\cdots, N$, are simple rearrangements of the coefficients of $t(\ba;x)$. Thus, we conclude that the quantum spectra of the XXX spin chain and the Gaudin model are also rearrangements of the same set of normalized vacuum expectation values of the local observables, $\lim_{\ve_2 \to 0} \left\langle \text{Tr}\, \phi^k \right\rangle_\ba$, $k=2,3,\cdots, N$. 

All in all, we prove the equivalence between the spectral problems of the XXX spin chain (defined on $N$ bi-infinite modules over $\fgl(2)$) and the Gaudin model (defined on two (co-)Verma modules and two bi-infinite modules over $\fsl(N)$), associated to the same $A_1$-quiver $\EN=2$ gauge theory. In particular, the vacuum expectation values $\psi(\ba;\mathbf{u})$ of the regular monodromy surface defect provide a basis of bispectral wavefunctions.

We stress that in both sides of the duality, the XXX spin chain and the Gaudin model are constructed on bi-infinite modules. In this sense, our $\EN=2$ gauge theoretical approach generalizes the bispectral duality between the two models only involving highest-weight (or lowest-weight) modules. See \cite{Mukhin2006,Mironov:2012uh,Bulycheva:2012ct,Gaiotto:2013bwa} for previous studies of the bispectral duality between the two integrable models. When the gauge theory parameters are specialized to trigger higgsing to a two-dimensional $\EN=(2,2)$ theory, the bispectral duality that we established above will reduce to the one for the integrable models on lowest-weight (and further finite-dimensional) modules. For instance, it would be possible to match the Bethe roots of the two models.

\section{Surface defect transition and separation of variables} \label{sec:SoV}
In the previous section, we proved that a common eigenfunction of the Hitchin integrable system--whether in the form of the Gaudin model or of the XXX spin chain--is given by the vacuum expectation value of the regular monodromy surface defect,
\begin{align}
    \left\langle \Psi (\bu) \right\rangle_\ba = \sum_{\boldsymbol\l} \qe^{\vert \boldsymbol\l  \vert} \m_{\boldsymbol\l}  \Psi(\bu)[\boldsymbol\l],
\end{align}
in the limit $\ve_2 \to 0$. Recall that the observable expression is given by (for general coloring) \eqref{eq:monoobser},
\begin{align} 
\begin{split}
    &\Psi _{\mathbf{c}} (\mathbf{u}) [\bl] =\sum_{{\bL} \in \r^{-1} (\bl)} \prod_{\o=0} ^{l-2} \qe_\o ^{k_\o - k_{l-1}}  \BE\left[ \frac{S \S_0 ^*}{P_1 ^*} + \sum_{\o=0} ^{l-2} \frac{- \S_\o (\S_\o - \S_{\o+1})^* - M^+ _\o \S_\o ^* + q_1 ^{-1} (M^- _{\o}) ^* \S_\o}{P_1 ^*} \right].
\end{split}
\end{align}
In this section, we will re-express this as a surface observable created by coupling a $\EN=(2,2)$ gauged linear sigma model and show that such a dual description leads to the separation of variables for the XXX spin chain and the Gaudin model.

\subsection{Dual description of monodromy surface defect as coupling to sigma model} \label{subsec:dualdes}
First, let us remind that the equivariant Chern characters of the framing bundle and the tautological bundle can be written as
\begin{align}
\begin{split}
    N_{\omega} & = \sum_{\omega=0}^{N-1} e^{a_{\omega}} \\
    K_{\omega} & = \sum_{\alpha=0} ^{N-1} e^{a_\alpha}  \sum_{\underset{c(\alpha)+j-1= \omega+l(J-1)}{(i,j)\in{\L}^{(\alpha)}}}q_1^{i-1}q_2^{J-1}
     \\
    & = \frac{1}{P_1} \sum_{\a=0 } ^{N-1} e^{a_\a} \sum_{J= J_{\text{min},\a,\o}} ^{J_{\text{max},\a,\o }} \left( 1- q_1 ^{\L^{(\a) t} _{\o+1 -c(\a) +l (J-1)    } } \right) q_2 ^{J-1} ,
\end{split}
\end{align}
where the lower and the upper bound of the $J$-summation are given by
\begin{align}
\begin{split}
 &J_{\text{min},\a,\o} = \begin{cases}
        1 \quad\quad \text{if} \quad 0\leq c(\a) \leq \o \\
        2 \quad\quad \text{if} \quad \o < c(\a) \leq l-1
    \end{cases} , \\
    &J_{\text{max},\a,\o} = 1+\left\lfloor \frac{c(\a) +{\L} ^ {(\a)} _1 -1 -\o}{l} \right\rfloor,\quad\quad J_{\text{max},\alpha}:= J_{\text{max},\a,l-1} = \left\lfloor \frac{c(\a)+{\L} ^{(\a)} _1 }{l} \right\rfloor.
\end{split}
\end{align}

At a fixed point in the moduli space of instantons labelled by the partitions $\boldsymbol\lambda= (\l^{(0)}, \l^{(1)}, \cdots. \l^{(N-1)}) $, the $\EY$-observable can be reorganized into
\begin{align}
\begin{split}
     \EY(x)[\boldsymbol\lambda] &= \prod_{\a=0}^{N-1} (x-a_\a) \prod_{i=1} ^{l(\l^{(\alpha)})} \prod_{j=1} ^{\l^{(\a)} _i } \frac{(x-c_\Box -\ve_1)(x-c_\Box -\ve_2) }{(x-c_\Box )(x-c_\Box -\ve)}  \\
     &=\prod_{\alpha=0}^{N-1} (x-A_\alpha) \prod_{i=1}^{l(\l^{(\alpha)})} \frac{x-(i-1)\ve_1 - \lambda^{(\alpha)}_i \ve_2}{x-i\ve_1-\lambda^{(\alpha)}_i\ve_2} \\
     & = \prod_{\a=0} ^{N-1} (x-A_\a) \prod_{J=1} ^{\l^{(\a)} _1} \frac{x-a_\a -\l^{(\a) t} _{J+1} \ve_1 - J\ve_2 }{x- a_\a  -\l^{(\a) t} _J \ve_1 - J\ve_2},
\end{split}
\end{align}
where we defined $A_\alpha := a_\alpha + l(\l^{(\alpha)}) \ve_1$. Equivalently, the equivariant Chern character of the universal sheaf can be expressed as
\begin{align}
\begin{split}
    S[{\boldsymbol\lambda}] 
    & = \sum_{\alpha=0}^{N-1} e^{A_\alpha} + \sum_{\alpha=0}^{N-1} e^{a_\alpha} \sum_{J=1}^{\lambda_{1}^{(\alpha)}} \left( q_1^{\lambda^{(\alpha)t}_{J+1}} - q_1^{\lambda^{(\alpha) t}_J} \right) q_2^J \\
    & := F_{l-1}[\boldsymbol\lambda] + P_1 W[\boldsymbol\lambda],
\end{split}
\end{align}
where we defined
\begin{align} \label{eq:charW}
\begin{split}
   &F_{l-1} [\bl] := \sum_{\a=0} ^{N-1} e^{A_\a} \\
   &W[\bl] := \frac{1}{P_1} \sum_{\alpha=0}^{N-1} e^{a_\alpha} \sum_{J=1}^{\lambda_{1}^{(\alpha)}} \left( q_1^{\lambda^{(\alpha)t}_{J+1}} - q_1^{\lambda^{(\alpha) t}_J} \right) q_2^J.
\end{split}
\end{align}
Note that $W[\boldsymbol\l]$ is in fact a pure character, since $\l^{(\a) t} _J \geq \l^{(\a) t} _{J+1} $ for all $1\leq J \leq \l_1 ^{(\a)}$ so that each summand is divisible by $P_1 = 1-q_1$. Thus, $W[\boldsymbol\l]$ is in fact the equivariant Chern character of a vector space, which we denote by the same letter $W[\boldsymbol\l]$ unless the notation incurs any confusion. It is straightforward to see that $w[\boldsymbol\l]:=\dim W[\boldsymbol\l] = \sum_{\a=0} ^{N-1} l(\l ^{(\a)})$.

\begin{figure}[h!]
    \centering
    \begin{subfigure}[b]{0.3\textwidth} \centering
    \begin{tikzpicture}[inner sep=0in,outer sep=0in]
\node (n) {
\begin{ytableau}
 *(blue!40)   & *(blue!40) & *(blue!40)\\
 *(blue!40)  & *(blue!40) \\
*(blue!40) \\
 *(blue!40)  \\
 *(blue!40) \\
\end{ytableau}};
\end{tikzpicture}  \vspace{1.5cm} \caption{A partition ${\l}^{(\a)}$} 
    \end{subfigure} 
    \hspace{1.5cm}
    \begin{subfigure}[b]{0.3\textwidth} \centering
    \begin{tikzpicture}[inner sep=0in,outer sep=0in]
\node (n) {
\begin{ytableau}
{} & & *(blue!40)   &  & & *(blue!40) & &  & *(blue!40) \bullet & *(red!40)\\
 &  &  *(blue!40)  &  &  & *(blue!40) \bullet & *(red!40) &  *(green!40)  \\
 &  &*(blue!40) \bullet & *(red!40)  &  *(green!40) \\
 & & *(blue!40) \bullet & *(red!40) \\
 &  & *(blue!40) \bullet \\
*(red!40) & *(green!40)\\
*(red!40) &  *(green!40) \\ 
*(red!40)  \\ 
*(red!40) 
\end{ytableau}};
\draw  (-0.4,-2.1)   node {Tail};
\draw[->,line width=0.4mm] (-0.9,-2.05) -- (-2.2,-1.7);
\draw (1.6,0) node {Jump};
\draw[->,line width=0.4mm] (0.9,0.1) -- (-0.4,0.7);
\draw[->,line width=0.4mm] (1.5,0.34) -- (1.3,1.38);
\draw[->,line width=0.4mm] (1.9,0.34) -- (2.7,1.95);
\draw  (2.5,-1)   node {\textcolor{red!40}{   \ytableausetup{boxsize=3mm}
    \begin{ytableau}
    *(red!40) 
\end{ytableau} : $V_0 = J_0 \oplus T_0 $}};
\draw  (2.5,-1.6)   node {\textcolor{green}{   \ytableausetup{boxsize=3mm}
    \begin{ytableau}
    *(green!40)
\end{ytableau} : $ V_1 = J_1 \oplus T_1$}};
\end{tikzpicture}  \vspace{1pt}  
\captionsetup{width=1.5\textwidth}
\caption{{A colored partition in the preimage \\ ${\L}^{(\a)} \in \r^{-1} (\l^{(\a)})$}}
    \end{subfigure}
    \caption{Monodromy defect observable encoded in tails and jumps of colored partitions, exemplified in the case of $l=3$. The colored partition ${\L}^{(\a)}$ (with $c(\a) = 0$ chosen in the example) is represented by a colored Young diagram depicted in (b). The projection $\r : {\L}^{(\a)} \mapsto \l^{(a)}$ maps ${\L}^{(a)}$ to the subdiagram $\l^{(a)}$ composed of the columns carrying $\BZ_l$-charge $l-1$ (the blue boxes carrying $\BZ_3$-charge $2$ in the example, $W[\boldsymbol\l]$ is the subspace with the bullets). Conversely, a colored partition ${\L} ^{(a)}$ in the preimage of a given $\l^{(\a)}$ can be reconstructed from $V_\o$, $\o=0,1,\cdots, l-2$ ($V_0$ and $V_1$ depicted by the red boxes and green boxes in the example). Each $V_\o $ is the direct sum of the jump $J_\o$, which is constructed based on $W[\boldsymbol\l]$, and the tail $T_\o$, which is independent of $\l^{(\a)}$. Note that the maximal number of boxes in each jump is constrained by the shape of $\l^{(\a)}$, while the tail can grow indefinitely toward downward.}
    \label{fig:colory}
\end{figure}
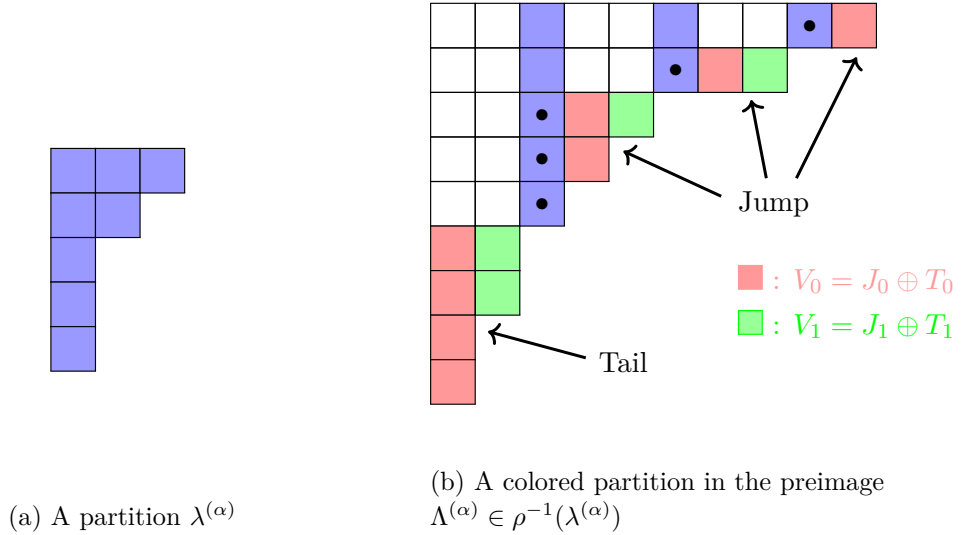

Now we reorganize the monodromy defect observable \eqref{eq:monoobser} in such a way that the dual description as coupling two-dimensional sigma model becomes more manifest. Recall that we defined
\begin{align} \label{def:sigmadef}
\begin{split}
   \S_\o [\boldsymbol\L] &:= S_{\o+1} [\boldsymbol\L] + S_{\o+2} [\boldsymbol\L] + \cdots + S_{l-1} [\boldsymbol\L] \\
   & = N_{\o+1} + \cdots + N_{l-1} +P_1 K_\o -P_1 K_{l-1}
\end{split},\quad\quad \o=0,1,\cdots, l-2.
\end{align}
This can be reorganized as
\begin{align}
    \S_\o = \sum_{\o <\o' \leq l-1} L_{\o'} +P_1 V_\o,\quad\quad \o=0,1,\cdots, l-2,
\end{align}
where we defined
\begin{align} \label{eq:charV}
\begin{split}
    &L_\o [\boldsymbol\l] = \sum_{c(\a) = \o} e^{A_\a} \\
    &V_\o [\boldsymbol\L] = J_\o [\boldsymbol\L] + T_\o [\boldsymbol\L], \\
    & J_\o [\boldsymbol\L] = \frac{1}{P_1} \sum_{\a=0} ^{N-1} e^{a_\a} \sum_{J=1} ^{\l_1 ^{(\a)}} \left( q_1 ^{\l_{J+1} ^{(\a) t} } - q_1 ^{\L ^{(\a) t} _{\o+1-c(\a) +l J} } \right) q_2 ^J, \\
    &   T_\o [\boldsymbol\L] = \frac{1}{P_1} \sum_{0\leq c(\a) \leq  \o} e^{A_\a} \left( 1- q_1 ^{{\L}^{(\a)t} _{\o - c(\a) +1} - {\l} ^{(\a)t } _{1} } \right).
\end{split}
\end{align}
Note that $L_\o [\boldsymbol\l]$ only depends on the four-dimensional instanton configuration represented by $\boldsymbol\l$. Also, it is crucial that $J_\o [\boldsymbol\L]$ and $T_\o [\boldsymbol\L]$ are in fact pure characters, and therefore they are $\mathsf{T}_{\mathsf{H}}$-equivariant Chern characters of vector spaces which we denote by the same letters. Indeed, there is a sandwich relation constraining the allowed colored partitions ${\boldsymbol\L}$ for given $\bl$,
\begin{align} \label{eq:sand}
    \lambda^{(\alpha)t}_J
    = {\L}_{l J-c(\alpha)}^{(\alpha)t}
    \geq {\L}_{\omega+lJ-c(\alpha)}^{(\alpha)t}
    \geq {\L}_{l+lJ-c(\alpha)}^{(\alpha)t}
    =\lambda_{J+1}^{(\alpha)t}, \quad\quad \o=0,1,\cdots, l-1,
\end{align}
which implies the summands of $J_\o$ and $T_\o$ are divisible by $P_1$. In turn, $V_\o$ is actually a direct sum of two vector spaces,
\begin{align} \label{eq:vjtchar}
    V_\o = J_\o \oplus T_\o, \quad\quad \o=0,1,\cdots, l-2,
\end{align}
with dimensions
\begin{align}
    v_\o : = \dim V_\o  = k_\o -k_{l-1} +\sum_{\o<c(\a)\leq l-1} l({\l} ^{(\a)}) \geq 0.
\end{align}
We refer to the subspace $J_\o$ as the \textit{jump}, and to the subspace $T_\o$ as the \textit{tail}. The jump and the tail encodes how the colored partition $\boldsymbol\L \in \r^{-1} (\boldsymbol\l)$ is reconstructed for a given $\boldsymbol\l$. The jump describes the restricted Young diagrams that are put in between the columns of $\boldsymbol\l$, while the tail describes the restricted Young diagrams attached to the left of the left-most column of $\boldsymbol\l$. See Figure \ref{fig:colory} for an example.

Note that if we decouple the two-dimensional degrees of freedom by turning off the bulk gauge coupling $\qe \to 0$, then there is no four-dimensional instanton so that we only have $\boldsymbol\l = \varnothing$, which leads to $W[\boldsymbol\l] = \varnothing$ and therefore all the jumps become empty $J_\o = \varnothing$, $\o = 0,1,\cdots, l-2$. The resulting two-dimensional partition function of the defect becomes a sum over restricted Young diagrams represented by the tail, which manifests the dual description of the defect as coupling to certain two-dimensional sigma model, as studied in \cite{Jeong:2018qpc}. We will now extend such a dual description of the monodromy defect to non-zero bulk coupling $\qe \neq 0$.

\begin{figure}[h!]
    \centering
    \begin{tikzpicture}[square/.style={regular polygon,regular polygon sides=4}]
    \node[circle, draw=black,fill=white, inner sep=5pt] at (-1,0) (N1) {$r_0$};
\node[circle, draw=black,fill=white, inner sep=5pt] at (1,0) (N2) {$r_1$};
\node[circle, draw=black,fill=white, inner sep=5pt] at (3,0) (N3) {$r_2$};
\node at (4.5,0) (NN) {$\cdots$};
\node[circle, draw=black,fill=white, inner sep=2.5pt] at (6,0) (N4) {$r_{l-2}$};
\node[square,inner sep=1.5mm,draw] at (8,0) (N5) {$N$};
\node[square,inner sep=0mm,draw] at (-1,-2) (M0) {$M_0 ^+$};
\node[square,inner sep=0mm,draw] at (1,-2) (M1) {$M_1 ^+$};
\node[square,inner sep=0mm,draw] at (3,-2) (M2) {$M_2 ^+$};
\node[square,inner sep=-1mm,draw] at (6,-2) (MM) {$M_{l-2} ^+$};

\node[square,inner sep=0mm,draw] at (-1,2) (MM0) {$M_0 ^-$};
\node[square,inner sep=0mm,draw] at (1,2) (MM1) {$M_1 ^-$};
\node[square,inner sep=0mm,draw] at (3,2) (MM2) {$M_2 ^-$};
\node[square,inner sep=-1mm,draw] at (6,2) (MMM) {$M_{l-2} ^-$};

\draw[->-=.5] (N1) -- (N2);
\draw[->-=.5] (N2) -- (N3);
\draw[->-=.5] (N3) -- (NN);
\draw[->-=.5] (NN) -- (N4);
\draw[->-=.5] (N4) -- (N5);
\draw[->-=.5] (M0) -- (N1);
\draw[->-=.5] (M1) -- (N2);
\draw[->-=.5] (M2) -- (N3);
\draw[->-=.5] (MM) -- (N4);
\draw[->-=.5] (N1) -- (MM0);
\draw[->-=.5] (N2) -- (MM1);
\draw[->-=.5] (N3) -- (MM2);
\draw[->-=.5] (N4) -- (MMM);

\draw (8.5,-1.5) node {Gauging by 4d};
\draw[->,line width=0.4mm] (8.2,-1.2) -- (8.1,-0.6);
    \end{tikzpicture}
    \caption{The $\EN=(2,2)$ quiver for the two-dimensional gauged linear sigma model coupled to the four-dimensional $\EN=2$ gauge theory, giving dual description of the monodromy surface defect.}
    \label{fig:2dquiver}
\end{figure}
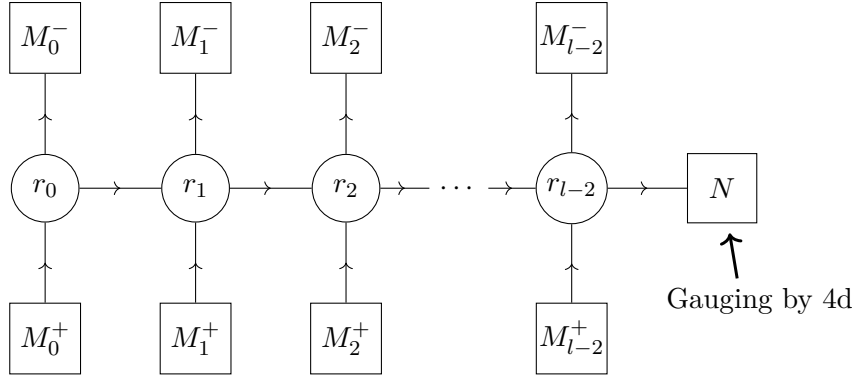

For this, let us re-express the monodromy defect observable \eqref{eq:monoobser} using $\Sigma_\o$ \eqref{def:sigmadef} as
\begin{align} \label{eq:monoobs}
\begin{split}
    &\Psi(\mathbf{u})[{\boldsymbol\lambda}]  =\sum_{\hat{\bl} \in \r^{-1} (\bl)} \prod_{\o=0} ^{l-2} \qe_\o ^{k_\o - k_{l-1}}  \BE\left[ \frac{S \S_0 ^*}{P_1 ^*} + \sum_{\o=0} ^{l-2} \frac{- \S_\o (\S_\o - \S_{\o+1})^* - M^+ _\o \S_\o ^* + q_1 ^{-1} (M^- _{\o}) ^* \S_\o}{P_1 ^*} \right] \\
    & = \sum_{{\boldsymbol\L} \in \r^{-1}(\boldsymbol\l) }\prod_{\omega=0}^{l-2} \kq_{\omega}^{k_{\omega} - k_{l-1}}  \mathbb{E}\left[\sum_{\omega=0}^{l-2}\left(S-\sum_{ \o< \o' \leq l-1 }L_{\o' } \right)\frac{  L_{\o+1} ^* }{P_1^*} - \frac{\sum_{ \o=0 } ^{l-2} \sum_{\o < \o' \leq l-1} M^+_{\omega}L_{\o'}^* - q_1 ^{-1} L_{\o'} (M^- _\o)^* )}{P_1^*}\right] \\
&\quad\quad\quad \times\mathbb{E}\left[P_1W V_0 ^*+\sum_{\omega=0}^{l-2}\left({ 
L_{\o} V_{\omega}^*  }+q_1V_\o 
  L_{\o+1} ^*-P_1V_\omega(V_\omega-V_{\omega+1})^*-M^+ _{\omega}V_{\omega}^* - (M^-_\o)^* 
V_\o \right)\right].
\end{split}
\end{align}
We will explain that this expression of the 
monodromy surface defect observable manifests the dual description of the surface defect, coupling of the two-dimensional $\EN=(2,2)$ gauged linear sigma model which is represented as a $\EN=(2,2)$ quiver gauge theory in Figure \ref{fig:2dquiver}. Here, the coupling is initiated by gauging the $GL(N)$ flavor symmetry for the right-most flavor node by the four-dimensional gauge field restricted to the surface. The Higgs branch of this gauged linear sigma model is the total space of a vector bundle,
\begin{align} \label{eq:vbflag}
    \bigoplus_{\o=0} ^{l-2} \mathcal{E}_\o \otimes \BC^{\vert c_ f ^{-1}(\o)\vert} \oplus \bigoplus_{\o=0} ^{l-2} \mathcal{E}^\vee _\o \otimes \BC^{\vert c_{af} ^{-1} (\o) \vert} \longrightarrow F(r_0,r_1,\cdots, r_{l-1}),
\end{align} 
over the partial flag variety $F(r_0,r_1,\cdots, r_{l-1})$. Here, $r_\o = \vert \{ \a  \, \vert \, 0 \leq c(\a) \leq \o  \} \vert$ and a partial flag in $F(r_0,r_1,\cdots, r_{l-1})$ is a filtration of the complex $N$-dimensional vector space $\BC^N$ by its subspaces of dimensions $0 \leq r_0 \leq r_1 \leq \cdots \leq r_{l-1} = N$. Finally, $\mathcal{E}_\o$ is the $\o$-th tautological vector bundle over the flag variety $F(r_0,r_1,\cdots, r_{l-1})$ with the rank $\text{rk}\,\mathcal{E}_\o = r_\o$, and $\mathcal{E} _\o ^\vee = \text{Hom}(\mathcal{E}_\o,\BC)$ is its dual vector bundle.

It has been well-known that the partition function of such a gauged linear sigma model, without coupling to the four-dimensional theory, localizes to an integral of equivariant Euler class of a vector bundle over 
(smooth resolution of) the moduli space of quasi-maps from $\BP^1$ to $F(r_0,r_1,\cdots, r_{l-1})$. The moduli space of these quasi-maps admits a construction from the hand-saw quiver \cite{FR2010,Nakajima:2011yq}. Now, we show that the surface defect observable \eqref{eq:monoobs} can be obtained by an integral of equivariant Euler class of a vector bundle over this hand-saw quiver variety modified by the effect of the coupling to the bulk four-dimensional theory. In this sense, we can regard the observable expression \eqref{eq:monoobs} as the manifestation of this dual description of the monodromy surface defect.

\subsubsection{Modified hand-saw quiver variety}
Let us now introduce our modified hand-saw quiver variety. Consider the following collection of vector spaces,
\begin{itemize}
    \item $l-1$ vector spaces $V_\o = \BC ^{v_\o}$, $\o=0,1,\cdots, l-2$. 
    \item $l$ vector spaces $L_\o [\boldsymbol\l] = \BC^{r_\o -r_{\o-1}}$, $\o=0,1,\cdots, l-1$. 
 \item A vector space $W[\boldsymbol\l] = \BC^{w [\boldsymbol\l]}$.
\end{itemize}
The vector spaces $(L_\o [\boldsymbol\l])_{\o=0} ^{l-1}$ and $W[\boldsymbol{\l}]$ are defined upon a given instanton configuration represented by $\boldsymbol\l$, in the following way. Firstly, their dimensions are determined by $\boldsymbol\l$ as $\dim L_\o = r_\o - r_{\o-1} = \vert \{\a \, \vert\, c(\a) = \o \} \vert$ and $\dim W = w = \sum_{\a=0} ^{N-1} l(\l^{(\a)} )$. Moreover, the equivariant weights of these vector spaces under the Cartan of the global symmetry group $\BC^\times _{\ve_1} \times GL(W) \times \bigtimes_{\o=0} ^{l-1} GL(L_\o)$ are imposed as \eqref{eq:charV} and \eqref{eq:charW}, namely,
\begin{align} \label{eq:charlw}
\begin{split}
    &L_\o [\boldsymbol\l] = \sum_{c(\a) = \o} e^{A_\a} = \sum_{c(\a) = \o} e^{a_\a} q_1 ^{l(\l^{(\a)})}\\
    &W[\boldsymbol\l] =\frac{1}{P_1} \sum_{\alpha=0}^{N-1} e^{a_\alpha} \sum_{J=1}^{\lambda_{1}^{(\alpha)}} \left( q_1^{\lambda^{(\alpha)t}_{J+1}} - q_1^{\lambda^{(\alpha) t}_J} \right) q_2^J,
\end{split}
\end{align}
where we recall that the character of $W[\boldsymbol\l]$ above is in fact a pure character. Note that $W$ is decomposed into column segments, $W[\boldsymbol\l] = \bigoplus_{\a=0} ^{N-1} \bigoplus_{J=1} ^{\l ^{(\a)}_1 } \left( \bigoplus_{n \geq 0} b^n w_{\a,J} \right)$, where $w_{\a,J} \in W$ is the basis vector carrying the character $e^{a_\a} q_1 ^{\l^{(\a) t } _{J+1} } q_2 ^J$ and we defined the map $b \in \text{End}(W)$ sending each basis vector to another basis vector carrying one more charge of $q_1 \in \BC^\times _{\ve_1}$.\footnote{$b \in \text{End}(W)$ is the restriction of $B_1: K \to K$, which is a part of the ADHM data for the instanton moduli space, to the subspace $W$.}

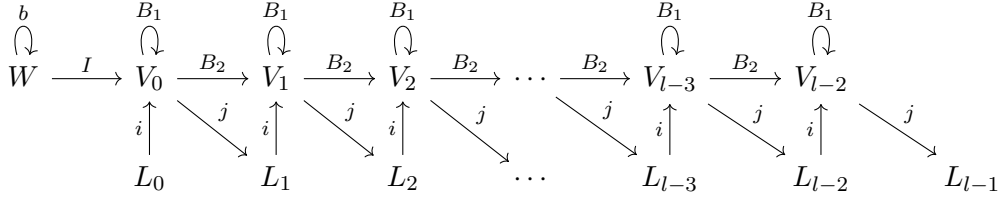
\begin{figure}[h!]\centering
\begin{tikzcd}
W  \arrow[r,"I"] \arrow[loop above,"b"] &   V_0 \arrow[loop above, "B_1"] \arrow[rd,"j"] \arrow[r, "B_2"] & V_1 \arrow[loop above, "B_1"] \arrow[r,"B_2"] \arrow[rd,"j"] & V_2 \arrow[loop above, "B_1"] \arrow[r,"B_2"] \arrow[rd,"j"] & \cdots \arrow[r,"B_2"] \arrow[rd,"j"] & V_{l-3} \arrow[loop above,"B_1"] \arrow[r,"B_2"] \arrow[rd,"j"] & V_{l-2} \arrow[loop above,"B_1"] \arrow[rd,"j"] \\
&L_0 \arrow[u,"i"] & L_1 \arrow[u,"i"] & L_2 \arrow[u,"i"] & \cdots & L_{l-3} \arrow[u,"i"] & L_{l-2} \arrow[u,"i"] & L_{l-1}
\end{tikzcd}
\caption{The modified hand-saw quiver provides the dual description of the monodromy defect observable: the $\mathcal{N}=(2,2)$ gauged linear sigma model coupled to the four-dimensional $\mathcal{N}=2$ gauge theory. Note that our modified hand-saw quiver includes a \textit{handle} from the newly introduced vector space $W$ and the map $I: W \to V_0$, in contrast to the original hand-saw quiver of \cite{FR2010, Nakajima:2011yq}.} \label{fig:handsaw}
\end{figure}

The modified hand-saw quiver is represented by Figure \ref{fig:handsaw}. 

Here, the arrows between the vector spaces are the linear maps,
\begin{align}
\begin{split}
    &B_1 \in \bigoplus_{\o=0} ^{l-2} \text{Hom}(V_\o, V_\o), \quad\quad B_2 \in \bigoplus_{\o=0} ^{l-3} \text{Hom}(V_\o,V_{\o+1}) \\
    & i \in \bigoplus_{\o=0} ^{l-2} \text{Hom}(L_\o, V_\o), \quad\quad j \in \bigoplus_{\o=0}^{l-2} \text{Hom}(V_\o,L_{\o+1}), \\
    & I  \in \text{Hom}(W,V_0).
\end{split}
\end{align}
We define the complex moment maps by
\begin{align} \label{eq:cmm1}
 \m_\o := (B_1)_{\o+1} (B_2)_{\o} - (B_2)_{\o} (B_1)_{\o} + i_{\o+1} j_\o : V_\o \longrightarrow V_{\o+1},\quad\quad \o=0,1,\cdots, l-3,
\end{align}
and
\begin{align} \label{eq:cmm2}
    \m_{-1} := (B_1)_0 I - I b : W \longrightarrow V_0. 
\end{align}
We also define the real moment maps by
\begin{align}
 \m^{\BR}_\o := \left[(B_1)_\o, (B_1)_\o ^\dagger \right] + (B_2)_\o (B_2) _{\o-1} ^\dagger - (B_2) _{\o+1} ^\dagger (B_2)_\o + i_\o i_\o ^\dagger - j_\o ^\dagger j_\o : V_\o \longrightarrow V_\o,
\end{align}
for $\o=0,1,\cdots, l-2$, where we used the notation $(B_2)_{-1} = I$.

The locus of the moment map equations $ \bigcap_{\o=-1} ^{l-3} \, \m_\o ^{-1} (0) \cap \bigcap_{\o=0} ^{l-2} (\m_\o ^\BR)^{-1} (\z\, \text{id}_{V_\o}) $, where we have chosen a stability chamber by $\z \in \BR$, is acted on by the gauge group
\begin{align}
    G = \bigtimes_{\o=0} ^{l-2} U(V_\o),
\end{align}
where the action is given by
\begin{align} \label{eq:gauget}
    \left((B_1)_\o,(B_2)_\o,i_\o,j_\o,I \right) \mapsto (g_\o  (B_1)_\o g_\o ^{-1}, g_{\o+1}  (B_2)_\o g_\o ^{-1}, g_\o  i_\o, j_\o g_\o^{-1}, g_0  I), \quad g_\o \in U(V_\o).
\end{align}
The modified hand-saw quiver variety, at the stability chamber given by $\z \in \BR$, is defined as the quotient by this gauge group action,
\begin{align} \label{eq:hsvar}
    \mathcal{M} _{\text{hand-saw}} ^\z [\boldsymbol\l] = \left. \bigcap_{\o=-1} ^{l-3} \, \m_\o ^{-1} (0) \cap \bigcap_{\o=0} ^{l-2} (\m_\o ^\BR)^{-1} (\z\, \text{id}_{V_\o}) \right/ G.
\end{align}
Note that imposing the moment map equation $\m _{-1} = 0 $ implies
\begin{align}
\begin{split}
    &(B_1)_0 I (W) \subset I(W).
\end{split}
\end{align}
 
The modified hand-saw quiver variety is acted on by the global symmetry group $\mathsf{H} = \BC^{\times} _{\ve_1} \times GL(W) \times \bigtimes_{\o=0} ^{l-1} GL(L_\o)$. The $\BC^\times _{\ve_1}$-action descends from the two-dimensional spacetime isometry, so that it acts as
\begin{align} \label{eq:q1act}
    \left((B_1)_\o,(B_2)_\o,i_\o,j_\o,I \right) \mapsto  \left( q_1 (B_1)_\o,(B_2)_\o,i_\o,q_1 j_\o,I \right) ,\quad\quad q_1 \in \BC^{\times }_{\ve_1},
\end{align}
and the action of $GL(W) \times \bigtimes_{\o=0} ^{l-1} GL(L_\o)$ framing is given by
\begin{align} \label{eq:glact}
 \left((B_1)_\o,(B_2)_\o,i_\o,j_\o,I \right) \mapsto
 \left((B_1)_\o,(B_2)_\o,i_\o h_{\o} ^{-1},h_{\o+1}  j_\o,I h_W ^{-1} \right), \quad h_\o \in GL(L_\o), \;h_W \in GL(W) .
 \end{align}
Recall that we want the modified hand-saw quiver variety to describe the supersymmetric localization locus of the two-dimensional gauged linear sigma model coupled to the four-dimensional gauge theory, so that this global symmetry group is embedded in the global symmetry group of the instanton moduli space,
\begin{align}
     \mathsf{H} \hookrightarrow \BC^\times _{\ve_1} \times \BC^{\times} _{\ve_2} \times GL(N) ,
\end{align}
where the $\BC^\times_ {\ve_1} \subset \mathsf{H}$ part is simply identified with $\BC_{\ve_1}^\times$ in the image. The framing part of $\mathsf{H}$ embeds $ GL(W) \times \bigtimes_{\o=0} ^{l-1} GL(L_\o) \hookrightarrow \BC^\times _{\ve_1} \times \BC^\times _{\ve_2} \times GL(N)$ in such a way that the equivariant Chern characters of $W$ and $(L\o){\o=0}^{l-1}$ are given by \eqref{eq:charlw}. In particular, the embedding breaks the global gauge symmetry group of the four-dimensional $\mathcal{N}=2$ theory to $\bigtimes_{\o=0} ^{l-1} GL(\vert c^{-1} (\o) \vert) \subset GL(N)$.

Since the Higgs branch of our gauged linear sigma model is the vector bundle \eqref{eq:vbflag} over the partial flag variety, we need to consider the corresponding matter bundle over the modified hand-saw quiver variety, 
\begin{align}
    {\mathcal{E}}_{\text{matter}} = \bigoplus_{\o=0} ^{l-2} \text{Hom} ( M^+ _\o, V_\o) \oplus \bigoplus _{\o=0} ^{l-2} \text{Hom}(V_\o , M^- _{\o}).
\end{align}
The flavor symmetry group $\mathsf{F} = \bigtimes_{\o=0}^{l-1} \left( GL(M^+_\o) \times GL(M^-_\o) \right)$ acts in an obvious way on the fiber of the bundle. Note that this flavor symmetry group embeds into the flavor symmetry group for the matter bundle over the bulk instanton moduli space. The embedding is determined by the flavor coloring functions, $\mathsf{F} \simeq \bigtimes_{\o=0}^{l-1} \left( GL(\vert c_f^{-1}(\o)\vert) \times GL(\vert c_{af}^{-1}(\o)\vert) \right) \subset GL(M^+) \times GL(M^-)$.

At this point, we stress again that the modified hand-saw quiver variety is defined based on a given instanton configuration in the four-dimensional bulk. If we turn off the four-dimensional gauge coupling ($\qe \to 0$), the instanton moduli space becomes empty, so we can only take $\boldsymbol\l = \varnothing$. This implies that $W[\boldsymbol\l] = \varnothing$ and that $I: W \to V_0$ is a trivial map. The equivariant Chern characters of the vector spaces $L_\o$ also simplify to $L_\o = \sum_{\a \in c^{-1} (\o)} e^{a_\a}$. In this limit, the \textit{handle} (the vector space $W$ and the map $I$) disappears, and we recover the original hand-saw quiver variety from \cite{FR2010, Nakajima:2011yq}. In this sense, our modification of the hand-saw quiver variety reflects the effect of coupling the two-dimensional sigma model to the four-dimensional bulk theory.

\subsubsection{Recovering monodromy defect observable in positive chamber}
We now show that the monodromy defect observable \eqref{eq:monoobs} can be recovered by an equivariant integral of the Euler class of a vector bundle over the modified hand-saw quiver variety \eqref{eq:hsvar}, in the positive stability chamber $\z >0$.

In this chamber, we can compensate the real moment map equations $\m_\o ^\BR = \z \, \text{id}_{V_\o}$ by the stability condition
\begin{align} \label{eq:stabp}
\begin{split}
    &V_0 =  \BC[B_1] I(W) \oplus \BC[B_1] i(L_0), \\
    &V_\o = \BC[B_1]B_2 (V_{\o-1}) \oplus \BC[B_1] i(L_\o), \quad\quad \o=1,2,\cdots, l-2.
\end{split}
\end{align}
and complexified gauge group action $G_\BC =  \bigtimes_{\o=0}^{l-2} GL(V_\o)$. Namely,
\begin{align}
    \mathcal{M}^{\z>0} _{\text{hand-saw}} [\boldsymbol\l] = \left. \left(\bigcap_{\o=-1} ^{l-3} \m_\o ^{-1} (0)\right)^{\text{stab}_{\z>0}} \right/ G_\BC.
\end{align}

To evaluate equivariant integral over this moduli space, we need to classify the fixed points under the action of the Cartan of the global symmetry group $\mathsf{T}_{\mathsf{H}} \subset \mathsf{H}$. At a fixed point, the $\mathsf{T}_{\mathsf{H}}$-action is compensated by a gauge transformation of $G_\BC = \bigtimes_{\o=0} ^{l-2} GL(V_\o) $. This implies that each $V_\o$ is a representation of $\mathsf{T}_\mathsf{H}$. Due to the stability condition \eqref{eq:stabp}, $V_\o$ is the direct sum of a subspace spanned by the vectors of the form $B_1 ^i B_2 ^\o I(W)$ and a subspace spanned by the vectors of the form $B_1 ^i B_2 ^{\o-\o'} i(L_{\o'})$ where $i\geq 0$ and $\o' \leq \o$. Each of them comprises a Young diagram with varying $\o$, due to the moment map equations. Consequently, the former subspace gives a jump $J_\o$, while the latter subspace produces a tail $T_\o$. To sum up, the fixed points are precisely classified by the colored partitions for the given $\boldsymbol\l$, i.e., the preimage ${\boldsymbol\L} \in \r^{-1} (\boldsymbol\l)$. At each fixed point ${\boldsymbol\L}$ we reproduce $V_\o$ as the representation of $\mathsf{T}_\mathsf{H}$ carried by the direct sum of the jump and the tail \eqref{eq:vjtchar}, $V_\o = J_\o \oplus T_\o$ (recall Figure \ref{fig:colory}). Their equivariant Chern characters are therefore given by \eqref{eq:charV},
\begin{align}
\begin{split}
    &V_\o[\boldsymbol\L] = J_\o [\boldsymbol\L] + T_\o [\boldsymbol\L] \\
    &J_\o [\boldsymbol\L]=  \frac{1}{P_1} \sum_{\a=0} ^{N-1} e^{a_\a} \sum_{J=1} ^{\l_1 ^{(\a)}} \left( q_1 ^{\l_{J+1} ^{(\a) t} } - q_1 ^{\L ^{(\a) t} _{\o+1-c(\a) +l J} } \right) q_2 ^J \\
    &T_\o [\boldsymbol\L] = \frac{1}{P_1} \sum_{0\leq c(\a) \leq  \o} e^{A_\a} \left( 1- q_1 ^{{\L}^{(\a)t} _{\o - c(\a) +1} - {\l} ^{(\a)t } _{1} } \right)
\end{split}, \quad\quad\quad \o=0,1,\cdots, l-2.
\end{align}

Also, we need the equivariant Euler classes of the vector bundles. It can be computed once the equivariant Chern character is obtained by taking the product of all the equivariant weights. For the tangent bundle, it is straightforward to compute
\begin{align}
    T \mathcal{M} _{\text{hand-saw}} [{\boldsymbol\L}] = P_1 W V_0 ^* + \sum_{\o=0} ^{l-2} \left(L_\o V_\o^* +q_1 V_\o L_{\o+1} ^* - P_1 V_\o (V_\o - V_{\o+1}) ^* \right),
\end{align}
from the equivariant Chern characters of the linear maps (\eqref{eq:q1act} and \eqref{eq:glact}), the linearized complex moment map equations (\eqref{eq:cmm1} and \eqref{eq:cmm2}), and the linearized complexified gauge transformations (\eqref{eq:gauget}). For the matter bundle, we also simply get
\begin{align}
    \mathcal{E}_{\text{matter}} [{\boldsymbol\L}] = \sum_{\o=0} ^{l-2} (M_\o^+ V_\o ^* + (M_\o ^-)^* V_\o ).
\end{align}

All in all, the integral of the equivariant Euler class of the matter bundle over the modified hand-saw quiver variety is computed by equivariant localization as
\begin{align} \label{eq:eqloc2d}
\begin{split}
   \Psi[\boldsymbol\l]^{\text{non-pert}} _{\z>0} &=  \sum_{v_0,v_1,\cdots, v_{l-2} = 0} ^{\infty} \prod_{\o=0} ^{l-2} \qe_\o ^{v_\o} \int_{\mathcal{M} ^{\z>0} _{\text{hand-saw}} [\boldsymbol\l] } e_{\mathsf{T}_{\mathsf{H}}} (\mathcal{E}_{\text{matter}}) \\
    &=
    \sum_{{\boldsymbol\L} \in \r^{-1} (\boldsymbol\l)} \prod_{\o=0} ^{l-2} \qe_\o ^{v_\o} \, \BE\left[ P_1 W V_0 ^* + \sum_{\o=0} ^{l-2} \left(L_\o V_\o^* +q_1 V_\o L_{\o+1} ^* - P_1 V_\o (V_\o - V_{\o+1}) ^* \right.\right. \\
    & \qquad\qquad\qquad\qquad \qquad \qquad \qquad \qquad \qquad  \left.\left. -M^+_\o V_\o ^* - (M^- _\o)^* V_\o \right)  \right],
\end{split}
\end{align}
where we wrote the complexified FI parameters of the $\EN=(2,2)$ gauged linear sigma model (which are the complexified K\"{a}hler parameters in the non-linear sigma model description) as $(\qe_\o)_{\o=0} ^{l-2}$. Thus, with the identification between the monodromy parameters and the complexified FI parameters $(\qe_\o)_{\o=0} ^{l-2}$, the above expression recovers the non-perturbative part of the monodromy defect observable \eqref{eq:monoobs}, after augmented by the following classical contribution of the two-dimensional sigma model,
\begin{align} \label{eq:classsigma}
    \mathbf{l}_\bu:=\prod_{\o=1} ^{l-1} \left(\frac{u_\o}{u_0} \right) ^{-\sum_{\a \in c^{-1} (\o)} l(\l^{(\a)})} =\prod_{\o=1} ^{l-1} \left(\frac{u_\o}{u_0} \right) ^{\sum_{\a \in c^{-1} (\o)} \frac{a_\a - A_\a}{\ve_1} }.
\end{align}

\subsubsection{Negative chamber and surface defect transition}
Now, let us consider the modified hand-saw quiver variety in the negative stability chamber $\z<0$. In this chamber, we can compensate the real moment map equations $\m_\o ^{\BR} = \z \, \text{id}_{V_\o}$ by another stability condition
\begin{align} \label{eq:Vneg}
\begin{split}
    &V_{l-2} = \BC[ B_1 ^\dagger ] j^\dagger (L_{l-1}) \\
    &V_\o= \BC[B_1 ^\dagger ]B_2 ^\dagger (V_{\o+1})  \oplus \BC[B_1 ^\dagger] j^\dagger (L_{\o+1}) ,\quad\quad \o=0,1,\cdots, l-3.
\end{split}
\end{align}
The modified hand-saw quiver variety in this chamber is thus given by
\begin{align}
    \mathcal{M}^{\z<0} _{\text{hand-saw}} = \left. \left(\bigcap_{\o=-1} ^{l-3} \m_\o ^{-1} (0)\right)^{\text{stab}_{\z<0}} \right/ G_\BC.
\end{align}

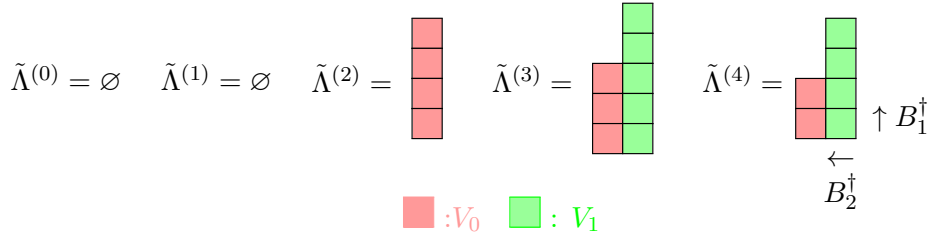
\begin{figure}[h!]
    \centering
    \ytableausetup
{mathmode, boxsize=1em}
    \begin{tikzpicture}
        \draw (-1,0)   node {$\tilde{\L}^{(0)}=\varnothing$};
        \draw (1,0)   node {$\tilde{\L}^{(1)}=\varnothing$};
        \draw (2.8,0) node {$\tilde{\L}^{(2)}=$};
        \draw (3.8,0) node {
            \begin{ytableau}
                *(red!40)  \\
                *(red!40)  \\
                *(red!40)  \\
                *(red!40)   \\
            \end{ytableau}
        };
        \draw (5.2,0) node {$\tilde{\L}^{(3)}=$};
        \draw (6.4,0) node (n) {
            \begin{ytableau}
                \none & *(green!40)   \\
                \none & *(green!40) \\
                *(red!40)  & *(green!40) \\
                *(red!40)  &  *(green!40) \\
                *(red!40)  & *(green!40) \\
            \end{ytableau}
        };
        \draw (8,0) node {$\tilde{\L}^{(4)}=$};
        \draw (9.1,0) node (n) {
            \begin{ytableau}
               \none & *(green!40) \\
                \none  & *(green!40) \\
                *(red!40)  &  *(green!40) \\
                *(red!40)  & *(green!40) \\
            \end{ytableau}
        };
        \draw (4,-1.8)   node {\textcolor{red!40}{ 
            \begin{ytableau}
                *(red!40) 
            \end{ytableau} :$ V_0 $}};
        \draw (5.5,-1.8)   node {\textcolor{green}{ 
            \begin{ytableau}
                *(green!40) 
            \end{ytableau} : $ V_1 $}};
        \draw (10.1,-0.5)   node {$\uparrow B_1 ^\dagger$};
        \draw (9.3,-1.1)   node {$\leftarrow$};
        \draw (9.3,-1.5)   node {$B_2 ^\dagger$};
\end{tikzpicture}
    \caption{A fixed point $\tilde{\boldsymbol\L}$ in the negative chamber represented by restricted 
colored Young diagrams, exemplified in the case of $N=5$ and $l=3$ with the coloring $c(0,1,2,3,4) = (0,0,1,2,2)$. Each Young diagrams $\tilde{\L} ^{(\a)}$ is built starting from the box in the lower-right corner, which carries the $\BZ_l$-charge $c(\a) - 1$ mod $l$. It grows toward the upward by the action of $B_1 ^\dagger$ and toward the left by the action of $B_2 ^\dagger$. Note the Young diagrams are growing in the opposite directions compared to the positive chamber. The action of $B_2 ^\dagger$ decreases the $\BZ_l$-charge by 1, and the Young diagrams are restricted in the way that there is no box carrying $\BZ_l$-charge $l-1$. In particular, note that $\tilde{\L}^{(\a)} =\varnothing $ if $c(\a) = 0$.}
    \label{fig:dualfixed}
\end{figure}

Let us classify the $\mathsf{T}_{\mathsf{H}}$-fixed points. By the same argument as above, the vector spaces $(V_\o)_{\o=0}^{l-2}$ become representations of $\mathsf{T}_{\mathsf{H}}$ at the fixed points. It should be noted that the jump disappears in this chamber, because there is no equivalent of $I$ map participating in the construction of the vector spaces $V_\o$. Thus the whole vector space $V_\o$ is spanned by the vectors of the form $(B_1 ^\dagger )^i (B_2 ^\dagger) ^{\o' -\o -1} j^\dagger (L_{\o'}) $ for $\o' >\o$ and $i \geq 0$, which can be represented by columns growing in the upward direction. Due to the moment map equations, there is an hierarchy between the heights of the columns with different $\o$ so that they form Young diagrams. These Young diagrams have restricted horizontal lengths, because of the vanishing condition $B_2 ^\dagger (V_0) = \varnothing$. Hence, the fixed points in the negative chamber are fully classified by $N$ tuples of such restricted colored Young diagrams, which we collectively denote by $\tilde{\boldsymbol\L} = (\tilde\L^{(0)} ,\cdots, \tilde\L^{(N-1)})$ (in fact, $\tilde{\L} ^{(\a)} = \varnothing$ if $c(\a) = 0$ due to the restriction). The equivariant Chern characters of $(V_\o)_{\o=0} ^{l-2}$ at the fixed point $\tilde{\boldsymbol\L}$ are thus given by
\begin{align}
    V_\o[\tilde{\boldsymbol\L}] = \frac{1}{P_1 ^*} \sum_{\o< c(\a) \leq l-1} e^{A_{\a} -\ve_1 } \left( 1- q_1 ^{ - \tilde{\L} ^{(\a) t}  _{c(\a)-\o}}  \right),\quad\quad\o=0,1,\cdots, l-2.
\end{align}
which is in fact a pure character since the term in the parenthesis is divisible by $P_1 ^* = 1-q_1 ^{-1}$. Note that 
\begin{align}
    \S_\o[\tilde{\boldsymbol\L}] = \sum_{\o <\o' \leq l-1} L_{\o'} +P_1 V_\o = \sum_{\o <c(\a) \leq l-1} e^{A_\a} q_1 ^{-\tilde{\L} ^{(\a)t} _{c(\a)-\o}}.
\end{align}
See Figure \ref{fig:dualfixed} for an example.

The integral of the equivariant Euler class of the matter bundle over the modified hand-saw quiver variety in this chamber is computed by the same formula \eqref{eq:eqloc2d}, where the fixed point set is replaced by $\{\tilde{\boldsymbol\L}\}$. Namely,
\begin{align} \label{eq:mononegative}
\begin{split}
     \Psi[\boldsymbol\l]^{\text{non-pert}} _{\z <0} &= \sum_{ \tilde{\boldsymbol\L}} \prod_{\o=0} ^{l-2} \qe_\o ^{v_\o} \, \BE\left[ P_1 W V_0 ^* + \sum_{\o=0} ^{l-2} \left(L_\o V_\o^* +q_1 V_\o L_{\o+1} ^* - P_1 V_\o (V_\o - V_{\o+1}) ^* \right.\right. \\
    & \qquad\qquad\qquad\qquad \qquad \qquad \qquad \qquad \qquad  \left.\left. -M^+_\o V_\o ^* - (M^- _\o)^* V_\o \right)  \right].
\end{split}
\end{align}
This expression is obviously different from \eqref{eq:eqloc2d}, because the fixed point sets, and therefore the equivariant Chern characters of vector bundles, are distinct. However, we will now show an equality relating the two partition functions up to a simple multiplicative factor.

To establish the equality, let us first introduce the contour integral presentation for the monodromy defect observable. Let us express the monodromy defect observable in the positive chamber as
\begin{equation}
\begin{aligned}
  \Psi[\boldsymbol\l]^{\text{non-pert}} _{\z>0} &=  \sum_{v_0,\dots,v_{l-2}=0}^{\infty} \prod_{\omega=0}^{l-2}\, \frac{\kq_{\omega}^{v_{\omega}}}{v_{\omega}!}\, \oint_{\CalC_\omega}\prod_{i=1}^{v_\omega}\frac{d\phi_i^{(\omega)}}{2\pi {\ii}}\frac{1}{\ve_1}
    \prod_{j<i}\frac{(\phi_i^{(\omega)}-\phi_j^{(\omega)})^2}{(\phi_i^{(\omega)}-\phi_j^{(\omega)})^2-\ve_1^2}\prod_{j=1}^{v_{\omega+1}}
    \frac{\phi_i^{(\omega)}-\phi_j^{(\omega+1)}+\ve_1}{\phi_i^{(\omega)}-\phi_j^{(\omega+1)}} \\
    & \quad\quad \times \frac{\prod_{\a \in c_f ^{-1} (\o)} ( m_{\a}^+ - \phi_i ^{(\omega)}) \prod_{\a \in c_{af}^{-1} (\o)} (\phi_i^{(\omega)}- m_{\a}^- )}{\prod_{\a \in c^{-1} (\o)} (A_{\a} - \phi_i^{(\omega)})\prod_{\a \in c^{-1} (\o+1) }(\phi_i^{(\omega)}-A_{\a}+\ve_1)}\times\prod_{s=1}^{w}\frac{\phi_i^{(0)}-w_s-\ve_1}{\phi_i^{(0)}-w_s},
    \label{eq:contourN}
\end{aligned}
\end{equation}
where we used the notation $W[\boldsymbol\l] = \sum_{s=1} ^{w} e^{w_s}$. The contours $\CalC_{\omega}$ are chosen precisely to reproduce \eqref{eq:eqloc2d}, as we explain now. The integration variables $\phi_i^{(\omega)}$ pick up poles at
\begin{enumerate}
    \item $A_{\a}$ where $\a \in c^{-1} (\o)$,
    \item $\phi_{i'}^{(\omega)}+\ve_1$ of another $i'$, or
    \item $\phi_{j}^{(\omega-1)}$ for some $j$.
    \item If $\omega=0$, $\phi_{i}^{(0)}$ can also pick up poles at $w_s$.
\end{enumerate}
We start by choosing poles for $\phi^{(0)} _i$, $i=1,2,\cdots, v_0$, and move on to $\phi^{(\omega+1)} _i$ only after we exhaust all $\phi^{(\omega)} _i$, $i=1,2,\cdots, v_\o$. The combination of 1 and 2 and the combination of 2 and 3 with $\phi^{(\omega-1)}_j = e^{A_l}q_1^{k'-1}$ generate a tail $T_\o$. Also, the combination of 3 and 4, where $\phi^{(\omega)}_i=w_s$ for some $s$, spans a jump $J_\o$. By taking all possible such combinations, we recover all the fixed points of the modified hand-saw quiver in the positive chamber enumerated by $\boldsymbol\L \in \r^{-1} (\boldsymbol\l)$ as the poles picked up. Moreover, the contour is chosen not to enclose poles elsewhere, and the rest of the integrand provides the correct factors to reproduce the monodromy defect observable \eqref{eq:eqloc2d} by the residue. For instance, $\prod_{\a\in c^{-1}(\o+1)}(\phi_i^{(\omega)}-A_{\a}+\ve_1)^{-1}$ term gives $\BE[q_1 V_{\omega} L_{\omega+1}^*]$, etc.

Now we explain that the contour integral \eqref{eq:contourN} can be evaluated in a different way, by picking up the exact complement set of the poles including the one at infinity. Namely,
\begin{enumerate}
    \item We start from $\omega=l-2$.
    \item We split $v_{\omega}$ integration variables $(\phi ^{(\o)} _i)_{i=1} ^{v_\o}$ into two groups, where the first group consists of $r_\o$ and the second group consists of $v_\o-r_\o$ of them, for some $0 \leq r_\o \leq v_\o$. The second group takes the pole at infinity, while the first group takes poles elsewhere. 
    \item The residue at infinity from the integration of the second group is computed as $\frac{(-1)^{v_\o-r_\o}}{(v_\o - r_\o)!} 
          \prod_{k=1} ^{ v_\o-r_\o}(R_{\infty}^{(\o)} +v_{\o-1} -r_{\o+1} - k+1 ) $, where $R_\infty^{(\o)} = \frac{1}{\ve_1}(m^+_{\o}+m^-_{\o}- \sum_{\a \in c^{-1}(\o) } A_{\a}-\sum_{\a \in c^{-1}(\o+1)} A_{\a}+\ve_1)  $. Here, we set $v_{-1} = w$ and $r_{l-1 }= 0 $.
    \item Change the order of summation by $\sum_{v_\o=0}^\infty \sum_{r_\o = 0} ^{v_\o} = \sum_{r_\o=0}^\infty \sum_{v_\o = r_\o} ^{\infty}  $. The second summation gives $c_\o ^{R_\infty ^{(\o)} +v_{\o-1} + r_{\o+1}}$, where $c_\o$ can be recursively determined by $c_{\o-1} = 1+ \qe_\o c_\o$ and $c_{l-2} = 1+ \qe_{l-2}$. Namely, 
\begin{align}\label{def:c_w}
    c_\omega = 1 + \kq_{\omega} + \kq_{\omega}\kq_{\omega+1} + \cdots +  \kq_{\omega} \qe_{\o+1} \cdots \kq_{l-2}.
\end{align}
    We leave the integration of the remaining $r_\o$ variables in the first group and the summation over $r_\o$ undone and proceed.
    \item Repeat the steps 2 to 4 with ${\omega} \to \omega-1$, up to $\o=0$. Then, we are left with the integration of $r_\o$ variables for each $\o=0,1,\cdots, l-2$, and the summation over $r_\o$. As a result of step 4, we get the prefactor $ c_0 ^{w} \prod_{\o=0} ^{l-2} c_\o ^{R_\infty ^{(\o)}}$, and also the mutation of the complexified FI parameters $\qe_\o \to \qe_\o \frac{c_{\o+1}}{c_{\o-1}}$, $\o=0,1,\cdots, l-2$ appearing in the remaining integrals, where we set $c_{-1} = c_{l-1} = 1$.
  \item The integration of the remaining $r_\o$ variables is performed by taking the residues at $A_{\alpha}-\ve_1 - (i-1)\ve_1$, $i \geq 0$, where $ \o < c(\a) \leq l-1 $. These poles gathered from all $\o=0,1,\cdots, l-2$ compose a restricted Young diagram $\tilde{\boldsymbol\L}$. The residue is precisely the contribution of $\tilde{\boldsymbol\L}$ to the equivariant integral \eqref{eq:mononegative} in the negative chamber, with the mutated $\qe_\o$.
  \item Sum over all $r_\o \geq 0$. Then we recover \eqref{eq:mononegative} with the mutated $\qe_\o$ and the prefactor explained in step 5.
\end{enumerate}
All in all, we achieve the following crossing formula relating the monodromy observable in the positive and the negative chamber:
\begin{align}\label{eq:Relation of chi's}
    {\Psi}(\qe_\o)[\boldsymbol\l] ^{\text{non-pert}} _{\z>0}
    &=c_0^{\frac{1}{\ve_1}\sum_{\alpha=0}^{N-1} (A_\alpha-a_\alpha )} \prod_{\o=0}^{l-2}c_{\o }^{R^{(\o)}_\infty} \times \Psi\left( \qe_\o \frac{c_{\o+1}}{c_{\o-1}} \right)[\boldsymbol\l]^{\text{non-pert}}_{\z<0}.
\end{align}
We remind that the classical contribution \eqref{eq:classsigma} of the sigma model should further be augmented to reproduce the original monodromy surface observable \eqref{eq:monoobs}.

This crossing formula is useful since the expression of the monodromy defect observable is simplified in the negative chamber. Furthermore, the crossing formula gives a direct exhibition of transition of the monodromy surface defect into (multiple) $\mathbf{Q}$-observables or $\mathbf{H}$-observables, as we investigate from now on.

\subsection{Transition between $\BZ_2$-monodromy defect and $\mathbf{H}$-observable}
Before studying the transition that leads to the separation of variables for the Gaudin model and the XXX spin chain, we will apply the crossing formula \eqref{eq:Relation of chi's} to show dualities among the monodromy surface defects defined by $\BZ_2$-orbifold and the $\mathbf{H}$-observables.

\subsubsection{Monodromy surface defect dual to $\mathbf{H}$-observable}
Let us consider the monodromy defect defined by the $\BZ_2$-orbifold, where the coloring function is given by
\begin{align} 
\begin{split}
    c(\alpha) & = \begin{cases}
        1& \alpha = \beta, \\
        0 & \text{otherwise},
    \end{cases} 
\end{split}
\end{align}
for a chosen $\b \in \{0,1,\cdots, N-1\}$. In other words, we have
\begin{align} \label{eq:gaugecolor}
\begin{split}
    &L_0 = \sum_{ 0 \leq  \a\neq \b \leq N-1 } e^{A_\a}, \qquad L_1 = e^{A_\b}.
\end{split}
\end{align}
We also assign the coloring for the flavor as
\begin{align} \label{eq:massz2}
\begin{split}
\begin{split}
    &M_0 ^+ = \sum_{\a=0} ^{N-1} e^{m_{\a} ^+},\qquad M_1 ^+ = 0 \\
    & M_0 ^- =  0    \quad\quad  M_1 ^- = \sum_{\a=0} ^{N-1} e^{m_{\a} ^-}.
\end{split}
\end{split}
\end{align}
In this case, the Higgs branch of the two-dimensional $\EN=(2,2)$ gauged sigma model is the total space of
\begin{align}
    \mathcal{E} \otimes \BC^N \longrightarrow F(N-1,N),
\end{align}
where $\mathcal{E}$ is the tautological vector bundle over the flag variety $F(N-1,N)$. This is dual to the total space of
\begin{align}
    \mathcal{O}(-1)\otimes \BC^N \longrightarrow \BP^{N-1}. 
\end{align}
This is precisely the gauged linear sigma model that we discussed in section \ref{sec:Qopssrf}. It is in the non-linear sigma model phase here by turning on the complexified FI parameter, rather than the vev of the adjoint scalar, which produces the $\mathbf{H}$-observable by coupling to the $\mathcal{N}=2$ $A_1$-quiver gauge theory \cite{Jeong:2018qpc}. Therefore, we expect that the monodromy surface defect defined by the $\BZ_2$-orbifold is dual to the $\mathbf{H}$-observable, where the prescribed singularity in the former can be thought of as a remnant of integrating out two-dimensional degrees of freedom in the latter. We will now explain that this duality can indeed be confirmed by the equality between their observable expressions, as a consequence of the crossing formula derived in \eqref{eq:Relation of chi's}.

The restricted Young diagrams enumerating the fixed points in the negative chamber are Young diagrams with single column of length, say, $s \geq 0$ (see Figure \ref{fig:dualfixed}). These Young diagrams represent the $V_0$ space generated by the action of $B_1 ^\dagger$ on the image of the one-dimensional space $L_1$ under $j^\dagger$, $V_0 = \BC [B_1 ^\dagger] j^\dagger (L_1)$. Its equivariant Chern character is therefore 
\begin{align}
    V_0 = \sum_{i=1} ^{s} e^{A_\b } q_1 ^{-i},
\end{align}
so that
\begin{align}
    \S_0 = L_1 +P_1 V_0 = e^{A_\b} q_1 ^{-s}.
\end{align}
The monodromy defect observable in the negative chamber is computed as 
\begin{align} \label{eq:z2neg}
\begin{split}
    \Psi_\b [\boldsymbol\l]_{\z<0} &= \sum_{s = 0} ^\infty \qe_0 ^{s} \; \BE\left[ \frac{(S- \S_0 -M_0 ^+)\S_0 ^* }{P_1 ^*} \right] \\
    & =  \frac{\prod_{\alpha} \sin \left( \frac{a_\beta - m_\alpha^+ }{\ve_1} \right) }{\pi \prod_{\alpha\neq \beta} \sin\left( \frac{a_\beta-a_\alpha}{\ve_1} \right) } \times \sum_{s=0}^\infty \kq_0^s \, 
\prod_{\a=0}^{N-1} \G\left( s+1 - \frac{A_\b -m^+ _\a}{\ve_1} \right) Q(A_\beta-(s+1))  \\
    &=   \frac{\prod_{\alpha} \sin \left( \frac{a_\beta - m_\alpha^+ }{\ve_1} \right) }{\pi \prod_{\alpha\neq \beta} \sin\left( \frac{a_\beta-a_\alpha}{\ve_1} \right) }  \qe_0 ^{\frac{A_\b}{\ve_1} - 1}  \mathbf{H}^{(\b)} (\qe_0 ) 
\end{split}
\end{align}
Here, $Q(x)[\boldsymbol\l ] = \BE \left[- \frac{e^{x} S ^*[\bl] }{P_1 ^*} \right]$ is the $\mathbf{Q}$-observable and $\mathbf{H}^{(\b)} (\qe_0)$ is precisely the $\mathbf{H}$-observable observable with the complexified FI parameter $\qe_0$, where $\b \in \{0,1,\cdots, N-1\}$ enumerates the choice of vacuum at infinity among $N$ discrete vacua. Recall that the monodromy defect observable is convergent in the domain $0< \vert \qe_0 \vert , \vert \qe_1 \vert < 1$, or equivalently, $ \vert \qe \vert < \vert \qe_0 \vert <1 $. This is precisely the domain where the series expansion of the emergent $\mathbf{H}$-observable converges. The $\mathbf{H}$-observable can in fact be analytically continued to other domains, $0< \vert \qe_0 \vert < \vert \qe \vert $ and $1 < \vert \qe_0 \vert$, by using its contour integral representation \cite{Jeong:2018qpc}. We will not discuss this analytic continuation here, but remark that the equality \eqref{eq:z2neg} will still hold across different convergence domains when the monodromy surface defect observable is also properly analytically continued.

The crossing formula \eqref{eq:Relation of chi's} simplifies in this case $(l=2)$ since the mutation of the complexified FI parameters is absent. It gives
\begin{align}
\begin{split}
    \Psi_\b (\qe_0) [\boldsymbol\l] &=   \qe_0 ^{-l(\l^{(\b)})} (1+\qe_0) ^{1+ \frac{1}{\ve_1} \sum_{\a=0} ^{N-1} (m_{\a} ^+ -a_\a) } \Psi [\boldsymbol\l]_{\z<0} \\
    & = \frac{\prod_{\alpha} \sin \left( \frac{a_\beta - m_\alpha^+ }{\ve_1} \right) }{\pi \prod_{\alpha\neq \beta} \sin\left( \frac{a_\beta-a_\alpha}{\ve_1} \right) }  \qe_0 ^{\frac{a_\b}{\ve_1} -1 } (1+\qe_0) ^{1+ \frac{1}{\ve_1} \sum_{\a=0} ^{N-1} (m_{\a} ^+ -a_\a) } \mathbf{H}^{(\b)} (\qe_0) [\bl].
\end{split}
\end{align}
The equality between the observable expressions manifests the duality between the monodromy surface defect defined by the $\BZ_2$-orbifold with the coloring functions \eqref{eq:massz2} and the $\mathbf{H}$-observable. The identity was derived and used for the cases of $N=2$ and $3$ in \cite{Jeong:2018qpc}, extended to arbitrary $N$ here.

\subsubsection{Transition between $\BZ_2$-monodromy defects and $\mathbf{H}$-observable} \label{eq:z2transition}
We emphasize here that the $\BZ_2$-monodromy defect defined by the coloring in \eqref{eq:gaugecolor} and \eqref{eq:massz2} is different from the regular $\BZ_N$-monodromy defect, even in the case of $N=2$ where the two defects can be compared on equal footing. The distinction comes from the flavor coloring; the flavor coloring \eqref{eq:massz2} of the $\BZ_2$-monodromy defect sets all the fundamental (resp. anti-fundamental) hypermultiplets to carry charge $0$ (resp. charge $1$), while the flavor coloring of the regular monodromy defect sets exactly one fundamental hypermultiplet and one anti-fundamental hypermultiplet to carry charge $\o$ for each $\o \in \{0,1,\cdots, N-1\}$. 

Let us recall that the flavor coloring determines the embedding of the flavor group of the two-dimensional sigma model into the flavor group of the four-dimensional bulk theory. In the case of $N=2$, where the complexified flavor group of the four-dimensional gauge theory is $GL(4)$ by treating all four hypermultiplets as fundamental, the two defects correspond to the same two-dimensional sigma model whose $GL(2)$ flavor group is identified with $GL(2)$ subgroup appearing in two different decompositions of $GL(4)$ into $GL(2) \times GL(2)$. Here, we show that such a discrete group action on the flavor group embedding is reflected in their observable expressions as an integral transformation between them. 

Let us consider the regular monodromy defect at $N=2$. The coloring for the Coulomb moduli is chosen as the same as in the case of the $\BZ_2$-monodromy defect \eqref{eq:gaugecolor} at $N=2$ with $\b=1$,
\begin{align}
    L_0 = e^{A_0}, \quad\quad L_1 = e^{A_1},
\end{align}
while the coloring for the flavor is chosen in a different manner as
\begin{align}
\begin{split}
    &M_0 ^+ = e^{m^+ _0}  ,\quad\quad M_1 ^+ = e^{m^+ _1}  \\
    &M_0 ^- = e^{m^- _0} ,\qquad M_1 ^- = e^{m^- _1}.
\end{split}
\end{align}
Compared to the $N=2$ case of \eqref{eq:massz2}, the $\BZ_2$-charge assignment for the masses got shuffled according to a discrete group action. The vacuum expectation value of the regular monodromy defect in $N=2$ can again be expressed using the $\mathbf{Q}$-observable as
\begin{align}
\begin{split}
    \Psi[\boldsymbol\l] &=  \qe_0 ^{\frac{a_1-A_1}{\ve_1}} (1+\qe_0) ^{1+ \frac{1}{\ve_1}  (m_{0} ^+ + m_0 ^- -a_0 - a_{1}) } \\
    & \quad  \times  \sum_{s=0} ^\infty \qe_0 ^{s} \,  \frac{  (-\ve_1)^{1+2s + \frac{m_0 ^+ + m_0 ^- - 2A_1}{\ve_1}}\G \left( 1+s + \frac{m^- _0 - A_1}{\ve_1} \right)}{\G\left( -s + \frac{A_1 - m_0 ^+}{\ve_1} \right)} \frac{\pi}{\sin \pi \frac{A_1 -A_0}{\ve_1}} Q(A_1 -(s+1)\ve_1) \\
    & = -\frac{   (-\ve_1)^{-1+\frac{m_0 ^- - m_1 ^+}{\ve_1}} e^{\pi i \frac{a_1 - m^+ _1}{\ve_1}} \sin\pi \frac{a_1 - m_0 ^+}{\ve_1}   }{ 4\sin \pi \frac{a_1 -A_0}{\ve_1} \sin \pi \frac{m_0 ^- -a_1}{\ve_1} \G\left( \frac{m_1 ^+ - m_0 ^-}{\ve_1}\right) }  \times  \qe_0 ^{\frac{a_1}{\ve_1} -1 } (1+\qe_0) ^{1+ \frac{1}{\ve_1}  (m_{0} ^+ + m_0 ^- -a_0 - a_1) }  \\
    & \quad \quad \times \sum_{s=0} ^\infty \qe_0 ^{-\frac{A_1}{\ve_1} +s+1} Q(A_1 -(s+1)\ve_1) \prod_{\a =0,1} (-\ve_1)^{1+s - \frac{A_1 - m_\a ^+}{\ve_1}} \G\left( 1+s - \frac{A_1 - m^+ _\a}{\ve_1} \right) \\ 
    & \quad \quad  \times\int_{\text{PC}} dt \, t^{ s + \frac{m_0 ^- - A_1}{\ve_1}} (1-t) ^{-1 + \frac{m_1 ^+ - m_0 ^-}{\ve_1}},
\end{split}
\end{align}
where we used the integral representation of the beta function,
\begin{align} \label{eq:betaint}
    B(a,b-a) = \frac{\Gamma(a) \Gamma(b-a)}{\Gamma(b)} = - \frac{e^{-\pi i b}}{ 4 \sin \pi a \sin \pi (b-a) } \int_\text{PC} dt\, t^{a-1} (1-t)^{b-a-1}.
\end{align}
Here, PC denotes the Pochhammer contour on the $t$-plane enclosing $t=0$ and $t=1$.

Now, we change the order of summation and integration in the last line. The summation converges in the domain $\vert \qe \vert <\vert \qe_0 t \vert <1$ and gives the $\mathbf{H}$-observable. Some part of the Pochhammer contour deviates from this domain, where the integrand can be properly analytically continued. As a result, after absorbing irrelevant functions by redefinition of the observable, we have
\begin{align}
\begin{split}
    \Psi(\qe_0)[\bl] &= \qe_0 ^{1 + \frac{a_1- m^+ _1}{\ve_1}  } (1+\qe_0) ^{1+ \frac{1}{\ve_1}  (m_{0} ^+ + m_0 ^- -a_0 - a_1) }  \\
    &\times\int_{\text{PC}} dy\, y^{-\frac{1}{2} + \frac{m_0 ^-- m^- _1}{2\ve_1}} (y-\qe_0) ^{-1 + \frac{m_1 ^+ - m^- _0}{\ve_1}} (y-\qe)^{\frac{\bar{m}^- -\bar{a}}{\ve_1}} (y-1)^{-1 -\frac{\bar{m}^+ - \bar{a}}{\ve_1}} \mathbf{H}^{(1)} (y)[\bl],
\end{split}
\end{align}
where $\text{PC}$ is now the Pochhammer contour in the $y$-plane enclosing $y=0$ and $y=\qe_0$. Note that even though the integrand has monodromies around $y=0$ and $y=\qe_0$, it is single-valued along the Pochhammer contour and therefore it is well-defined. 

Note that the integral transformation has been written for observable expressions at each $\bl$. Thus, it describes the transition of the regular $\BZ_2$-monodromy surface defect into a $\mathbf{H}$-observable. Upon taking the expectation value, we get
\begin{align}
    \Big\langle \Psi(\qe_0) \Big\rangle_\ba = \int_{\text{PC}} dy \,   \EuScript{K}(\qe_0;y) \Big\langle \mathbf{H}^{(1)} (y) \Big\rangle_\ba,
\end{align}
where the integral kernel $\EuScript{K}(\qe_0 ;y)$ is given by 
\begin{align} \label{eq:kernelsl2}
\begin{split}
    \EuScript{K}(\qe_0;y) &= \qe_0 ^{1 + \frac{a_1- m^+ _1}{\ve_1}  } (1+\qe_0) ^{1+ \frac{1}{\ve_1}  (m_{0} ^+ + m_0 ^- -a_0 - a_1) } \\
    & \quad \times y^{-\frac{1}{2} + \frac{m_0 ^-- m^- _1}{2\ve_1}} (y-\qe_0) ^{-1 + \frac{m_1 ^+ - m^- _0}{\ve_1}} (y-\qe)^{\frac{\bar{m}^- -\bar{a}}{\ve_1}} (y-1)^{-1 -\frac{\bar{m}^+ - \bar{a}}{\ve_1}}.
\end{split}
\end{align}
Note that the relation is valid for any value of $\ve_1,\ve_2$. The results of \cite{Nekrasov:2021tik, Jeong:2021bbh, Jeong:2023qdr} (at $N=2$) identifies the left hand side to a twisted coinvariant of four $\widehat{\fsl}(2)$-modules (two (co-)Verma modules and two bi-infinite modules). Also, the results of \cite{Nikita:IV,Jeong:2018qpc} (at $N=2$) show that the vacuum expectation value $\langle \mathbf{H}(y) \rangle$ of the $\mathbf{H}$-observable is identified with a Virasoro conformal block, with four generic vertex operators and one simplest degenerate operator (see also \cite{Jeong:2017mfh,Jeong:2020uxz}). Thus, the above equation can be interpreted as a transformation between conformal blocks of two different vertex algebras. Differently put, it reformulates the simplest example of the KZ/BPZ correspondence \cite{Frenkel95,Stoyanovsky2000ARB,Ribault:2005wp} as a surface defect transition in the $\EN=2$ gauge theory. Note that the integral kernel \eqref{eq:kernelsl2} is guaranteed to match with the ones in \cite{Stoyanovsky2000ARB,Frenkel:2015rda}, after reducing the latter to the invariants under the action of global $\fsl(2)$ on $\widehat{\fsl}(2)$-modules and global $SL(2)$ on $\BP^1$ with four marked points, due to the KZ equation solved by the left hand side and the BPZ equation solved by the integrand on the right hand side.

\subsection{Separation of variables for XXX spin chain}
Finally, let us apply the crossing formula \eqref{eq:Relation of chi's} to the regular monodromy surface defect defined by the $\BZ_N$-orbifold. We show that the crossing formula manifests a transition to $N-1$ $\mathbf{Q}$-observables or $N-1$ $\mathbf{H}$-observables, which leads to the separation of variable transformations for the XXX spin chain and the Gaudin model, respectively.

\subsubsection{Separated expression for eigenstates}
\begin{figure}[h!]
    \centering
        \begin{tikzpicture}        
\draw  (-1,0)   node {$\tilde{\L}^{(0)}=\varnothing$};
\draw  (1,0)   node {$\tilde{\L}^{(1)}=$};

\draw (2,0) node {
\ytableausetup
{mathmode, boxsize=1em}
\begin{ytableau}
*(red!40)  \\
  *(red!40)  \\
  *(red!40)  \\
  *(red!40)   \\
\end{ytableau} 
 };

 \draw  (2.8,0) node {$\begin{rcases} \\ \\ \\ \end{rcases} s_{1,1}$};

\draw  (4.5,0) node {$\tilde{\L}^{(2)}=$};

\draw (5.8,0) node (n) {
\begin{ytableau}
  \none & *(green!40)   \\
  \none & *(green!40) \\
 *(red!40)  & *(green!40) \\
 *(red!40)  &  *(green!40) \\
 *(red!40)  & *(green!40) \\
\end{ytableau} };

\draw  (6.8,-0.4) node {$\begin{rcases}  \\ \\  \end{rcases} s_{2,1}$};
\draw  (6.8,0.6) node {$\begin{rcases} \\  \end{rcases} s_{2,2}$};

\draw  (8.5,0) node {$\tilde{\L}^{(3)}=$};
\draw (9.8,0) node (n) {

\begin{ytableau}
\none &  \none & *(blue!40)  \\
\none & \none & *(blue!40)   \\
\none & *(green!40)  & *(blue!40)  \\
\none &*(green!40)  & *(blue!40)  \\
*(red!40) & *(green!40)  &  *(blue!40)  \\
*(red!40) & *(green!40)  & *(blue!40)  \\
\end{ytableau}};

\draw  (4,-1.8)   node {\textcolor{red!40}{ $ \begin{varwidth}{5cm} \ytableausetup{boxsize=3mm}
    \begin{ytableau}
    *(red!40) 
\end{ytableau}\end{varwidth} : V_0 $}};
\draw  (5.5,-1.8)   node {\textcolor{green}{ $ \begin{varwidth}{5cm} \ytableausetup{boxsize=3mm}
    \begin{ytableau}
    *(green!40) 
\end{ytableau}\end{varwidth} : V_1 $}};
\draw  (7,-1.8)   node {\textcolor{blue!40}{ $ \begin{varwidth}{5cm} \ytableausetup{boxsize=3mm}
    \begin{ytableau}
    *(blue!40) 
\end{ytableau}\end{varwidth} : V_2 $}};

\draw  (11,-0.82) node {$\begin{rcases} \\   \end{rcases} s_{3,1}$};
\draw  (11,0) node {$\begin{rcases} \\   \end{rcases} s_{3,2}$};
\draw  (11,0.82) node {$\begin{rcases} \\   \end{rcases} s_{3,3}$};

\end{tikzpicture}
    \caption{A collection $\tilde{\boldsymbol\L}$ of restricted Young diagrams for the regular monodromy defect observable in the negative chamber, exemplified in the case of $N=4$. Note that $\tilde{\L} ^{(\o)}$ is composed of $\o$ columns each of which carries different representations of $\BZ_N$. The restricted Young diagrams in this case can be parametrized by the differences of lengths of these columns (in the example, $\frac{4 \times 3}{2} = 6$ non-negative integers $(s_{1,1},s_{2,1},s_{2,2},s_{3,1},s_{3,2},s_{3,3})$).}
    \label{fig:regularnegative}
\end{figure}
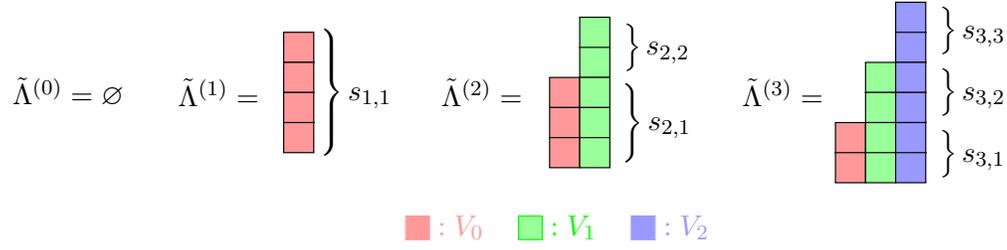

Let us recall that, in the negative chamber, the fixed points of the modified hand-saw quiver variety is enumerated by the collections $\{\tilde{\boldsymbol\L} \}$ of restricted Young diagrams. The vector spaces $(V_\o)_{\o=0} ^{N-2}$ are generated as in \eqref{eq:Vneg},
\begin{align}
\begin{split}
    &V_{N-2} = \BC[ B_1 ^\dagger ] j^\dagger (L_{N-1}) \\
    &V_\o= \BC[B_1 ^\dagger ]B_2 ^\dagger (V_{\o+1})  \oplus \BC[B_1 ^\dagger] j^\dagger (L_{\o+1}) ,\quad\quad \o=0,1,\cdots, N-3,
\end{split}
\end{align}
where each $(L_\o)_{\o=0} ^{N-1}$ is one-dimensional in this case. The columns in the restricted Young diagram $\tilde{\boldsymbol\L}$ represent the basis vectors in $V_\o$ obtained by repetitive $B_1 ^\dagger$-action on $j^\dagger (L_{\o+1}) $ or $B_2 ^\dagger (V_{\o+1})$. We can parametrize the restricted Young diagrams by introducing $\frac{N(N-1)}{2}$ non-negative integers $\mathbf{s} = \left(s_{n,j} \right)_{1\leq j \leq n \leq N-1}$ for the differences of lengths of the neighboring columns (see Figure \ref{fig:regularnegative}).

Then the equivariant Chern characters of $V_\o$ can be written in terms of $\mathbf{s}$ by
\begin{align}
\begin{split}
     & V_\o = \sum_{n=\o+1} ^{N-1}  \sum_{i=1} ^{\sum_{j=1} ^{\o+1} s_{n,j} } e^{A_{n}} q_1 ^{-i}, \quad\quad \o=0,1,\cdots, N-2.
\end{split}
\end{align}
This implies
\begin{align}
\begin{split}
    \S_\o &= \sum_{n= \o+1} ^{N-1} L_{n} + P_1 V_\o = \sum_{n = \o+1} ^{N-1} e^{A_{n}} q_1 ^{-\sum_{j=1} ^{\o+1} s_{n,j}} \\
    & =: \sum_{i=\o+1} ^{N-1} e^{x_{\o+1,i} +\ve_1}  \end{split}, \quad\quad \o=0,1,\cdots, N-2,
\end{align}
where we defined 
\begin{align} \label{eq:xpoles}
    x_{n,i} := A_i - \left(1 + \sum_{j=1} ^{n} s_{i,j} \right)\ve_1,\quad\quad 1 \leq n \leq i \leq N-1.
\end{align}
Note that we can write this in a recursive way as $x_{n,i} = x_{n-1,i} -s_{i,n} \ve_1$, with $2 \leq n \leq  i \leq N-1$, and $x_{1,i} = A_i - (1+ s_{i,1})\ve_1$ for $i=1,2,\cdots, N-1$. The monodromy defect observable in the negative chamber is expressed as
\begin{align} \label{eq:psimiddle}
\begin{split}
    \Psi(\bu)[\boldsymbol\l]_{\z<0} &= \sum_{\mathbf{s} = 0} ^\infty \prod_{\o=0} ^{N-2} \qe_\o ^{\sum_{n =\o+1} ^{N-1} \sum_{j=1} ^{\o+1} s_{n,j}} \mathbb{E}\left[\frac{S \S_0 ^*+\sum_{\omega=0} ^{N-2} \left( -  \S_\o (\S_{\omega}-\S_{\omega+1})^* -M_{\omega} ^+ \S_{\omega}^* + q_1 ^{-1} (M_\o ^-)^* \S_\o \right) }{P_1^*}\right] \\
    & = \sum_{\mathbf{s} = 0} ^\infty \prod_{\o=0} ^{N-2} \qe_\o ^{\sum_{i =\o+1} ^{N-1} 
 \left( -1 + \frac{A_i-x_{\o+1,i}}{\ve_1} \right) }   \frac{ \prod_{ \substack{ \o+1 \leq i \leq N-1 \\  \o+2 \leq j \neq i \leq N-1  } } (-\ve_1) ^{\frac{x_{\o+2,j}- x_{\o+1,i} }{\ve_1}} \G\left( \frac{ x_{\o+2,j} -x_{\o+1,i} }{\ve_1} \right) }{ \prod_{\o +2 \leq i \leq N-1 } s_{i,\o+2}! \ve_1 ^{s_{i,\o+2}} \prod_{ \o +1 \leq i\neq j \leq N-1 } \G\left( \frac{x_{\o+1,i}-x_{\o+1,j}}{\ve_1} \right)  } \\
 & \quad \times  \prod_{1\leq n \leq i \leq N-1}   \frac{ (-\ve_1) ^{ \frac{m^- _{n-1} -x_{n,i}} {\ve_1}} \G\left( \frac{ m^- _{n-1}- x_{n,i} }{\ve_1} \right) }{ 
 (-\ve_1) ^{1 + \frac{x_{n,i} -m^+ _{n-1} }{\ve_1}} \G\left( 1+ \frac{ x_{n,i} -m^+ _{n-1}}{\ve_1} \right)}  \times \prod_{i=1}^{N-1} (-1)^{s_{i,1}+1}  Q(x_{1,i})[\boldsymbol\lambda] \prod_{\substack{\a=0 \\ \a \neq i } } ^{N-1} \frac{\pi}{ \sin \pi  \frac{x_{1,i} - A_\a}{\ve_1} }  .
\end{split}
\end{align}
We can express the summation over non-negative integers $\mathbf{s}$ as a $\frac{N(N-1)}{2}$-dimensional contour integral, which reproduces the summation over $\mathbf{s}$ by picking up the residues at simple poles. For this, we promote $\bx= (x_{n,i})_{1 \leq n \leq i \leq N-1}$ to integration variables, and write
\begin{align} \label{eq:qSov}
\Psi[{\boldsymbol\lambda}]_{\z<0} 
    & = \int_{\CalC}  \prod_{i=1} ^{N-1} \frac{dx_{1,i}}{2\pi \rm{i} \ve_1} \; \mathbf{V}(\bx_1) \mu(\bx_1 )C(\bx_1), 
\end{align}
where $\m(\bx_1)$ and $C(\bx_1)$ only depends on $\bx_1 = (x_{1,i})_{i=1} ^{N-1}$, and $\mathbf{V}(\bx_1)$ is a $\frac{(N-1)(N-2)}{2}$-dimensional contour integral of remaining variables $(x_{n,i})_{2\leq n  \leq i \leq N-1}$. The contour $\mathcal{C}$ is chosen precisely in the way that $\bx$ picks up the simple poles at \eqref{eq:xpoles}, as we now explain.

Firstly, ${\bf V}(\bx_1)$ is a $\frac{(N-1)(N-2)}{2}$-dimensional contour integral given by
\begin{align}\label{eq: Mellin fund}
\begin{split}
    {\bf V}(\bx_1)=\int_\mathcal{C}
    &\prod_{\o=1}^{N-2} \frac{1}{(N-\o-1)!} \prod_{i=\o+1}^{N-1}  \frac{dx_{\o+1,i}}{2\pi {\ve_1}{\ii}} {\kq}_{\o}^{\sum\limits_{i=\o+1}^{N-1} \left( -1+ \frac{A_i - x_{\o+1,i}}{{\ve_1}}  \right)} 
    \frac{ \prod_{ \substack{ \o \leq i \leq N-1 \\  \o+1 \leq j \leq N-1 } } (-\ve_1) ^{ \frac{x_{\o+1,j}-x_{\o,i}}{\ve_1}} \G\left( \frac{x_{\o+1,j}- x_{\o,i} }{\ve_1} \right) }{ \prod_{ \o +1 \leq i\neq j \leq N-1 } \G\left( \frac{x_{\o+1,i}-x_{\o+1,j}}{\ve_1} \right)  } \\
    & \quad\quad \times \prod_{2\leq n \leq i \leq N-1}  \frac{(-\ve_1) ^{ \frac{ m^- _{n-1}  - x_{n,i}}{\ve_1}} \G\left( \frac{  m^- _{n-1} - x_{n,i} }{\ve_1} \right)}{  
 (-\ve_1) ^{1 + \frac{x_{n,i}- m^+ _{n-1} }{\ve_1}} \G\left( 1+ \frac{ x_{n,i}- m^+ _{n-1} }{\ve_1} \right)} ,
\end{split}
\end{align}
where the contour $\mathcal{C}$ picks up the simple poles of the $\G$-functions in the numerator in the following way.
\begin{enumerate}
    \item Start with $x_{N-1,N-1}$. Due to the $\G$-functions in the numerator, there are two semi-infinite lattices of simple poles in the $\frac{x_{N-1,N-1}}{\ve_1}$-plane starting from
    $\frac{x_{N-2,N-1}}{\ve_1}$ and $\frac{x_{N-2,N-2}}{\ve_1}$, growing toward the left. We choose the contour that encloses all the poles at $x_{N-1,N-1} = x_{N-2,N-1} - s_{N-1, N-1} \ve_1$ and $x_{N-1,N-1} = x_{N-2,N-2} - s_{N-1, N-1} \ve_1$ with $s_{N-1,N-1} \geq 0$. The residue is given by $\frac{(-1)^{s_{N-1,N-1}}}{ s_{N-1,N-1}! \ve_1 ^{s_{N-1,N-1}}}$.
    
    \item Proceed recursively. For the integral of $x_{\o+1,j}$, the relevant $\G$-functions in the numerator are $\G \left( \frac{x_{\o+1,j } - x_{\o,i}}{\ve_1} \right)$ for $1 \leq \o <i \leq N-1$, $\G\left( \frac{x_{\o+2,i} - x_{\o+1,j}}{\ve_1} \right)$ for $3 \leq \o+2 < i \leq N-1$, and finally $\G\left( \frac{m^- _{\o} - x_{\o+1,j}}{\ve_1} \right)$ for $3 \leq \o+1 \leq i \leq N-1 $. Correspondingly, there are three groups of semi-infinite lattices of simple poles. We choose the contour that only encloses the poles at $x_{\o+1,j} = x_{\o,i} - s_{j,\o+2} \ve_1 $ with $s_{j,\o+2} \geq 0 $. The residue is given by $\frac{(-1)^{s_{j,\o+2}}}{ s_{j,\o+2}! \ve_1 ^{s_{j,\o+2}}}$ and $\o < i \leq N-1$.

    \item The duplication made by rearranging $(x_{\o+1,i})_{\o+1 \leq i\leq N-1}$ is cancelled by the $(N-\o-1)!$ in the denominator.
\end{enumerate}
By taking the residues in the described way, $\mathbf{V}(\bx_1)$ precisely reproduce \eqref{eq:psimiddle} except the terms involving $\bx_1$. To get the rest, we perform the $(N-1)$-dimensional contour integral for $\bx_1 = (x_{1,i})_{i=1} ^{N-1}$. Here, $\mu(\bx_1)$ is given by
\begin{align}\label{def:transform measure}
\begin{split}
    \mu(\bx_1)
    &=\frac{\kq_{0}^{\sum_{i=1}^{N-1} \left( -1 + \frac{A_i -x_{1,i}}{{\ve_1}} \right)}}{(N-1)!} \prod_{i=1} ^{N-1} \frac{{(-\ve_1) ^{ \frac{ m^- _0 -x_{1,i} }{\ve_1}} \G\left(  \frac{ 
m^- _0- x_{1,i} }{\ve_1} \right) }}{  
  (-\ve_1) ^{1 + \frac{x_{1,i} -m^+ _0 }{\ve_1}} \G\left( 1+ \frac{ x_{1,i} -m^+ _0}{\ve_1} \right) }  \\
& \quad\quad\quad\quad\quad \times \frac{1}{\prod_{1 \leq j\neq k \leq N-1} \Gamma\left(\frac{x_{1,j}-x_{1,k}}{{\ve_1}}\right) } \prod_{i=1}^{N-1} \prod_{\a=0} ^{N-1} \frac{\pi}{\sin\pi\left(\frac{x_{1,i} -a_\a }{{\ve_1}}\right)}
\end{split}
\end{align}
Finally, $C(\bx_1)$ consists of $\mathbf{Q}$-observables,
\begin{align}\label{def:C(y) transfer}
    C(\bx_1)[\boldsymbol\lambda]= \prod_{i=1}^{N-1}  Q(x_{1,i})[\boldsymbol\lambda] .
\end{align}
The contour for $x_{1,i}$ is chosen in the following way. There are simple poles in the integrand at $x_{1,i} = A_\a - (s_{i,1}+1)\ve_1$ with $s_{i,1} \geq 0$ and $\a=0,1,\cdots, N-1$ from the sine functions in the denominator of $\m(\bx_1)$ (the semi-infinite lattices of poles toward the other direction are cancelled by the zeros of the inverse $\G$-functions in the $\mathbf{Q}$-observables in $C(\bx_1)$), and at $x_{1,i} = m_0 ^- + s_{i,1} \ve_1$ with $s_{i,1} \geq 0$ from the $\G$-functions in the numerator of $\m(\bx_1)$. The contour is chosen to enclose the poles at $x_{1,i} = A_\a - (s_{i,1}+1)\ve_1$ with $s_{i,1} \geq 0$ and $1\leq \a \leq N-1$. The residue is $(-1)^{s_{i,1}+1}$. The duplication of rearranging $(x_{1,i})_{i=1} ^{N-1}$ is cancelled by the $(N-1)!$ in the denominator of $\m(\bx_1)$. Performing the $\bx_1$-integral in this way, we precisely reproduce the observable expression \eqref{eq:psimiddle} in the negative chamber at last. 

To fully recover the observable expression in the original chamber, we need to take account for the crossing formula \eqref{eq:Relation of chi's} and the classical contribution \eqref{eq:classsigma}. Thus we get
\begin{align}
\begin{split}
    \Psi(\bu)[\bl] & = c_0 ^{\frac{\sum_{\a=0} ^{N-1} (A_\a -a_\a)}{\ve_1}} \prod_{\o=0} ^{N-2} c_\o ^{\frac{m^+ _\o + m^- _\o -A_\o -A_{\o+1} +\ve_1}{\ve_1}} \prod_{\o=1} ^{N-1} \left( \frac{u_\o}{u_0} \right)^{\frac{a_\o - A_\o}{\ve_1}} \\
    & \quad \times \int_{\CalC} \prod_{i=1} ^{N-1} \frac{dx_{1,i}}{2\pi \rm{i} \ve_1} \; \left( V(\bx_1) \m(\bx_1)\right)_{\qe_\o \to \qe_\o \frac{c_{\o+1}}{c_{\o-1}}}  C(\bx_1) [\bl],
\end{split}
\end{align}
The $\qe_\o$ only appears in elementary functions in the integrand $V(\bx_1) \m(\bx_1)$. We can recombine the effect of mutation $\qe_\o \to \qe_\o \frac{c_{\o+1}}{c_{\o-1}}$ with the terms in the first line. Importantly, all the $A_\o$'s in the exponents cancel with each other, yielding
\begin{align} \label{eq:SoVobs}
    \Psi(\bu) [\bl] = \int_{\CalC} \prod_{i=1} ^{N-1} \frac{dx_{1,i}}{2\pi \rm{i} \ve_1} \; \mathcal{K}(\bu;\bx_1) C(\bx_1 )[\bl],
\end{align}
where the integral kernel $\mathcal{K} (\bu;\bx_1) = \mathbf{c} \mathbf{V}' (\bx_1) \m' (\bx_1)$ is given by the product of
\begin{align} \label{eq:cconst}
    \mathbf{c} = c_0 ^{ N + \frac{x_{0,0} + x_{0,1} 
- a_0 -a_1 }{\ve_1} - \sum_{i=2} ^{N-1} \frac{a_i - x_{2,i}}{\ve_1} } \prod_{\o=0 }^{N-2} c_\o ^{-1 + \frac{m^+ _\o + m^- _\o -x_{\o,\o} -x_{\o,\o+1}}{\ve_1}} ,
\end{align}
\begin{align} \label{eq:vp}
\begin{split}
    {\bf V'}(\bx_1)=\int_\mathcal{C}
    &\prod_{\o=1}^{N-2} \frac{1}{(N-\o-1)!} \prod_{i=\o+1}^{N-1}  \frac{dx_{\o+1,i}}{2\pi {\ve_1}{\ii}} {\kq}_{\o}^{\sum\limits_{i=\o+1}^{N-1} \left( -1+ \frac{a_i - x_{\o+1,i}}{{\ve_1}}  \right)} 
    \frac{ \prod_{ \substack{ \o \leq i \leq N-1 \\  \o+1 \leq j \leq N-1 } } (-\ve_1) ^{ \frac{x_{\o+1,j}-x_{\o,i}}{\ve_1}} \G\left( \frac{x_{\o+1,j}- x_{\o,i} }{\ve_1} \right) }{ \prod_{ \o +1 \leq i\neq j \leq N-1 } \G\left( \frac{x_{\o+1,i}-x_{\o+1,j}}{\ve_1} \right)  } \\
    & \quad\quad \times \prod_{2\leq n \leq i \leq N-1}  \frac{(-\ve_1) ^{ \frac{ m^- _{n-1}  - x_{n,i}}{\ve_1}} \G\left( \frac{  m^- _{n-1} - x_{n,i} }{\ve_1} \right)}{  
 (-\ve_1) ^{1 + \frac{x_{n,i}- m^+ _{n-1} }{\ve_1}} \G\left( 1+ \frac{ x_{n,i}- m^+ _{n-1} }{\ve_1} \right)} ,
\end{split}
\end{align}
and 
\begin{align}\label{eq:mp}
\begin{split}
    \mu'(\bx_1)
    &=\frac{\kq_{0}^{\sum_{i=1}^{N-1} \left( -1 + \frac{a_i -x_{1,i}}{{\ve_1}} \right)}}{(N-1)!} \prod_{i=1} ^{N-1} \frac{{(-\ve_1) ^{ \frac{ m^- _0 -x_{1,i} }{\ve_1}} \G\left(  \frac{ 
m^- _0- x_{1,i} }{\ve_1} \right) }}{  
  (-\ve_1) ^{1 + \frac{x_{1,i} -m^+ _0 }{\ve_1}} \G\left( 1+ \frac{ x_{1,i} -m^+ _0}{\ve_1} \right) }  \\
& \quad\quad\quad\quad\quad \times \frac{1}{\prod_{1 \leq j\neq k \leq N-1} \Gamma\left(\frac{x_{1,j}-x_{1,k}}{{\ve_1}}\right) } \prod_{i=1} ^{N-1} \prod_{\a=0} ^{N-1} \frac{\pi}{\sin\pi\left(\frac{x_{1,i} -a_\a }{{\ve_1}}\right)},
\end{split}
\end{align}
and the contour $\CalC$ is chosen exactly as explained above. In particular, all the $A_\o [\bl] = a_\o + l (\l^{(\o)})\ve_1$ in the exponents were cancelled as we emphasized, so that the $\bl$-dependence of the integrand is solely in the product of $\mathbf{Q}$-observables, $C(\bx_1)[\bl]$. We stress that the equality \eqref{eq:SoVobs} holds at the level of observable expressions at each $\boldsymbol\l$. Thus, it is a direct manifestation of the transition of the regular monodromy surface defect into the $N-1$ $\mathbf{Q}$-observables.

Upon taking the vacuum expectation value, the integral transformation \eqref{eq:SoVobs} relates the vacuum expectation value of the regular monodromy surface defect and the correlation function of the $\mathbf{Q}$-observables. Namely, we get
\begin{align} \label{eq:SoVvac}
    \Big\langle \Psi(\bu) \Big\rangle_\ba = \int_{\CalC} \prod_{i=1} ^{N-1} \frac{dx_{1,i}}{2\pi \rm{i} \ve_1} \; \mathcal{K}(\bu;\bx_1) \left\langle \prod_{i=1} ^{N-1} Q(x_{1,i}) \right\rangle_\ba.
\end{align}
Note that the integral transformation is valid for any value of $\ve_1,\ve_2$. 

In particular, the kernel $\mathcal{K}(\bu,\bx_1)$ does not involve $\ve_2$, and the limit $\ve_2 \to 0$ can be taken without harming the validity of the integral transformation with the same kernel. In the limit $\ve_2 \to 0$, the normalized vacuum expectation value $\psi(\ba;\bu)$ of the regular monodromy surface defect gives a common eigenstate in $\CalH^{\BC^\times}$ of the $Q$-operators, as we showed in section \ref{subsubsec:paralleldefs}. We also showed in the companion paper \cite{Jeong:2024hwf} that this $Q$-eigenstate $\psi(\ba;\bu)$ is also a common eigenstate of the quantum Hamiltonians of the $\fgl(2)$ XXX spin chain. Meanwhile, the correlation function of the $\mathbf{Q}$-observables in the right hand side of the transition formula \eqref{eq:SoVvac} factorizes into the product of the vacuum expectation values of individual $\tilde{\mathbf{Q}}$-observables,
\begin{align}
    \lim_{\ve_2 \to 0} \left\langle \prod_{i=1} ^{N-1} Q(x_{1,i}) \right\rangle_\ba = e^{\frac{\widetilde{\EuScript{W}}(\ba;\qe)}{\ve_2}} \prod_{i=1} ^{N-1} Q(\ba; x_{1,i}),
\end{align}
since the $\mathbf{Q}$-observables, being local on the topological plane $\BC_2$, can be arbitrarily separated from each other in this limit. Therefore, this cluster decomposition of the $\mathbf{Q}$-observables yields an integral transformation between the $Q$-eigenstate and the product of $N-1$ $Q$-functions,
\begin{align} \label{eq:SoVsc}
    \psi(\ba;\bu) = \int_{\CalC} \prod_{i=1} ^{N-1} \frac{dx_{1,i}}{2\pi \rm{i} \ve_1} \; \mathcal{K}(\bu;\bx_1)  \prod_{i=1} ^{N-1} Q(\ba;x_{1,i}) .
\end{align}
Recall that each $Q$-function $Q(\ba;x_{1,i})$ is separately a solution to the same $\hbar$-oper difference equation \cite{Jeong:2024hwf},
\begin{align}
       0 = \left[ 1 - t(\ba;x)e^{-\ve_1 \p_x} + \qe P (x) e^{-2 \ve_1 \p_x} \right] Q(\ba;x+\ve_1),
\end{align}
which is none other than the Baxter TQ equation for the $Q$-function $Q(\ba;x_{1,i})$. Therefore, the integral transformation \eqref{eq:SoVsc} translates the $N-1$ spectral equations for the XXX spin chain (single $(N-1)$-body problem) into $(N-1)$-copies of identical second-order difference equations ($N-1$ single-body problems). Thus, we conclude \eqref{eq:SoVsc}, which expresses the transition between the regular monodromy surface defect and the $N-1$ $\mathbf{Q}$-observables in the limit $\ve_2 \to 0$, establishes the quantum separation of variables for the XXX spin chain associated to the $A_1$-quiver $\EN=2$ gauge theory. It was also studied in \cite{Lee:2020hfu} without making $\ve_2 \neq 0$ refinement.

We stress that our XXX spin chain associated to the four-dimensional $A_1$-quiver $\EN=2$ gauge theory is defined on bi-infinite $\fgl(2)$-modules, for which the algebraic Bethe ansatz method is not applicable by construction. In this sense, the separation of variables established in \eqref{eq:SoVsc} generalizes the one on highest-weight modules (see \cite{Derkachov:2002tf}, for example, for a previous study of the separation of variables for the $\fgl(2)$ spin chain for highest-weight modules). When the gauge theory parameters are specialized to trigger a higgsing to two-dimensional $\EN=(2,2)$ theory, we recover the standard Bethe/gauge correspondence where the XXX spin chain is defined on lowest-weight modules, and we will recover from \eqref{eq:SoVsc} the separation of variables for XXX spin chain on lowest-weight modules.

\subsubsection{Matching with Sklyanin's prescription}
We confirm here that the separation of variables for the XXX spin chain that we achieved above from the surface defect transition precisely reproduces the prescription by Sklyanin \cite{sklyanin1995separation}. 

The gauge theoretical construction the R-matrices was established in \cite{Jeong:2024hwf}. 
\begin{align}
    R_\o (x) = x - \theta_\o - \ve_1 \begin{pmatrix}
        u_\o \p_{u_\o} - s_\o  & 2s_\o \p_{u_\o} - u_\o \p_{u_\o}^2 \\ u_\o & -u_\o \p_{u_\o} + s_\o 
    \end{pmatrix} \in \text{End}(\BC^2) \otimes \text{End}(\CalH_\o),
\end{align}
where $\CalH_\o = u_\o ^{\frac{m^+ _\o -a_\o}{\ve_1}} \BC(\!(u_\o)\!)$ is a space of Laurent polynomials in a formal variable $u_\o$, realizing an evaluation module over the Yangian of $\fgl(2)$, $Y(\fgl(2))$. The spin $s_\o$ is given in terms of the mass parameters by
\begin{align} \label{eq:spin}
   s_\o = \frac{m^+_\o - m_\o^- -\ve_1}{2\ve_1}, \qquad \o=0,1,\cdots, N-1.
\end{align}
The trace part acts trivially. The evaluation parameter $\th_\o$ is encoded in the rest of the degrees of freedom of the mass parameters by
\begin{align} \label{eq:evalpa}
    \th_\o = \frac{m^+ _\o + m^- _\o+\ve_1 }{2 }, \qquad \o=0,1,\cdots, N-1.
\end{align}
Note that the $\fgl(2)$-module $\CalH_\o$ is irreducible for generic values of gauge theory parameters. Moreover, it is a bi-infinite module, possessing neither a highest-weight state nor a lowest-weight state.

The monodromy matrix is obtained by concatenating these R-matrices,
\begin{align}
    T(x)\vert_{\CalH} =  R_{N-1}(x) R_{N-2} (x) \cdots R_0 (x) =:\begin{pmatrix}
        A(x) & B(x) \\ C(x) & D(x)
    \end{pmatrix},
\end{align}
and we denoted the entries of the monodromy matrix by $A(x),B(x),C(x),D(x) \in \text{End}(\CalH)$. We verify that the separated variables $\bx_1 = (x_{1,i})_{i=1} ^{N-1}$ are given by the zeros of the lower-left entry $C(x)$ of the monodromy matrix, deriving the prescription of Sklyanin in our $\EN=2$ gauge theoretical setup.

In fact, we will show that a stronger version of the statement is true. Let us define \textit{partial} monodromy matrices by
\begin{align}
    T(x) \vert_{\otimes_{\o'=\o} ^{N-1} \CalH_{\o'}} =  R_{N-1}(x) \cdots R_\o (x) =: \begin{pmatrix}
        A_\o (x) & B_\o (x) \\ C_\o (x) & D_\o (x)
    \end{pmatrix},\qquad \o=0,1,\cdots, N-1,
\end{align}
where we denoted the entries by $A_\o (x),B_\o (x), C_\o (x), D_\o (x) \in \text{End}\left( \otimes_{\o'=\o} ^{N-1} \CalH_{\o'} \right)$. In particular, we recover the monodromy matrix at $\o = 0$, $ T_0(x) =  T (x) \vert_{\CalH}$. It will turn out that the variables $\bx_{\o+1} = (x_{\o+1,i})_{i=\o+1} ^{N-1}$ are provided by the zeros of $C_\o (x)$ for each $\o = 0,1,\cdots, N-2$ ($C_{N-1} (x)$ does not depend on $x$ by construction, so that there is no zero).

In the convention of the section 5.2 of \cite{Jeong:2024hwf}, we undo the transformation of $g_\o (x)$ and consider
\begin{align}
    g_N (x) ^{-1} \left( T(x) \vert_{\otimes_{\o'=\o} ^{N-1} \CalH_{\o'}}\right) g_\o (x) = L_{N-1} (x) L_{N-2}(x) \cdots L_\o (x) =: \begin{pmatrix}
        \a(x) & \b(x) \\ \g (x) & \d (x)
    \end{pmatrix},
\end{align}
since $C_\o(x)$ and $\g(x)$ are related by a simple relation:
\begin{align}
\begin{split}
     T(x) \vert_{\otimes_{\o'=\o} ^{N-1} \CalH_{\o'}}
    & = g_N (x) {L}_{N-1}(x) \cdots {L}_{\o}(x) g_\o (x) ^{-1} \\
    & =  
    \frac{1}{u_0} \begin{pmatrix}
        1 & b_N (x) \\ 0 & a_N (x)
    \end{pmatrix} 
    \begin{pmatrix}
        \a_\o (x) & \b_\o (x) \\ \g_\o (x) & \d_\o (x)
    \end{pmatrix}
    \begin{pmatrix}
        a_\o (x)& - b_\o (x) \\ 0 & 1
    \end{pmatrix} \frac{u_\o}{a_\o} \\
    & = \begin{pmatrix}
        * & * \\ \frac{a_N (x)}{u_0} \g_\o(x) u_\o & *
    \end{pmatrix},
\end{split}
\end{align}
namely, $C_\o(x) = a_N (x) u_0 ^{-1} \g_\o (x) u_\o = \frac{1}{P_0 ^+ (x+\ve_1)} u_0 ^{-1} \g_\o (x) u_\o $, where we denoted $g(x) = \frac{1}{u_{\overline{\o}}} \begin{pmatrix}
    1 & b_\o (x) \\ 0 & a_\o (x)
\end{pmatrix}$. An important feature of the fractional TQ equation written in the matrix form (cf. (5.11) in \cite{Jeong:2024hwf}) is that the lower-right entry of $L_\o (x)$ is zero, yielding
\begin{align}
\begin{split}
    \begin{pmatrix}
        \a_{\o}(x) & \b_{\o}(x) \\ \g_{\o}(x) & \d_{n}(x)
    \end{pmatrix} 
    & = 
    \begin{pmatrix}
        \a_{\o+1}(x) & \b_{\o+1}(x) \\ \g_{\o+1}(x) & \d_{\o+1}(x)
    \end{pmatrix} 
    \begin{pmatrix}
        T_{\o}(x) & -\kq_\o P^-_\o(x) \\ P^+_{\o+1}(x) & 0
    \end{pmatrix} \\
    & = \begin{pmatrix}
        * & * \\ 
        \g_{\o+1}(x) T_\o(x) + P^+_{\o+1}(x) \d_{\o+1}(x) & - \g_{\o+1}(x) \kq_\o P^-_\o(x)
    \end{pmatrix},
\end{split}, 
\end{align}
which implies that $\g_\o (x)$ satisfies a second-order recursion relation,
\begin{align} \label{eq:recursion}
    \g_{\o-1}(x) = \g_\o(x) T_{\o-1}(x) -  \g_{\o+1}(x) \kq_\o P_\o(x).
\end{align}
Thus, having $\g_\o (x) = (x- m^+ _0 + \ve_1) u_0 C_{\o} (x) u_\o ^{-1} $, we get a recursive formula for the degree $N-\o-1$ polynomial $C_\o (x)$ as
\begin{align} \label{eq:recursionx}
\begin{split}
    C_{\o-1} (x) + C_{\o+1}(x) \frac{u_{\o-1}}{u_{\o}} P_{\o} (x) &= C_\o (x) u_\o ^{-1} T_{\o-1} (x) u_{\o-1}  \\
    &=: C_\o (x) \left(\left(1+ \frac{u_{\o-1}}{u_\o}  \right)x - W_\o \right),
\end{split}
\end{align}
where the differential operator $W_\o$ is determined from
\begin{align}
T_\o(x) = (1+\kq_\o)x + u_{\o+1}(\p_{u_{\o+1}}-\p_{u_\o}) - m_{\o+1}^+ - \kq_\o m_\o^-
\end{align}
as
\begin{align}
\begin{split}
    W_\o &= u_\o ^{-1} \left(-\ve_1 u_{\o} (\p_{u_\o} - \p_{u_{\o-1}}) + m^+ _\o + \qe_{\o-1}  m^- _{\o-1} \right) u_{\o-1} \\
    &= -\ve_1 \left( \qe_{\o+1} \p_{\qe_{\o+1}} - \qe_{\o} \p_{\qe_{\o}}  -\qe_\o ( \qe_{\o} \p_{\qe_{\o}} - \qe_{\o-1} \p_{\qe_{\o-1}} )   \right) + m^+ _\o \frac{u_{\o-1}}{u_{\o}} + m^- _{\o-1} +\ve_1.
\end{split}
\end{align}

To show that the zeros of $C_\o (x)$ give the variables $\bx_{\o+1} = (x_{\o+1,i})_{i=\o+1} ^{N-1}$ in the separated expression for the eigenstate, let us act the recursion formula \eqref{eq:recursionx} for the operators $C_\o (x)$ to the integrand of the separated expression \eqref{eq:SoVvac}. The differential operator $W_\o$ acts in a simple way, since the all the $\bu$-dependent terms in the integrand are just elementary functions in $\bu$. Specifically, the $\bu$-dependent terms in the integrand are gathered into
\begin{align}
\begin{split}
    \mathcal{K}_\bu &= \mathbf{c}\times \prod_{\o=0} ^{N-2} \qe_\o ^{\sum_{i=\o+1}^{N-1} \left(-1+ \frac{a_i - x_{\o+1,i}}{\ve_1} \right) } \times \prod_{\o=0} ^{N-1} u_\o ^{\frac{m^+ _\o -a_\o}{\ve_1}} ,
\end{split} 
\end{align}
where the first term is from \eqref{eq:cconst}, the second term is from $\mathbf{V}' (\bx_1)$ \eqref{eq:vp} and $\m'(\bx_1)$ \eqref{eq:mp}, and the third term is the classical part \eqref{eq:classcal} of the regular monodromy defect observable. Thus, the differential operator $W_\o$ is converted into a number upon acting on the separated expression as
\begin{align}
\begin{split}
    w_\o &:= \mathcal{K}_\bu ^{-1} W_\o \mathcal{K}_\bu \\
    & = \qe_\o (m^+_\o + m^- _\o + \ve_1) + \sum_{i=\o+2} ^{N-1} x_{\o+2,i} - \sum_{i=\o+1} ^{N-1} x_{\o+1,i} -\qe_\o \left( \sum_{i=\o+1} ^{N-1} x_{\o+1,i} - \sum_{i=\o} ^{N-1} x_{\o,i} \right).
\end{split}
\end{align}
After acting on the separated expression, the recursion formula \eqref{eq:recursionx} reads
\begin{align}
     C_{\o-1} (x) + C_{\o+1}(x) \frac{u_{\o-1}}{u_{\o}} P_{\o} (x) = C_\o (x) \left(\left(1+ \frac{u_{\o-1}}{u_\o}  \right)x -w_\o \right).
\end{align}
It can be solved by a straightforward computation, starting from $\o = N-2$ all the way to $\o=0$. The solution is simply
\begin{align}
    C_\o (x) = \prod_{i=\o+1} ^{N-1} (x- x_{\o+1,i}),\qquad \o=0,1,\cdots, N-2.
\end{align}
Therefore, we prove that the zeros of $C_\o (x)$ provide the variables $\bx_{\o+1} = (x_{\o+1,i})_{i=\o+1} ^{N-1}$ in the separated expression \eqref{eq:SoVsc} for the XXX spin chain eigenstate. In particular, the zeros of $C(x) = C_0 (x)$ give the separated variables $\bx_1 = (x_{1,i})_{i=1} ^{N-1}$. This confirms the separation of variables that we established from the surface defect transition recovers the Sklyanin's prescription.

\subsection{Separation of variables for Gaudin model}
In section \ref{sec:Qopssrf}, we discussed that the $\mathbf{Q}$-observable and the $\mathbf{H}$-observable arise from the same two-dimensional gauged linear sigma model coupled to the four-dimensional gauge theory. The former has non-zero vev of the complex adjoint scalar, while the latter has non-zero the complexified FI parameter. In section \ref{sec:hbarlang}, we elucidated that the Fourier transformation between the two observables implements a transition between the $Q$-operator and the Hecke operator.

In the separated expression \eqref{eq:SoVsc} for the spin chain eigenstate, the integrand contains a product of $N-1$ normalized vacuum expectation values $Q(\ba;x_{1,i})$ of the $\mathbf{Q}$-observables. The separated variables $\bx_1 = (\bx_{1,i})_{i=1} ^{N-1}$, which obviously act diagonally on $Q(\ba;x_{1,i})$, are precisely the Coulomb parameters of the $\mathbf{Q}$-observables. Noting that the transition between the $\mathbf{Q}$-observable and the $\mathbf{H}$-observable manifests as a integral transformation for their observable expressions \eqref{eq:transition}, we expect to have another separated expression for the eigenstate where the integrand is given by a product of $N-1$ normalized vacuum expectation values of the $\mathbf{H}$-observable instead.

Each of the normalized vacuum expectation values of the $\mathbf{H}$-observable in the product is separately a solution to the same oper differential equation, where the oper specifies a twisted $\ED$-module on $\text{Bun}_{PGL(N)} (\BP^1;S)$, whose section is the normalized vacuum expectation value $\psi(\ba;\bu)$ of the regular monodromy surface defect, representing the spectral equations for the Gaudin model. Thus, the resulting integral transformation for the twisted coinvariants precisely gives a separation of variables for the Gaudin model, where the separated variables are the complexified FI parameters of the surface defect theory for the $\mathbf{H}$-observables. In this way, we observe that the separation of variables for the XXX spin chain and the Gaudin model are connected to each other by the surface defect transition between the $\mathbf{Q}$-observable and the $\mathbf{H}$-observable. This is indeed in accordance with the bispectral duality between the two models that we established in section \ref{subsubsec:bispectral}.

For this, let us revisit the monodromy defect observable in the negative chamber \eqref{eq:psimiddle} expressed as a summation over non-negative integers. Instead of turning the summation into a contour integral of Barnes type as we did for the separation of variables for the XXX spin chain, we can actually perform the summation to transform the $\mathbf{Q}$-observable into the $\mathbf{H}$-observable, after properly expressing some combinations of $\G$-functions as contour integrals of Pochhammer type. To be explicit, we will exemplify the case of $N=2$ and $3$ first, and then present the prescription for general $N$.

\subsubsection{$\fsl(2)$ case}
We have already studied the regular monodromy defect at $N=2$ in section \ref{eq:z2transition}, where its transition to a $\mathbf{H}$-observable was shown by an integral transformation between their observable expressions. For completeness, let us restate here the integral transformation:
\begin{align} \label{eq:sovsl2}
\begin{split}
    \Big\langle \Psi(\qe_0) \Big\rangle_\ba  = \int_{\text{PC}} dy\, \EuScript{K}(\qe_0;y) \Big\langle \mathbf{H}^{(1)} (y) \Big\rangle_\ba,
\end{split}
\end{align}
where the integral kernel is 
\begin{align}
\begin{split}
    \EuScript{K}(\qe_0;y) &= \qe_0 ^{1 + \frac{a_1- m^+ _1}{\ve_1}  } (1+\qe_0) ^{1+ \frac{1}{\ve_1}  (m_{0} ^+ + m_0 ^- -a_0 - a_1) } \\
    & \quad \times y^{-\frac{1}{2} + \frac{m_0 ^-- m^- _1}{2\ve_1}} (y-\qe_0) ^{-1 + \frac{m_1 ^+ - m^- _0}{\ve_1}} (y-\qe)^{\frac{\bar{m}^- -\bar{a}}{\ve_1}} (y-1)^{-1 -\frac{\bar{m}^+ - \bar{a}}{\ve_1}}.
\end{split}
\end{align}
We remind that $\text{PC}$ is the Pochhammer contour enclosing $y=0$ and $y=\qe_0$. The integral transformation is valid for any value of $\ve_1,\ve_2$. In particular, $\ve_2$ is not involved in the kernel $\EuScript{K}(\qe_0;y)$, so that the limit $\ve_2\to 0$ can be safely taken without harming the validity of the transformation with the same kernel. The limit $\ve_2 \to 0$ of the formula gives
\begin{align}
    \psi(\ba; \qe_0 ) =   \int_{\text{PC}} dy\, \EuScript{K}(\qe_0;y) \chi_1 (\ba;y),
\end{align}
where $\psi(\ba;\qe_0)$ is the normalized vacuum expectation value of the regular monodromy surface defect (at $N=2$) and $\chi_1 (\ba;y)$ is the normalized vacuum expectation value of the $\mathbf{H}$-observable, in the limit $\ve_2 \to 0$.

Recall that $\psi(\ba;\qe_0)$ is a section of twisted $\ED$-module on $\text{Bun}_{PGL(2)} (\BP^1;S)$ corresponding to the oper $\r_\ba$, namely, an eigenfunction of the Gaudin Hamiltonian with the eigenvalue fixed by the chosen oper.\footnote{There is only one Gaudin Hamiltonian for this case of $N=2$.} Also note that $\chi_1 (\ba;y)$ is a solution to the second-order oper differential equation for $\r_\ba$. Therefore, the integral transformation is precisely the separation of variable transformation for the Gaudin model. We also confirmed our construction recovers Sklyanin's prescription of the separation of variables by using the Gaudin Lax matrix expressed in the $\EN=2$ gauge theoretical terms, obtained in \cite{Jeong:2023qdr}. Namely, a straightforward computation shows that the zero of the lower-left entry of the $2 \times 2$ Lax matrix $L(z)$, when acting on the separation of variable kernel $\EuScript{K}(\qe_0;y)$, is precisely at $z=y$.

Calling it separation of variables is somewhat misleading in this particular case $N=2$, since the Gaudin model is a single-body spectral problem even before the transformation. Nevertheless, it clarifies the distinction between the eigenfunction $\psi(\ba;\qe_0)$ of the Gaudin Hamiltonian and the oper solution $\chi(\ba;y)$, where the two relevant second-order differential equations are almost identical except a shuffling of monodromy parameters. Such a coincidence will not persist in the separation of variables for higher-rank Gaudin models that we discuss now.

\subsubsection{$\fsl(3)$ case}
The regular monodromy defect observable in the negative chamber is written as a summation over $\frac{3 \times 2}{2} =3$ non-negative integers $(s_{1,1},s_{2,1},s_{2,2})$,
\begin{align} \label{eq:psi3}
\begin{split}
    \Psi[\bl]_{\z<0} &=\sum_{s_{1,1},s_{2,1},s_{2,2} =0} ^\infty \qe_0 ^{s_{1,1}+ s_{2,1}} \qe_1 ^{s_{2,1}+ s_{2,2}} \\
    & \quad \times \BE \left[ \frac{S \S_0 ^* -\S_0 \S_0 ^* - \S_1 \S_1 ^* + \S_0 \S_1 ^* - M_0^+ \S_0 ^* + q_1 ^{-1} (M_0 ^- )^* \S_0 - M_1 ^+ \S_1 ^* + q_1 ^{-1} (M_1 ^- )^* \S_1 }{P_1 ^*} \right],
\end{split}
\end{align}
where
\begin{align} \label{eq:sig3}
\begin{split}
    &\S_0 = e^{A_1 - s_{1,1} \ve_1} + e^{A_2 - s_{2,1} \ve_1} \\
    &\S_1 = e^{A_2 - (s_{2,1} + s_{2,2}) \ve_1}.
\end{split}
\end{align}
By substituting \eqref{eq:sig3} into \eqref{eq:psi3}, we get
\begin{align}
\begin{split}
    \Psi[\bl]_{\z<0} &= \sum_{s_{1,1},s_{2,1},s_{2,2} = 0} ^\infty \qe_0 ^{s_{1,1} + s_{2,1}} \qe_1 ^{s_{2,1} + s_{2,2}} \frac{  \G\left( s_{1,1} -s_{2,1} - s_{2,2} + \frac{A_2 - A_1}{\ve_1} \right) }{  \G\left( s_{2,1}-s_{1,1} + \frac{A_1 -A_2}{\ve_1} \right) \G\left( -s_{2,1}+s_{1,1} + \frac{A_2-A_1}{\ve_1} \right) } \\
    & \times \frac{(-\ve_1) ^{ \frac{m_1 ^- - m_1 ^+ + 2m_0 ^- - 2 m_2 ^+}{\ve_1} } (-1)^{s_{2,2}}}{s_{2,2}!} \times \frac{\prod_{\a=1 } ^2 \prod_{i=1} ^2  \G\left( -s_{i,1} + \frac{A_i - m^+ _\a}{\ve_1} \right) }{ \prod_{i=1} ^2  \G \left( -s_{i,1} + \frac{A_i - m_0 ^-}{\ve_1} \right) \prod_{\pm}  \G\left( -s_{2,1}-s_{2,2} + \frac{A_2-m_1 ^\pm}{\ve_1} \right) } \\
    &\times Q(A_1 - (s_{1,1}+1)\ve_1) Q(A_2 - (s_{2,1}+1)\ve_1)  \prod_{i=1,2} \BE \left[\frac{ e^{A_i - (s_{i,1}+1)\ve_1 } (M^+)^* }{P_1 ^*} \right].
\end{split}
\end{align}
The summations can be converted into contour integrals as we now explain. First, note that the summation over $s_{2,2}$ produces a hypergeometric function. We use the Euler integral representation of the hypergeometric function
\begin{align} \label{eq:euler}
\begin{split}
       {}_2 F _1 (a,b;c;z) = \frac{e^{-c\pi i } \G(c) \G(1-b) \G(1+b-c) }{4\pi^2} \int_{\text{PC}} dt \, t^{b-1} (1-t)^{c-b-1} (1-zt) ^{a},
\end{split}
\end{align}
where $\text{PC}$ denotes the Pochhammer contour enclosing the singularities at $t=0$ and $t=1$. Then, we can express the ratios of remaining $\G$-functions having $s_{1,1}$ or $s_{2,1}$ in their arguments as contour integrals using \eqref{eq:betaint}. In this way, all the $s_{1,1}$ and $s_{2,1}$ only appear in the exponents of simple rational functions of integration variables and in the arguments of the two $\mathbf{Q}$-observables. Finally, we change the order of summation and integration. Each of $s_{1,1}$ and $s_{2,1}$ summations yields a $\mathbf{H}$-observable, which is analytically continued from the original domain if needed. The net result is a four-dimensional contour integral given by
\begin{align}
\begin{split}
    \Psi(\bu)[\bl]_{\z<0} &=  \qe_0 ^{-1 + \frac{A_1 - \bar{a}}{\ve_1}} (\qe_0 \qe_1 )^{-1 + \frac{A_2 -\bar{a}}{\ve_1}} \\
& \times  \int_{\text{PC}^{\times 4}} \prod_{i=1} ^4 dt_i \, t_1 ^{-1 + \frac{m_1 ^- +  \bar{m}^-}{\ve_1}} (1-t_1)^{-\frac{m_1 ^-}{\ve_1}} (1- \qe_1 t_1) ^{\frac{m_1 ^+ + \bar{m}^-}{\ve_1}} t_2 ^{-\frac{m_1 ^+ + \bar{m}^-}{\ve_1}} (1-t_2 )^{-1 + \frac{m_1 ^+ - m_1 ^- - \bar{m}^-}{\ve_1}} \\
&\quad\quad\times  t_3 ^{- \frac{m_2 ^+ + \bar{m}^-}{\ve_1}} (1-t_3 )^{-2 + \frac{m_2 ^+ - m_0 ^- }{\ve_1}} t_4 ^{- \frac{m_2 ^+ + \bar{m}^-}{\ve_1}} (1-t_4)^{-2 + \frac{m_2 ^+ - m_0 ^-}{\ve_1}} (t_3-t_4) \\
& \qquad \times \left( (1-t_1)t_2 t_4 -\qe_1 \qe_2 \right)^{\frac{\bar{m}^- -\bar{a}}{\ve_1}} \left(\frac{\qe_0 \qe_1 t_3}{t_1 (1-\qe_1 t_1)} -1 \right)^{-1 -\frac{\bar{m}^+ - \bar{a}}{\ve_1}}  \\
& \qquad \times \left( \frac{t_3}{t_1 (1-\qe_1t_1)} +  \qe_2 \right)^{\frac{\bar{m}^- - \bar{a}}{\ve_1}} \left( \frac{\qe_0 \qe_1 t_3}{t_1 (1-\qe_1 t_1)} +1 \right)^{-1 -\frac{\bar{m}^+ - \bar{a}}{\ve_1}} \\
&\quad\quad  \times \mathbf{H}^{(1)} \left( \qe_0 (1-t_1) t_2 t_4 \right)[\bl] \, \mathbf{H}^{(2)} \left( -\frac{\qe_0 \qe_1 
 t_3 }{t_1 (1- \qe_1 t_1)} \right)[\bl],
\end{split}
\end{align}
where we absorbed irrelevant prefactor into the observable expression and omitted it. Now, applying the crossing formula \eqref{eq:Relation of chi's} we recover the original monodromy surface defect observable,
\begin{align} \label{eq:sl3gsov1}
\begin{split}
    \Psi[\bl] &= \qe_0 ^{-l (\l ^{(1)}) } (\qe_0\qe_1) ^{- l (\l ^{(2)})} c_0 ^{R_\infty ^{(0)} + \frac{1}{\ve_1} \sum_{\a=0}^2 (A_\a -a_\a) }  c_1 ^{R_\infty ^{(1)} } \times \Psi \left( \qe_\o \frac{c_{\o+1}}{c_{\o-1}} \right) [\bl]_{\z<0} \\
    & =  \qe_0 ^{-1 + \frac{a_1 - \bar{a}}{\ve_1}} (\qe_0 \qe_1 )^{-1 + \frac{a_2 -\bar{a}}{\ve_1}} (1+\qe_0 + \qe_0 \qe_1)^{2+ \frac{m_0 ^+ + m_0 ^- 2 \bar{a}}{\ve_1}} (1+\qe_1)^{-1 + \frac{m_0 ^+ + m_0 ^- -2\bar{a}}{\ve_1}} \\
    & \quad \times \int_{\text{PC}^{\times 4}} \prod_{i=1} ^4 dt_i \, t_1 ^{-1 + \frac{m_1 ^- +  \bar{m}^-}{\ve_1}} (1-t_1)^{-\frac{m_1 ^-}{\ve_1}} \left(1- \frac{\qe_1 t_1}{1+ \qe_0 + \qe_0 \qe_1}\right) ^{\frac{m_1 ^+ + \bar{m}^-}{\ve_1}} \\
    &\quad\quad \times t_2 ^{-\frac{m_1 ^+ + \bar{m}^-}{\ve_1}} (1-t_2 )^{-1 + \frac{m_1 ^+ - m_1 ^- - \bar{m}^-}{\ve_1}} \\
&\quad\quad\times  t_3 ^{- \frac{m_2 ^+ + \bar{m}^-}{\ve_1}} (1-t_3 )^{-2 + \frac{m_2 ^+ - m_0 ^- }{\ve_1}} t_4 ^{- \frac{m_2 ^+ + \bar{m}^-}{\ve_1}} (1-t_4)^{-2 + \frac{m_2 ^+ - m_0 ^-}{\ve_1}} (t_3-t_4) \\
& \qquad \times \left( (1-t_1)t_2 t_4 - \frac{\qe_1 \qe_2}{(1+\qe_0+\qe_0 \qe_1)(1+\qe_1)} \right)^{\frac{\bar{m}^- -\bar{a}}{\ve_1}} \left(\frac{\qe_0 \qe_1 (1+\qe_1) t_3}{t_1  ( 1+\qe_0+\qe_0 \qe_1 -\qe_1 t_1)} -1 \right)^{-1 -\frac{\bar{m}^+ - \bar{a}}{\ve_1}}  \\
& \qquad \times \left( \frac{t_3}{t_1 \left(1- \frac{\qe_1t_1}{1+\qe_0 +\qe_0 \qe_1}  \right) } +  \frac{\qe_2}{1+\qe_1} \right)^{\frac{\bar{m}^- - \bar{a}}{\ve_1}} \left( \frac{\qe_0 \qe_1 (1+\qe_1) t_3}{t_1  (1+\qe_0 + \qe_0 \qe_1 -\qe_1 t_1)} +1 \right)^{-1 -\frac{\bar{m}^+ - \bar{a}}{\ve_1}} \\
&\quad\quad  \times \mathbf{H}^{(1)} \left( \qe_0 (1+\qe_1) (1-t_1) t_2 t_4 \right)[\bl] \, \mathbf{H}^{(2)} \left( -\frac{\qe_0 \qe_1 (1+\qe_1) 
 t_3 }{t_1 (1+ \qe_0 +\qe_0 \qe_1- \qe_1 t_1)} \right)[\bl],
\end{split}
\end{align}
expressed as an integral transformation of a product of two $\mathbf{H}$-observables. Note that the relation holds at the level of observable expressions at each $\bl$, and the $\bl$-dependence is solely at the $\mathbf{H}$-observables in the right hand side. Therefore, upon taking the vacuum expectation value, we arrive at
\begin{align}
    \Big\langle\Psi(\bu) \Big\rangle_\ba &=  \int_{\text{PC}} dy_1 dy_2\; \EuScript{K}(\bu;\by) \Big\langle \mathbf{H}^{(1)} (y_1) \mathbf{H}^{(2)} (y_2) \Big\rangle_\ba,
\end{align}
where we split the four-dimensional contour integral \eqref{eq:sl3gsov1} into two parts; the two-dimensional contour integral for $y_1 = \qe_0 (1+\qe_1)(1-t_1)t_2t_4$, $y_2 = -\frac{q_0 q_1 (1+q_1) t_3}{t_1 (1+\qe_0 + \qe_0 \qe_1 -\qe_1 t_1)}$ and the integral of the rest two. The integral of the rest and the elementary functions of $\qe_0$ in front are gathered into the separation of variable kernel $\EuScript{K}(\mathbf{u};\by)$. Note that the relation is valid for any value of $\ve_1, \ve_2$.

The vacuum expectation value $\langle \Psi(\bu) \rangle _\ba$ of the monodromy surface defect is identified with a twisted coinvariant of four $\widehat{\fsl}(3)$-modules (two (co-)Verma modules (at $0$ and $\infty$) and two bi-infinite modules (at $\qe$ and $1$)) \cite{Nekrasov:2021tik, Jeong:2021bbh}. Meanwhile, the correlation function of the two $\mathbf{H}$-observables $\langle \mathbf{H}^{(1)} (y_1) \mathbf{H}^{(2)} (y_2) \rangle_\ba$ gives a $\CalW_3$-algebra conformal block with two generic vertex operators (at $0$ and $\infty$), two semi-degenerate vertex operators (at $\qe$ and $1$), and two simplest fully degenerate vertex operators (at $y_1$ and $y_2$) \cite{Jeong:2018qpc, Jeong:2021bbh}. In this sense, we may interpret this integral transformation as a generalization of the KZ/BPZ correspondence \cite{Frenkel95,Stoyanovsky2000ARB,Ribault:2005wp} to a higher-rank.

Moreover, the kernel $\EuScript{K}(\bu;\by)$ does not involve $\ve_2$ so that the limit $\ve_2 \to 0$ can be taken to the integral transformation. In the integrand, the correlation function of two $\mathbf{H}$-observables in the integrand factorizes in the limit $\ve_2 \to 0$,
\begin{align}
    \lim_{\ve_2 \to 0} \Big\langle \mathbf{H}^{(1)} (y_1) \mathbf{H}^{(2)} (y_2)\Big\rangle_\ba = e^{\frac{\widetilde{\EuScript{W}}(\ba;\qe)}{\ve_2}} \chi_1 (\ba;y_1) \chi_2 (\ba;y_2),
\end{align}
since the two $\mathbf{H}$-observables are local on the topological $\BC_2$-plane, and thus can be arbitrarily separated from each other. Thus, the integral transformation reads
\begin{align} \label{eq:gsov3}
\begin{split}
    \psi(\ba;\bu) &=  \int_{\text{PC}} dy_1 dy_2\; \EuScript{K}(\bu;\by) \, \chi_1 (\ba;y_1) \chi_2 (\ba;y_2).
\end{split}
\end{align}

As briefly reviewed in section \ref{sec:hbarlang}, it was shown in \cite{Jeong:2023qdr} that the normalized vacuum expectation value $\psi(\ba;\bu)$ of the regular monodromy surface defect in the limit $\ve_2 \to 0$ is identified with a section of the twisted $\ED$-module corresponding to an oper $\r_\ba$. It is also a common eigenfunction of the Gaudin Hamiltonians with the spectra fixed by the chosen oper. Meanwhile, the integrand on the right hand side contains a product of two normalized vacuum expectation values $\chi(\ba;y)$ of the $\mathbf{H}$-observable in the limit $\ve_2 \to 0$, which are solutions to the third-order oper differential equation associated to $\r_\ba$. Therefore, the integral transformation converts the Gaudin spectral equations (a two-particle system) into two copies of identical third-order differential equations (two independent single-particle systems). Therefore, we conclude that \eqref{eq:gsov3} gives a separation of variable transformation for the $\fsl(3)$ Gaudin model.

\subsubsection{General $\fsl(N)$ case} \label{eq:subsubsec:slngaudinsov}
For general $\fsl(N)$ Gaudin model, we can achieve the integral transformation for the separation of variables by the same procedure. The only difference is that we have to deal with a higher multi-dimensional contour integral to sort out the kernel $\EuScript{K}(\bu;\by)$. Here, we will outline how this computation can actually be done.

Just as in the cases of $N=2$ and $3$, the separation of variable kernel $\EuScript{K}$ for general $N$ is obtained by simply rewriting the following expression in \eqref{eq:psimiddle},
\begin{align}
\begin{split}
 & \sum_{\mathbf{s}} \prod_{\o=0}^{N-2} \kq_{\o}^{\sum_{i=\o+1}^{N-1} \left( -1 + \frac{A_i-x_{\o,i}}{\ve_1}\right) } \\
    & \quad \times \BE \left[ -\frac{\Sigma_0q_1^{-1} (M^+)^*}{P_1^*} - \sum_{\o=0}^{N-2} \frac{ \Sigma_{\o}(\Sigma_\o-\Sigma_{\o+1})^* + {M}_\o^+ \Sigma_{\o}^* - q_1^{-1} (M^-_\o)^*\Sigma_\o -(N-1) }{P_1^*} \right].
\end{split}
\end{align}
The second line gives complicated ratios of $\G$-functions, given by
\begin{align}
\begin{split}
   &\prod_{i=1}^{N-1} \frac{ (-\ve_1)^{\frac{m_0^--x_{1,i} }{\ve_1}} \Gamma\left( \frac{ m_0^- - x_{1,i} }{\ve_1} \right)}{ (-\ve_1)^{1+\frac{x_{1,i}-m_0^+}{\ve_1}} \Gamma\left( 1+\frac{x_{1,i}-m_0^+}{\ve_1} \right)} 
    \prod_{\alpha=0}^{N-1} \frac{(-\ve_1)^{\frac{x_{1,i}-m_\alpha^+-\ve_1}{\ve_1}} }{\Gamma \left( - \frac{x_{1,i}-m_\alpha^+}{\ve_1} \right)} 
    \frac{1}{ \prod_{1\leq i \neq j \leq N-1} \Gamma\left( \frac{x_{1,i}-x_{1,j}}{\ve_1} \right) } \\
    & \quad \times \prod_{\o=0}^{N-2} \prod_{i=\o+1}^{N-1} \frac{\ve_1^{-s_{i,\o+2}}}{s_{i,\o+2}!} \frac{(-\ve_1)^{\frac{ m_\o^- -x_{\o+1,i} }{\ve_1}} \Gamma\left( \frac{ m_\o^- -x_{\o+1,i} }{\ve_1} \right)}{ (-\ve_1)^{1+\frac{x_{\o+1,i}-m_\o^+}{\ve_1}} \Gamma\left( 1+\frac{x_{\o+1,i}-m_\o^+}{\ve_1} \right)}
    \frac{\prod\limits_{\substack{\o+1\leq i \leq N-1 \\ \o+2 \leq j \neq i \leq N-1 } } (-\ve_1)^{\frac{x_{\o+2,j}-x_{\o+1,i}}{\ve_1}} \Gamma \left( \frac{x_{\o+2,j}-x_{\o+1,i}}{\ve_1} \right) }{\prod\limits_{\o+1\leq i\neq j\leq N-1} \Gamma\left( \frac{x_{\o+1,i} - x_{\o+1,j} }{\ve_1} \right) }.
\end{split}
\end{align}
Let us start with plugging in $x_{N-1,N-1} = x_{N-2,N-1} - s_{N-1,N-1}\ve_1$. Collecting all the terms involving $x_{N-1,N-1}$, the summation over $s_{N-1,N-1}$ gives a hypergeometric function. Then we use the Euler integral representation \eqref{eq:euler} to rewrite the hypergeometric function. Namely,
\begin{align}
\begin{split}
    & \sum_{s_{N-1,N-1}=0}^\infty \frac{(-\kq_{N-2})^{s_{N-1,N-1}} }{s_{N-1,N-1}!} 
    \frac{\Gamma\left( \frac{ m_{N-2}^- -x_{N-1,N-1} }{\ve_1} \right) \Gamma \left( \frac{x_{N-1,N-1}-x_{N-2,N-2}}{\ve_1} \right) }{ \Gamma\left( 1+\frac{x_{N-1,N-1}-m_{N-2}^+}{\ve_1} \right)} \\ 
    = & \frac{ \Gamma \left( \frac{m_{N-2}^--x_{N-2,N-1}}{\ve_1} \right) \Gamma \left( \frac{x_{N-2,N-1}-x_{N-2,N-2} }{\ve_1} \right) }{ \Gamma \left( 1 + \frac{x_{N-2,N-1}-m_{N-2}^+ }{\ve_1} \right) } \\
    & \times {}_2F_1 \left( \frac{m_{N-2}^\pm-x_{N-2,N-1}}{\ve_1} ; 1+\frac{x_{N-2,N-2}-x_{N-2,N-1} }{\ve_1} ; -\kq_{N-2} \right) \\
    = & \frac{e^ {-\pi\ri \left( 1+\frac{x_{N-2,N-2}-x_{N-2,N-1} }{\ve_1} \right) } }{4\pi} 
    \frac{ \Gamma \left( \frac{m_{N-2}^--x_{N-2,N-1}}{\ve_1} \right) \Gamma \left( \frac{m_{N-2}^+-x_{N-2,N-2}}{\ve_1} \right) }{\sin\pi \left( \frac{x_{N-2,N-1}-x_{N-2,N-2} }{\ve_1} \right)} \\
    & \times \int_\text{PC} dt_{N-1}\, t_{N-1} ^{\frac{m_{N-2}^+-x_{N-2,N-1}}{\ve_1}-1} (1-t_{N-1})^{\frac{m_{N-2}^+-x_{N-2,N-2}}{\ve_1} } (1+\kq_{N-2}t_{N-1})^{- \frac{m_{N-2}^--x_{N-2,N-2}}{\ve_1}} .
\end{split}
\end{align}
The sine function in the denominator can be combined with the cross terms in the next level,
\begin{align}
\begin{split}
    & \frac{\pi}{\sin\pi\left( \frac{x_{N-2,N-1}-x_{N-2,N-2} }{\ve_1} \right)} \frac{1}{\Gamma\left( \frac{x_{N-2,N-1}-x_{N-2,N-2} }{\ve_1} \right) \Gamma\left( \frac{x_{N-2,N-2}-x_{N-2,N-1} }{\ve_1} \right) } \\
    & = \frac{\Gamma\left( \frac{x_{N-2,N-2}-x_{N-2,N-1} }{\ve_1} +1 \right)}{\Gamma\left( \frac{x_{N-2,N-2}-x_{N-2,N-1} }{\ve_1} \right)}=  \frac{x_{N-2,N-2}-x_{N-2,N-1} }{\ve_1}.
\end{split}
\end{align}
We combine this term with relevant $\G$-functions in the next level using
\begin{align}\label{eq:useful-gamma-3}
    & \frac{\Gamma(a)\Gamma(b)\Gamma(c-a)\Gamma(d-b)}{\Gamma(c)\Gamma(d)} (c-d) \\
    & =  \int_{\text{PC}^{\times 2}}dt_1 dt_2 \, t_1^{a-1}(1-t_1)^{c-a-2} t_2^{b-1} (1-t_2)^{d-b-2} \left[ (c-a-1)(1-t_2) - (d-b-1)(1-t_1) \right]  \nonumber
\end{align}
We can apply this procedure repetitively, leaving only ratios of $\G$-functions which can be converted into Pochhammer contour integrals by \eqref{eq:betaint}. Then the summation over the non-negative integers $s_{\o,1}$, $\o=1,2,\cdots, N-1$ can be computed inside the integral, yielding a product of $N-1$ $\mathbf{H}$-observables. By leaving the $N-1$ Pochhammer contour integrals for the arguments of the $\mathbf{H}$-observables undone while performing the rest of the integrals, we obtain the separation of variable kernel $\EuScript{K}(\bu;\by)$ for the $\fsl(N)$ Gaudin model. See appendix \ref{sec:sovkernel} for more detail.

\section{Discussion} \label{sec:discussion}
We explained that the $Q$-operator and the Hecke operator originate from the same two-dimensional $\EN=(2,2)$ sigma model coupled to the four-dimensional gauge theory, but in two different parameter regimes. In the former, the vev of the complex adjoint scalar is turned on, while in the later, the complexified FI parameter is turned on. The transition between the two can be initiated by specializing the adjoint scalar vev (resp. the complexified FI parameter) and turning on the complexified FI parameter (resp. the adjoint scalar vev). Such a transition of the surface defect was interpreted as a Fourier transformation between the $Q$-operator and the Hecke operator, which then led to the bispectral duality between the XXX spin chain and the Gaudin model associated with the same $\EN=2$ gauge theory.

Moreover, we elucidated that there is a dual two-dimensional sigma model description of the monodromy surface defect. Using this description, its observable expression was computed in a different stability chamber, which describes the transition into multiple $(N-1)$ $\mathbf{Q}$-observables or $\mathbf{H}$-observables. The transition of the surface observables can be expressed as two different integral transformations, which were shown to yield the separation of variables for both the XXX spin chain and the Gaudin model, respectively.

\paragraph{Separation of variables for $\fgl(n)$ XXX spin chain and Gaudin model with $n+2$ sites}
A direct generalization of our formulation would involve the higher-rank linear $A_{n-1}$-quiver gauge theory, with $n>2$. We expect that our prescription could be further generalized to the XXX spin chain associated with $Y(\fgl(n))$, steeming from the surface defect transition of the monodromy surface defect to $(n-1)(N-1)$ $\mathbf{Q}$-observables. It would be interesting to connect our construction to more recent studies (see \cite{Derkachov:2018ewi, Ryan:2020rfk} for instance) of the quantum separation of variables for XXX spin chain, and thereby to generalize the prescription to bi-infinite modules. In the bispectral dual perspective, we expect that our prescription would generalize the quantum separation of variables for the $\fsl(N)$ Gaudin model to the one defined on $n+2$ sites. Due to technicalities, we postpone this to future work.

\paragraph{Quantum analytic Langlands correspondence and KZ/BPZ correspondence}
In a recent study \cite{Gaiotto:2024tpl}, it was suggested that the separation of variables for the common eigenfunction of the quantum Hitchin Hamiltonians provides a unitary map between properly defined $L^2$-spaces, which is well-suited for the study of analytic Langlands correspondence \cite{Etingof:2019pni,Etingof:2021eub, Etingof:2021eeu}. Further, in their study of quantum deformation of the analytic Langlands correspondence \cite{Gaiotto:2024tpl}, which in our setting corresponds to the deviation $\ve_2 \neq 0$ along a certain line, the twisted $\widehat{\fsl}(2)$-coinvariants at $\k \neq 0$ form a Hilbert space of states, on which the KZ/BPZ correspondence is applied to analyze the wavefunction by using Liouville correlation functions with degenerate fields. Our integral transformations for the vev of the regular monodromy surface defect (i.e., twisted $\widehat{\fsl}(N)$-coinvariants at $\k = - \frac{\ve_2}{\ve_1} \neq 0$) provide two generalizations of the KZ/BPZ correspondence: the $\hbar$-deformation and the higher $N$ generalization. It would be interesting to apply these generalized integral transformations to investigate the (quantum) analytic ($\hbar$-)Langlands correspondence in our $\EN=2$ theory setup. This would require placing the $\EN=2$ theory on $\BP^1 \times \BR^2$ with two (as opposed to one) fixed points of the isometry. See also \cite{Gaiotto:2021tsq} for the GL-twisted $\EN=4$ theory formulation of the analytic Langlands correspondence.

\paragraph{Separation of variables and 5d Chern-Simons theory}
In the twisted M-theory setup \cite{Costello:2016nkh}, the monodromy surface defect in the $\EN=2$ theory corresponds to having an $A_{N-1}$-singularity in the topological part of the worldvolume, which can be deformed to the $N$-centered Taub-NUT space. Reducing along the Taub-NUT circle and localizing, we arrive at the 5d non-commutative $\fgl(N)$ Chern-Simons theory \cite{Costello:2016nkh} with intersecting holomorphic surface defects. Our integral transformation suggests that this is dual to the 5d non-commutative $\fgl(1)$ Chern-Simons theory with intersecting holomorphic surface defects and additional line defects. It would be interesting to re-examine our integral transformations from this perspective. 

\appendix

\section{Gauge origami constructions} \label{sec:ori}

The partition function of 4d $A_1$-quiver $U(N)$ gauge theory, the $\mathbf{Q}$-observables, and the monodromy surface defect can be constructed using the \emph{gauge origami} \cite{Nikita:III}, which is a configuration of intersecting D3-branes on $X \times \BC$ in the IIB theory, where $X$ is a Calabi-Yau four-fold. Here, we briefly recall this construction.

\subsection{$A_1$-quiver gauge theory}
We consider the gauge origami defined on $X = \BC^2_{12} \times \BC^2_{34}/\BZ_3$, where the subscripts enumerates the four $\BC$-planes. In particular, we take $3N$ D3-branes supported on $\BC^2 _{12}$, represented by the character of their Chan-Paton bundle by
\begin{align}\label{def:GO-setup}
    N_{12} = \sum_{\alpha=1}^N e^{a_\alpha} \cdot \CalR_0 + \sum_{\alpha=1}^N e^{m_\alpha^- - \ve_4 } \cdot \CalR_1 + \sum_{\alpha=1}^{N} e^{m_\alpha^+-\ve_3} \cdot \CalR_2,
\end{align}
where $\CalR_i$ denotes the $i$-th representation of $\BZ_3$ assigned to the D3-branes. Note that there are $N$ D3-branes for each representation $\CalR_i$, $i=0,1,2$. The spiked instanton partition function reads 
\begin{align}
\begin{split}
    \CalZ_S & = \sum_{{\boldsymbol\lambda}} \prod_{i=0,1,2} \kq_i^{|{\boldsymbol\lambda_i}|} \BE \left[ -\frac{P_3S_{12}S_{12}^*}{P_{12}^*} \right]^{\BZ_3} \\
    & = \sum_{{\boldsymbol\lambda}} \prod_{i=0,1,2} \kq_i^{|{\boldsymbol\lambda_i}|} 
    \BE \left[ \frac{-S_0S_0^*-S_1S_1^*-S_2S_2^*+ S_0S_1^* + q_3^3 S_1S_2^* + q_{12}^{-1}S_2S_0^*}{P_{12}^*} \right]
\end{split}
\end{align}
where we used the notation
\begin{align}
\begin{split}
    &q_a = e^{\ve_a}, \quad P_a = 1-q_a, \\
    &P_{ab} = P_a P_b,
\end{split} \qquad a,b=1,2,3,4,
\end{align}
and defined the universal sheaf $S_i = N_i - P_{12}K_i$ for the instanton associated to the $i$-th node, $i=0,1,2$, where $K_i$ is the vector space encoding the fixed points of the moduli space of instantons for each stack of $N$ D3-branes. For instance,
\begin{align}
    K_0 = \sum_{\Box_{(i,j)} ^{(\a)} \in \bl_0} e^{a_\a + (i-1)\ve_1 + (j-1)\ve_2},
\end{align}
with each $\bl_i$, $i=0,1,2$, being an $N$-tuple of Young diagrams. $\qe_i$, $i=0,1,2$ are gauge couplings for each stack. We used the symbol $\BE$ to convert characters into products of equivariant weights,
\begin{align}
    \BE\left[ \sum_i e^{x_i} - \sum_{j} e^{y_j} \right] = \frac{\prod_j y_j}{\prod_i x_i}.
\end{align}
This is precisely the partition function of the $\hat{A}_2$-quiver $U(N)$ gauge theory. 

We may now turn off the gauge coupling $\kq_1=\kq_2=0$ while keeping $\kq_0=\kq$. 
This is known as the \emph{freezing limit}.
Some of the universal bundle becomes the flavor bundle $S_1=M^+$, $S_2=M^-$, $S_0 = S$, so that the spike instanton partition function becomes the partition function of $U(N)$ gauge theory with $N$ fundamental and $N$ anti-fundamental hypermultiplets: 
\begin{align}
\begin{split}
    \CalZ & = \sum_{\boldsymbol\lambda} \kq^{|{\boldsymbol\lambda}|} \BE \left[ \frac{-SS^* + M^+S^* + q_{12}^{-1}S(M^-)^* }{P_{12}^*} \right] \\\end{split}
\end{align}
with $M=M^++M^-$. Here, $S$ is the universal sheaf for the instantons, and $M^\pm$ are the flavor bundles. We sometimes use the notation
\begin{align}
    \bar{a} = \frac{1}{N} \sum_{\a=0} ^{N-1} a_\a,\qquad \bar{m} ^\pm = \frac{1}{N}\sum_{\a=0} ^{N-1} m^\pm _\a,
\end{align}
to denote the average of the Coulomb parameters and the masses.

\subsection{$\mathbf{Q}$-observable} \label{subsec:qobsapp}
We consider the gauge origami set up:
\begin{align}
\begin{split}
    N_{12} & = \sum_{\alpha=1}^N e^{a_\alpha} \cdot \CalR_0 + \sum_{\alpha=1}^N e^{m_\alpha^- - \ve_4 } \cdot \CalR_1 + \sum_{\alpha=1}^{N} e^{m_\alpha^+ -\ve_3} \cdot \CalR_2 \\
    N_{14} & = e^{x+\ve_1+\ve_4} \cdot \CalR_2,
\end{split}
\end{align}
by adding one more D3-brane. The spike instanton partition function is 
\begin{align}
    \CalZ = \sum_{{\boldsymbol\lambda}} \prod_{i=0,1,2} \kq_i^{|{\boldsymbol\lambda}_{12,i}|+|{\boldsymbol\lambda}_{14,i}|} 
    \BE\left[ - \frac{P_3S_{12}S_{12}^*}{P_{12}^*} - \frac{P_2S_{13}S_{13}^*}{P_{13}^*} + q_2 P_3 \frac{S_{14}S_{12}^*}{P_1^*} \right]^{\BZ_3}
\end{align}
We take the freezing limit $\kq_1=\kq_2=0$, $\kq_0 = \kq$ which yields the $A_1$-quiver: 
\begin{align}
\begin{split}
    \CalZ = & \sum_{{\boldsymbol\lambda}} \kq^{|{\boldsymbol\lambda}|} \BE \left[ - \frac{SS^* - M^+S^* - q_{12}^*S(M^-)^*}{P_{12}^*} \right] \BE \left[ -\frac{e^{x}(S-M^+)^*}{P_1^*}  \right] \\
    & = \sum_{{\boldsymbol\lambda}} \CalZ[{\boldsymbol\lambda}] \EQ(x)[\boldsymbol\lambda]
\end{split}
\end{align}
The $\mathbf{Q}$-observable associated to an instanton configuration $\boldsymbol{\lambda}$ is given by
\begin{align} \label{eq:qobsappp}
    \EQ(x)[{\boldsymbol\lambda}] =  \BE \left[ \frac{e^{x}(M^+-S[\bl])^*}{P_1^*}  \right].
\end{align}
We often separate the 1-loop part with the $\G$-functions,
\begin{align}
    M^+ (x) = \BE\left[ -\frac{e^x M^+}{P_1 ^*} \right] = \prod_{\a=0} ^{N-1} \frac{(-\ve_1)^{\frac{x-m^+ _\a}{\ve_1}} }{ \G\left(-\frac{x-m^+ _\a}{\ve_1} \right)},
\end{align}
and write
\begin{align}
\begin{split}
    \mathbf{Q}(x) &= \BE \left[ - \frac{e^x S[\bl] ^*}{P_1 ^*} \right] = \prod_{{\alpha} = 0}^{N-1} \left( \frac{{\ve}_{1}^{\frac{x-a_{\alpha}}{\ve_1}}}{{\Gamma} \left( - \frac{x-a_{\alpha}}{{\ve}_1} \right)} 
\prod_{i=1}^{\infty} \frac{x - a_{\alpha} - (i-1) {\ve}_{1} -  {\lambda}^{({\alpha})}_{i} {\ve}_{2}}{x - a_{\alpha} - (i-1) {\ve}_{1}} \right) \\
& = \prod_{{\alpha} = 0}^{N-1} \left( \frac{{\ve}_{1}^{\frac{x-a_{\alpha}}{\ve_1}}}{{\Gamma} \left( l( \l^{(\a)}  ) - \frac{x-a_{\alpha}}{{\ve}_1} \right)} 
\prod_{i=1}^{ l(\l^{(\a)}) } \frac{x - a_{\alpha} - (i-1) {\ve}_{1} -  {\lambda}^{({\alpha})}_{i} {\ve}_{2}}{\ve_1} \right) .
\end{split}
\end{align}
Namely, $\EQ(x) = \frac{\mathbf{Q}(x)}{M^+ (x)}$. Somewhat loosely, we call both $\EQ(x)$ and $\mathbf{Q}(x)$ the $\mathbf{Q}$-observable.

\subsection{$\tilde{\mathbf{Q}}$-observable} \label{subsec:dualqobsapp}
We consider the following gauge origami setup where the $\BZ_3$ charge assignment for the $\hat{N}_{14}$ is slightly different: 
\begin{align}
\begin{split}
    & \hat{N}_{12} = \sum_{\alpha=1}^N e^{a_\alpha} \cdot \CalR_0 + \sum_{\alpha=1}^N e^{m_\alpha^--\ve_4} \cdot \CalR_1 + \sum_{\alpha=1}^N e^{m_\alpha^+-\ve_3} \CalR_2,  \\
    & \hat{N}_{14} = e^{x}q_1 \cdot \CalR_0.
\end{split}
\end{align}
We impose freezing $\kq_1=\kq_2=0$. Then $\hat{K}_{14}$ grows only along the direction of $\BC_1$, yielding 
\begin{align}
    P_1 \hat{K}_{14} = e^{x} q_1 (1-q_1^d),
\end{align}
where $d = |\hat{K}_{13}|$. The spiked instanton partition function gives
\begin{align}
\begin{split}
    \CalZ_S 
    & = \sum_{{\boldsymbol\lambda}} \sum_{d=0}^\infty \kq^{|{\boldsymbol\lambda}| + d} \BE \left[ - \frac{\hat{P}_3\hat{S}_{12} \hat{S}_{12}^* }{\hat{P}_{12}^*} - \frac{\hat{P}_2 \hat{S}_{14}\hat{S}_{14}^* }{\hat{P}_{14}^*} + \hat{q}_2\hat{P}_3 \frac{ \hat{S}_{14}\hat{S}_{12}^* }{\hat{P}_1^*} \right] \\
    & = \sum_{\boldsymbol\lambda} \kq^{|\boldsymbol\lambda|} \CalZ[{\boldsymbol\lambda}] \sum_{d=0}^\infty \prod_{j=1}^k \frac{j\ve_1+\ve_2}{j\ve_2} \frac{\EQ(x) M(x+d\ve_1)}{Q(x+d\ve_1)Q(x+d\ve_1+\ve_+)} \\
    & =: \sum_{\boldsymbol\lambda} \kq^{|\boldsymbol\lambda|} \CalZ[{\boldsymbol\lambda}] {\tilde{Q}(x)[\lambda]}.
\end{split}
\end{align}
The $\tilde{Q}$-observable associated to instanton configuration $\boldsymbol\lambda$ is
\begin{align}
\begin{split}
    \tilde\EQ(x) &= \qe^{\frac{x}{\ve_1}}\sum_{d=0}^\infty \qe^d \prod_{j=1}^k \frac{j\ve_1+\ve_2}{j\ve_2} \frac{\EQ(x)[\boldsymbol\lambda] M(x+d\ve_1)}{Q(x+d\ve_1)[\boldsymbol\lambda]Q(x+(d+1)\ve_1+\ve_2)[\boldsymbol\lambda]} \\
    & =\qe^{\frac{x}{\ve_1}} \sum_{d=0} ^\infty \qe^d \left( \prod_{j= 1}^d \frac{j\ve_1 + \ve_2}{j \ve_1} \frac{P^+ (x+j\ve_1)}{\EY(x+j\ve_1)} \right)  \frac{M^- (x+d\ve_1)}{Q(x+ \ve_2 + (d+1)\ve_1)}.
\end{split}
\end{align}
We often separate the 1-loop part $M^+ (x) = \BE\left[ -\frac{e^x M^+}{P_1 ^*} \right]$ and write
\begin{align}
    \tilde{\mathbf{Q}} (x) = \qe ^{\frac{x}{\ve_1}} \sum_{d=0} ^\infty \qe^d \left( \prod_{j= 1}^d \frac{j\ve_1 + \ve_2}{j \ve_1} \frac{1}{\EY(x+j\ve_1)} \right)  \frac{M (x+d\ve_1)}{Q(x+ \ve_2 + (d+1)\ve_1)}.
\end{align}
Namely, $ \tilde{\EQ}(x)= \frac{\tilde{\mathbf{Q}} (x)}{M^+ (x)}$. Somewhat loosely, we call both $\tilde{\EQ}(x)$ and $\tilde{\mathbf{Q}}(x)$ the $\tilde{\mathbf{Q}}$-observable.

\subsection{Monodromy surface defect} \label{subsec:orimono}

Consider the gauge origami \eqref{def:GO-setup} in the presence of regular monodromy defect: 
\begin{align}
    \hat{N}_{12} = \sum_{\o=0}^{N-1} e^{a_\o} \hat{q}_2^\o \CalR_0 \otimes \fR_\o + q_{13} M_\o^- \hat{q}_2^{\o+1} \CalR_1 \otimes \fR_{\o+1} + q_3^{-1} M_\o^+ \hat{q}_2^\o \CalR_2 \otimes \fR_\o
\end{align}
The universal sheaf of the instanton on the orbifold, carrying $\BZ_N$ representation, can be written as
\begin{align}
    \hat{S}_{12} = \sum_{\o=0}^{N-1} S_\o \hat{q}_2^\o \CalR_0 \otimes \fR_\o + q_{13} M_\o^- \hat{q}_2^{\o+1} \CalR_1 \otimes \fR_{\o+1} + q_3^{-1} M_\o^+ \hat{q}_2^{\o} \CalR_2 \otimes \fR_{\o}
\end{align}
where we defined $S_\o = N_\o - P_1K_\o + q_2^{\d_{\o,0}} P_1K_{\o-1}$, $M_\o^\pm = e^{m^\pm_\o}$. We get the universal
sheaf for the instantons in the absence of the orbifold and the flavor bundle by
\begin{align}
    S = \sum_{\o=0}^{N-1} S_\o = N - P_{12} K_{N-1}, \quad M^\pm = \sum_{\o=0}^{N-1} M_\o^\pm. 
\end{align}

The spiked instanton partition function is given by
\begin{align}
\begin{split}
    \hat\CalZ & = \sum_{\hat{\boldsymbol\lambda}} \prod_{\o=0}^{N-1} \kq_\o^{|K_\o|} \BE \left[ - \frac{\hat{P}_3\hat{S}_{12}\hat{S}_{12}^*}{\hat{P}_{12}^*} \right]^{\BZ_3\times\BZ_N} \\
    & = \sum_{{\boldsymbol\lambda}} \kq^{|{\boldsymbol\lambda}|} \BE \left[ \frac{-SS^*+M^+S^*}{P_{12}^*} + \frac{S(M^-)^*}{P_{12}} \right] \\
    & \qquad \times \sum_{\bL \in \rho^{-1}({\boldsymbol\lambda})} \prod_{\o=0}^{N-2} \kq_\o^{k_\o-k_{N-1}} \BE \left[ \sum_{\o_1<\o_2} \frac{S_{\o_1}S_{\o_2}^*}{P_{1}^*} - \frac{M_{\o_1}^+S_{\o_2}^*}{P_1^*} - \frac{S_{\o_2}(M_{\o_1}^-)^*}{P_1} \right] \\
    & = \langle \Psi_{\mathbf{c}}(\bu) \rangle \CalZ
\end{split}
\end{align}
Here, $\rho$ is a projection from colored partitions $\bL=(\Lambda^{(0)},\dots,\Lambda^{(N-1)})$ to ordinary partitions $\boldsymbol\lambda=(\lambda^{(0)},\dots,\lambda^{(N-1)})$, $\r : \{\hat{\bl}\} \to \{\bl\}$, defined by 
\begin{equation}
     \rho(\Lambda^{(\alpha)}_i) =\left\lfloor\frac{\Lambda^{(\alpha)}_i+c(\alpha)}{N}\right\rfloor,
\end{equation}
where $\lfloor\cdots\rfloor$ is the floor function. Define
\begin{align}
    \Sigma_\o = S_{\o+1} + \cdots + S_{N-1}, \quad  \Sigma_{\o-1} - \Sigma_{\o} = S_\o. 
\end{align}
The surface observable is obtained as
\begin{align} \label{eq:monoapp}
\begin{split}
    \Psi_{\mathbf{c}}(\bu) = & \sum_{\bL \in \rho^{-1}({\boldsymbol\lambda})} \prod_{\o=0}^{N-2} \kq_\o^{k_\o-k_{N-1}} \BE \left[ \sum_{\o=0}^{N-2} \frac{ (S - \Sigma_\o) (\Sigma_{\o}-\Sigma_{\o+1})^* }{P_{1}^*} - \frac{M_{\o}^+\Sigma_{\o}^*}{P_1^*} - \frac{\Sigma_{\o}(M_{\o}^-)^*}{P_1} \right] .
\end{split}
\end{align}

\section{Separation of variable kernel $\EuScript{K}(\bu;\by)$ for general $N$} \label{sec:sovkernel}
It will be convenient for us to set $x_{N-\o,N-j}$, $\o=1,\dots,N-1$, $j=1,\dots,\o$, and proceed starting from $\o=1$ up to $\o=N-1$.

We showed explicit computations for $\o=1$ in section \ref{eq:subsubsec:slngaudinsov}. Now, there will be $\o+1$ $\G$-functions in the numerator for $x_{N-\o,N-j}$, $\o=2,\dots,N-1$ taking the form of
$$
    \prod_{i=1}^{\o+1} \Gamma \left( \frac{x_{N-\o,N-j}-x_{N-\o-1,N-k}}{\ve_1} \right) 
$$
The assignment can be represented by a $\o \times (\o+1)$ rectangle matrix $M_{j,k}^{(\o)}$, whose matrix components are either $0$ or $1$, satisfying the following condition:
\begin{itemize}
    \item $M_{j,j}^{(\o)}=0$ for $j=1,\dots,\o$. 
    \item The number of $1$ in each column must be even. 
    \item The number of $1$ in each row must be $\o-1$. 
    \item The total number of non-zero entries in $M_{j,k}^{(\o)}$ is $\o(\o-1)$. 
    \item There can not be two columns in $M^{(\o)}_{j,k}$ that are identical when $\o>2$. 
\end{itemize}
This will leave us with the only two zero-entries on each row of $M_{j,k}^{(\o)}$: $k=j, \ k^{(\o)}_j$ for some $k^{(\o)}_j$. 

We list the first few $M^{(\o)}$:

\begin{subequations}
\begin{align}
    M^{(2)} : &  \begin{pmatrix}
        0 & 0 & 1 \\ 0 & 0 & 1
    \end{pmatrix}; \\
    M^{(3)} : & \begin{pmatrix}
        0 & 1 & 1 & 0 \\ 1 & 0 & 1 & 0 \\ 1 & 1 & 0 & 0
    \end{pmatrix}, \quad \begin{pmatrix}
        0 & 1 & 0 & 1 \\ 1 & 0 & 0 & 1 \\ 1 & 1 & 0 & 0
    \end{pmatrix}, \quad \begin{pmatrix}
        0 & 0 & 1 & 1 \\ 1 & 0 & 0 & 1 \\ 1 & 0 & 1 & 0
    \end{pmatrix}, \quad \begin{pmatrix}
        0 & 1 & 1 & 0 \\ 0 & 0 & 1 & 1 \\ 0 & 1 & 0 & 1
    \end{pmatrix}; \\
    M^{(4)} : & 
    \begin{pmatrix}
        0 & 0 & 1 & 1 & 1 \\ 
        1 & 0 & 0 & 1 & 1 \\ 
        1 & 1 & 0 & 0 & 1 \\ 
        0 & 1 & 1 & 0 & 1
    \end{pmatrix}, \quad \begin{pmatrix}
        0 & 1 & 1 & 0 & 1 \\ 
        0 & 0 & 1 & 1 & 1 \\ 
        1 & 0 & 0 & 1 & 1 \\ 
        1 & 1 & 0 & 0 & 1
    \end{pmatrix};
\end{align}
\end{subequations}
Notice that when $\o$ is even, all components of $M_{j,\o+1}^{(\o)}$ must be 1. The argument is the following: In the sub-square matrix, the maximal number of 1's are $\o(\o-2)$ as the number of $1$ in a single column must be even.

Note that a matrix satisfying all the first 4 conditions but not the last one does not give a legitimate pairing. The first of such matrix occurs at $\o=4$:
\begin{align}
    \begin{pmatrix}
        0 & 1 & 0 & 1 & 1 \\ 
        1 & 0 & 1 & 0 & 1 \\
        0 & 1 & 0 & 1 & 1 \\
        1 & 0 & 1 & 0 & 1
    \end{pmatrix}.
\end{align}
The third/fourth column is the same as the first/second column. 

We want to use \eqref{eq:useful-gamma-3} such that all $x_{N-\o-1,N-k}$ should only show up in the $\G$-function and powers of integration variables, which would require $c-a=d-b$. To guarantee it always happens, we multiply 
\begin{align}
    1 = \prod_{j=1}^{\o} \left[ \frac{\Gamma\left( \frac{x_{N-\o,N-j}-f^{(\o)}}{\ve_1} \right)}{\Gamma\left( \frac{x_{N-\o,N-j}-f^{(\o)}}{\ve_1} \right)} \right]^{\o-1}.
\end{align}
The pairing is done in the following way: 
\begin{itemize}
    \item Define a set $\mathbb{S}^{(\o)}_{k} := \{j| M^{(\o)}_{j,k} = 1 \}$ that collects the rows in the $k$-th column of the matrix $M_{j,k}^{(\o)}$ element being 1. The size of the set $\mathbb{S}^{(\o)}_{k}$ is always even according to how $M^{(\o)}_{j,k}$ is defined.
    \item We pair $(j,j')_k$ , $j<j'$, of all elements $j,j' \in \mathbb{S}_k^{(\o)}$ for all $k=1,\dots,\o+1$. (Notice that by the construction of $M^{(\o)}_{j,k}$, $j\neq k \neq j'$)
    \item One can not make a pair $(j,j')_{k}$ for $\mathbb{S}^{(\o)}_{k}$ if a pairing $(j,j')_{k'}$ already exist for another $\mathbb{S}^{(\o)}_{k'}$, $k \neq k'$. We obtain a collection of paring $\mathbb{B}^{(\o)}$ for each $\o$. $|\mathbb{B}^{(\o)}| = \frac{\o(\o-1)}{2}$.
    \item For a pair $(j,j')_k$ defined, we will perform the integral transformation \eqref{eq:useful-gamma-3} on 
    \begin{align}
    \begin{split}
        & \frac{\Gamma \left( \frac{x_{N-\o,N-j}-x_{N-\o-1,N-k}}{\ve_1} \right) \Gamma \left( \frac{x_{N-\o,N-j'}-x_{N-\o-1,N-k}}{\ve_1} \right)}{\Gamma \left( \frac{x_{N-\o,N-j}-f^{(\o)}}{\ve_1} \right)\Gamma \left( \frac{x_{N-\o,N-j'}-f^{(\o)}}{\ve_1} \right)} \frac{x_{N-\o,N-j}-x_{N-\o,N-j'}}{\ve_1} \\
        = & \frac{1}{\Gamma\left( \frac{x_{N-\o-1,N-k}-f^{(\o)}}{\ve_1} \right) \Gamma\left( \frac{x_{N-\o-1,N-k}-f^{(\o)}}{\ve_1} - 1 \right) } \\
        & \times \int_{\text{PC}^{\times 2} } t_{(j,j')_k,1}^{\frac{x_{N-\o,N-j}-x_{N-\o-1,N-k} }{\ve_1} - 1} t_{(j,j')_k,2}^{\frac{x_{N-\o,N-j'}-x_{N-\o-1,N-k} }{\ve_1}} \\
        & \qquad \times [(1-t_{(j,j')_k,1})(1-t_{(j,j')_k,2})]^{\frac{x_{N-\o-1,N-k}-f^{(\o)}}{\ve_1} - 2} (t_{(j,j')_k,1}-t_{(j,j')_k,2}) dt_{(j,j')_k,1} dt_{(j,j')_k,2}
    \end{split}
    \end{align}
\end{itemize}

By performing \eqref{eq:useful-gamma-3} to all the pairing generated by $M^{(\o)}_{j,k}$ in $\mathbb{B}^{(\o)}$
\begin{align}
\begin{split}
    & \prod_{k=1}^{\o+1} \left[ \Gamma\left( \frac{x_{N-\o-1,N-k}-f^{(\o)}}{\ve_1} \right) \Gamma\left( \frac{x_{N-\o-1,N-k}-f^{(\o)}}{\ve_1} - 1 \right) \right]^{-\frac{1}{2} \sum_{j=1}^\o M^{(\o)}_{j,k} } \\
    & \times \prod_{(j,j')_k \in \mathbb{B}^{(\o)} } \int_{\text{PC}^{\times 2} } t_{(j,j')_k,1}^{\frac{x_{N-\o,N-j}-x_{N-\o-1,N-k} }{\ve_1} - 1} t_{(j,j')_k,2}^{\frac{x_{N-\o,N-j'}-x_{N-\o-1,N-k} }{\ve_1}} \\
    & \qquad \qquad \times [(1-t_{(j,j')_k,1})(1-t_{(j,j')_k,2})]^{\frac{x_{N-\o-1,N-k}-f^{(\o)}}{\ve_1} - 2} (t_{(j,j')_k,1}-t_{(j,j')_k,2}) dt_{(j,j')_k,1} dt_{(j,j')_k,2}
\end{split}
\end{align}
The integration degree is $2 \times |\mathbb{B}^{(\o)}|= \o (\o-1)$.

Finally we will use the ensemble for $x_{N-\o,N-j}=x_{N-\o-1,j}-s_{N-j,N-\o}\ve_1$ dependence, we denote $F_{\o,j}(\bt_{\o,j})$ of the function in all previous $t_{(j,j')_k,l}$, $l=1,2$ that has $x_{N-\o,N-j}$ as its power:
\begin{align}
\begin{split}
    & \prod_{j=1}^{\o} \sum_{s_{N-j,N-\o}=0}^\infty (\kq_\o\cdots\kq_j F_{\o,j}(\bt_{\o,j}))^{ - \frac{x_{N-\o-1,N-j}}{\ve_1} + s_{N-j,N-\o}} \\
    & \qquad \frac{\Gamma\left( 1 + s_{N-j,N-\o} - \frac{x_{N-\o-1,N-j}-{m}_{N-\o-1}^-}{\ve_1} \right) \Gamma\left( - s_{N-j,N-\o} + \frac{x_{N-\o-1,N-j}-x_{N-\o-1,N-k^{(\o)}_j}}{\ve_1} \right) }{\Gamma\left( 1 +s_{N-j,N-\o} \right) \Gamma\left( - n_{\o,j} + \frac{x_{N-\o-1,N-j}-{m}_{N-\o-1}^+}{\ve_1} \right) } \\
    & = \prod_{j=1}^{\o} \Gamma \left( 1 - \frac{y_{\o+1,j}-{m}_\o^+}{\ve_1} \right) \Gamma \left( 1 - \frac{y_{\o+1,k^{(\o)}_j}-{m}_\o^-}{\ve_1} \right) \\ 
    &\qquad \times \int_{\text{PC}} t_{\o,j,0}^{-\frac{y_{\o+1,j}-{m}_\o^-}{\ve_1}}  (1-t_{\o,j,0})^{-\frac{y_{\o+1,k^{(\o)}_j}-{m}_j^+}{\ve_1}-1} (1-\kq_\o\dots\kq_j F_{\o,j}(\bt_{\o,j}) t_{\o,j,0})^{\frac{y_{\o+1,j}-{m}_\o^+}{\ve_1}-1} \ d t_{\o,j,0} \\
    & = \frac{e^ {-\pi\ri \left( 1+\frac{x_{N-2,N-2}-x_{N-2,N-1} }{\ve_1} \right) } }{4\pi} 
    \frac{ \Gamma \left( \frac{m_{N-\o-1}^--x_{N-\o-1,N-k^{(\o)}_j }}{\ve_1} \right) \Gamma \left( \frac{m_{N-\o-1}^+-x_{N-\o-1,N-j}}{\ve_1} \right) }{\sin\pi \left( \frac{x_{N-\o-1,N-j}-x_{N-\o-1,N-k_j^{(\o)}} }{\ve_1} \right)} \\
    &\times \int_{\text{PC}} t_{\o,j,0}^{-\frac{x_{N-\o-1,N-j}-{m}_{N-\o-1}^-}{\ve_1}-1}  (1-t_{\o,j,0})^{-\frac{x_{N-\o-1,N-k^{(\o)}_j}-{m}_{N-\o-1}^+}{\ve_1}} (1-\kq_\o\dots\kq_jF_{\o,j}(\bt_{\o,j}) t_{\o,j,0})^{\frac{x_{N-\o-1,N-j}-{m}_{N-\o-1}^+}{\ve_1}} \ d t_{\o,j,0}
\end{split} 
\end{align}
The $\G$-factor can be organized to
\begin{align}
\begin{split}
    & \prod_{j=1}^{\o} \Gamma \left( - \frac{x_{N-\o-1,N-j}-{m}_{N-\o-1}^+}{\ve_1} \right) \Gamma \left( - \frac{x_{N-\o-1,k^{(\o)}_j}-{m}_{N-\o-1}^-}{\ve_1} \right) \\
    & = \prod_{j=1}^{\o} \Gamma \left( - \frac{x_{N-\o-1,N-j}-{m}_{N-\o-1}^+}{\ve_1} \right) \prod_{i=1}^{\o+1} \Gamma \left( - \frac{x_{N-\o-1,N-i} - {m}_{N-\o-1}^-}{\ve_1} \right)^{\sum_{j=1,j\neq i}^{\o} (1-M_{j,i}^{(\o)}) } 
\end{split}
\end{align}

Up to this moment, we still have $\G$-factors generated from the last level $\o-1$: 
\begin{align}
\begin{split}
    & \frac{ \prod\limits_{j=1}^{\o-1} \Gamma \left( - \frac{x_{N-\o,N-j}-{m}_{N-\o}^+}{\ve_1} \right) \prod\limits_{k=1}^{\o} \Gamma \left( - \frac{x_{N-\o,N-k} - {m}_{N-\o}^-}{\ve_1} \right)^{\sum\limits_{{j=1, j\neq k}}^{\o-1} (1-M_{j,k}^{(\o-1)}) } } 
    { \prod\limits_{k=1}^{\o} \left[ \Gamma\left( \frac{x_{N-\o,N-k}-f^{(\o-1)}}{\ve_1} \right) \Gamma\left( \frac{x_{N-\o,N-k}-f^{(\o-1)} }{\ve_1} - 1 \right) \right]^{\frac{1}{2} \sum\limits_{j=1}^{\o-1} M^{(\o-1)}_{j,k} } } 
    \prod_{k=1}^\o \Gamma \left( \frac{x_{N-\o,N-k}-f^{(\o)}}{\ve_1} \right)^{\o-1},
\end{split}
\end{align}
which finally can be rewritten in $\o \times (\o-1)$ integrals using the Pochhammer contour integral \eqref{eq:betaint}. This can be seen by counting the numbers of $\G$-function in both the numerator (one need to use the reflection formula for the $\G$-functions with negative arguments in $x_{N-\o,N-j}$ before proper counting). The total number of $\G$-functions in the denominator with positive argument of $x_{N-\o,N-j}$ and $\G$-functions with negative argument of $x_{N-\o,N-j}$ in the numerator is
\begin{align}
\begin{split}
    & 2 \times \sum_{j=1}^{\o-1} \sum_{k=1}^{\o} \frac{1}{2} M_{j,k}^{(\o-1)} + \sum_{k=1}^\o \sum_{j= 1, j\neq k}^{\o-1} (1-M_{j,k}^{(\o-1)} ) + \o-1 \\
    & = (\o-1)(\o-2) + (\o-1)^2 - (\o-1)(\o-2) + \o - 1 \\
    & = \o(\o-1).
\end{split}
\end{align}
It equals to the total number of $\G$-function in the numerator with positive argument of $x_{N-\o,N-k}$. The so-generated $\G$-functions are not $x_{N-\o,N-j}$ dependent.

\bibliographystyle{utphys}
\bibliography{reference}

\end{document}